%% file: template.tex
\documentclass{article}

\usepackage{arxiv}

\usepackage[utf8]{inputenc} % allow utf-8 input
\usepackage[T1]{fontenc}    % use 8-bit T1 fonts
\usepackage{hyperref}       % hyperlinks
\usepackage{url}            % simple URL typesetting
\usepackage{booktabs}       % professional-quality tables
\usepackage{amsfonts}       % blackboard math symbols
\usepackage{nicefrac}       % compact symbols for 1/2, etc.
\usepackage{microtype}      % microtypography
\usepackage{lipsum}		% Can be removed after putting your text content
\usepackage{graphicx}
\usepackage{doi}
\usepackage[round]{natbib}
\usepackage[]{appendix} % Include appendix package, add entries to ToC and restart page numbering
\usepackage{multirow}
\title{Nowcasting R\&D Expenditures:\\ \emph{A Machine Learning Approach}}

\author{ \href{https://orcid.org/0009-0008-3537-6768}{\includegraphics[scale=0.06]{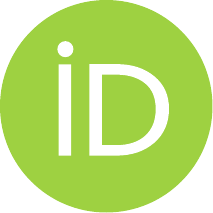}\hspace{1mm}Atin ~Aboutorabi} \\
	EPFL\\
	\texttt{atin.aboutorabi@epfl.ch} \\
	\And
	\href{https://orcid.org/0000-0002-7862-0918}{\includegraphics[scale=0.06]{orcid.pdf}\hspace{1mm}Gaétan ~de Rassenfosse} \\
	EPFL\\
	\texttt{gaetan.derassenfosse@epfl.ch} \\
}

\renewcommand{\headeright}{\textit{arXiv}}
\renewcommand{\undertitle}{}
\renewcommand{\shorttitle}{Nowcasting R\&D Expenditures}

\hypersetup{
pdftitle={Nowcasting R\&D},
pdfsubject={q-bio.NC, q-bio.QM},
pdfauthor={Atin ~Aboutorabi, Gaétan ~de Rassenfosse},
pdfkeywords={Economic Nowcasting, Machine Learning, Google trends, R\&D Expenditures},
}

\begin{document}
\maketitle

%%%%%%%%%%%%%%%%%%%%%%%%%%%%%%%%%%%%%%%%%%%%%%%%%%%
%% Abstract
%%%%%%%%%%%%%%%%%%%%%%%%%%%%%%%%%%%%%%%%%%%%%%%%%%%
\begin{abstract}
\label{sec:abstract}
Macroeconomic data are crucial for monitoring countries' performance and driving policy. However, traditional data acquisition processes are slow, subject to delays, and performed at a low frequency. We address this `ragged-edge' problem with a two-step framework. The first step is a supervised learning model predicting observed low-frequency figures. We propose a neural-network-based nowcasting model that exploits mixed-frequency, high-dimensional data. The second step uses the elasticities derived from the previous step to interpolate unobserved high-frequency figures. We apply our method to nowcast countries' yearly research and development (R\&D) expenditure series. These series are collected through infrequent surveys, making them ideal candidates for this task. We exploit a range of predictors, chiefly Internet search volume data, and document the relevance of these data in improving out-of-sample predictions. Furthermore, we leverage the high frequency of our data to derive monthly estimates of R\&D expenditures, which are currently unobserved. We compare our results with those obtained from the classical regression-based and the sparse temporal disaggregation methods. Finally, we validate our results by reporting a strong correlation with monthly R\&D employment data.

\end{abstract}

%%%%%%%%%%%%%%%%%%%%%%%%%%%%%%%%%%%%%%%%%%%%%%%%%%%
%% Introduction
%%%%%%%%%%%%%%%%%%%%%%%%%%%%%%%%%%%%%%%%%%%%%%%%%%%
\section{Introduction}
\label{sec:intro-00}
Three megatrends are transforming the forecasting literature: the burgeoning of machine learning (ML) methods, the expanding volume of available data, and the increasing computing ability. These trends give rise to new classes of prediction models that can process more and new types of data. Notable examples of new data include satellite data \citep{econ_growth_satellite, DIEBOLD20211509}, textual data \citep{bollen2011twitter, de2020incorporating}, and price data from online merchants to forecast inflation \citep{cavallo2016billion}. Combined, these trends lead to more accurate predictions or enable the prediction of new attributes \citep{DL_stock_price_2023, news_financial_market_2023, knowledge_discovery_2023}.

\begin{figure}[!htb]
    \centering
    \caption{Proposed Framework for Nowcasting R\&D Expenditures.}
     {\includegraphics[width=0.7\linewidth]{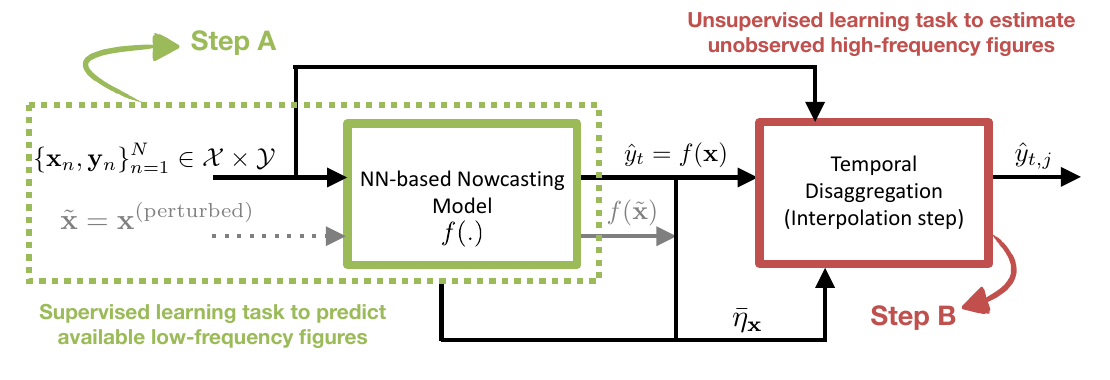}}
    \label{fig:framework}
\end{figure}

Our paper contributes to this research line. We introduce a framework, illustrated in Figure \ref{fig:framework}, in response to the so-called `ragged-edge' problem---publication delays of headline variables in official statistics \citep{mosley2022sparse}, and their low-frequency. This two-step framework includes one step for a supervised learning task to predict observed low-frequency figures (\textit{step A}) and an interpolation step for an unsupervised learning task to estimate unobserved high-frequency figures (\textit{Step B}). The \textit{step A} is a neural network-based model that incorporates low-frequency data and high-frequency data---in the form of web-search data---to nowcast countries’ R\&D expenditures. As innovation is the prime engine of economic growth \citep{romer1990endogenous, aghion2008economics}, R\&D investments are a central policy metric. However, they are poorly monitored, being based on expensive surveys and released yearly—sometimes even biennially—with a publication lag of two to three years. This delay hinders effective policy-making. As government policies seek to stimulate R\&D investments \citep{edler2017innovation} and institutions track countries' innovation performances \citep{WIPO2023}, our inability to produce accurate estimates of recent R\&D investments is particularly prejudicial \citep{oecd2009}. How can countries pilot the EU’s target of reaching R\&D investments amounting to 3 percent of GDP by 2030 \citep{ECommission2020} if the data have such shortcomings?

The \textit{step B} involves interpolating the R\&D series at a higher frequency than currently available. We exploit the high granularity of our input data to allocate yearly investments into monthly figures. Such data can be used to provide a better understanding of how R\&D investments react to economic shocks \citep{oecd2009} or policy stimuli \citep{guellec2003policy, gonzalez2008policy, tajaddini2021economicpolicyuncertain}. The method that we propose is also an important first step towards the nowcasting of monthly series on R\&D expenditures. 

Our work builds on a handful of studies that have sought to nowcast economic activities using ML methods, including \citet{sokolov2016MLwNonGT}, \citet{dai2017MLwNonGT}, and \citet{TUMER2018MLwNonGT}.
It also relates to \citet{preis2013quantifying}, \citet{BORUP20231122}, and \citet{woloszko2020tracking} that exploit web-search data, especially Google trends data, for prediction purposes. 
Finally, it relates to the literature that has sought to model R\&D dynamics \citep{bloom2007uncertainty} or forecast R\&D and innovation activities \citep{cheng2005rd_forecast_quantitative_china} using traditional econometric techniques. To the best of our knowledge, this study is the first to introduce a neural network-based nowcasting model for R\&D expenditures. It also sets a new foundation for advancing the nowcasting of R\&D expenditures at a higher frequency. Our method exhibits performance at least as good as existing regression-based temporal disaggregation methods \citep{chowlin1971best, mosley2022sparse}.

As the model leverages the web-search data, we first need to identify the relevant search terms. For that purpose, we conscientiously map the actors in the innovation ecosystem by surveying the `systems of innovation' literature \citep{lundvall2010innosys, edquist2013innosys}. Next, we run horse race models of nowcasting for yearly R\&D series, exploiting both traditional and web-based data. To our surprise, the neural-network-based model exploiting solely web-based data outperforms similar models exploiting solely traditional data, in terms of expressive power (prediction accuracy) and generalization (out-of-sample performance). We then move to high-frequency estimation relying on the output of the neural network model. To be more specific, we propose an interpolation approach, that involves corrupting the input and computing an elasticity value for each input feature based on output distribution. We compare our estimates to the well-known regression-based temporal disaggregation methods. Also, we validate them by using monthly figures for employment in scientific R\&D services, reporting significant correlation with our R\&D series.

The rest of the paper is organized as follows. Section \ref{sec:intro} discusses the background literature. Section \ref{sec:model_annual} outlines the neural network-based nowcasting model for yearly frequency. Section \ref{sec:data} describes the data used for the empirical analysis, and Section \ref{sec:results} discusses the results. Section \ref{sec:model_bem} extends the nowcasting model using temporal interpolation techniques, and reports the results and its respective validations. Finally, Section \ref{sec:conclusion} concludes and points out possible extensions.

%%%%%%%%%%%%%%%%%%%%%%%%%%%%%%%%%%%%%%%%%%%%%%%%%%%
%% Background
%%%%%%%%%%%%%%%%%%%%%%%%%%%%%%%%%%%%%%%%%%%%%%%%%%%
\section{Background}
\label{sec:intro}
Macroeconomic variables play a crucial role in guiding global economic decisions. Covering key facets of economic activity, they offer invaluable insights to policymakers and businesses about an economy's health and direction. Despite their importance, traditional methods of data acquisition have limitations. They rely on expensive surveys and low-frequency updates, potentially obscuring rapidly changing economic conditions.

Acknowledging these limitations, scholars have sought to `nowcast' economic activity. Nowcasting aims to predict the present, the near future, or the recent past, thereby addressing the growing demand for real-time economic insights \citep{banbura2010nowcasting}. The first generation of nowcasting models were classical statistical models exploiting macroeconomic variables \citep{evans2005we, giannone2008nowcasting, banbura2010nowcasting}. For instance, the seminal work by \cite{giannone2008nowcasting} proposes a factor model predicting economic activity as captured by GDP. While these models have pioneered this line of work, their reliance on macroeconomic data, to some extent, defeats their purpose by failing to capture the swiftly changing economic landscape, let alone potential economic shocks.

A subsequent line of inquiry has addressed this limitation by utilizing high-frequency data sources to complement the traditional ones and developing newer, more agile prediction models. Concerning data, scholars have shown that the use of high-frequency Internet search volume, such as Google trends, improves economic forecasting \citep{choi2009predicting, choi2012predicting, wu2015future, ferrara2019google, gotz2019google}.\footnote{Google trends is an analytical tool that quantifies search intensities for any given search term(s) by geographical area and over any selected time period. See Section \ref{sec:GT} for details.}

Advances in econometric modeling have seen the rise of (restricted) Mixed Data Sampling (MIDAS) approaches \citep{ghysels2004midas} and unrestricted MIDAS (U-MIDAS) approaches \citep{foroni2015unrestricted}, which offer novel ways of incorporating data of varying frequencies into macroeconomic analysis. MIDAS models employ distributed lag-polynomials to integrate high-frequency predictors into a low-frequency modeling framework \citep{ghysels2004midas, clements2008midasmacroeconomic, BORUP20231122}. U-MIDAS models, on the other hand, do not impose a lag-polynomial structure and do not require the alignment of the frequency of predictors and target variables, thereby allowing for more direct inclusion of high-frequency data  \citep{foroni2015unrestricted, BORUP20231122}.

Finally, recent advances in artificial intelligence (AI) have reinforced the shift from classical econometric models to machine learning techniques for forecasting \citep{BORUP20231122, borup2022search, woloszko2023nowcasting}. Building on these developments, we propose a neural-network-based nowcasting model that leverages high-frequency (search volume) data and low-frequency (macroeconomic) data. By constructing different configurations for the input space, we blend the same-frequency data sampling and U-MIDAS approaches.

%%%%%%%%%%%%%%%%%%%%%%%%%%%%%%%%%%%%%%%%%%%%%%%%%%%
%% Nowcasting MODEL
%%%%%%%%%%%%%%%%%%%%%%%%%%%%%%%%%%%%%%%%%%%%%%%%%%%
\section{A mixed-frequency neural network-based nowcasting model}
\label{sec:model_annual}

\subsection{General set-up}
\label{sec:theoretical_model}
We consider a target ($y \in \mathbb{R}$) sampled at a yearly frequency and a vector of predictors ($\mathbf{x} \in \mathbb{R}^d$) containing mixed frequency input, \textit{i.e.}, some variables are sampled monthly and others yearly. The pair $(\mathbf{x}, y)$ represents any given data point (observation) for which the general prediction model is given by:

\begin{equation*}
y = f(\mathbf{x}) + \mathbf{\varepsilon}
\end{equation*}

in which $ \varepsilon $ is a zero-mean error term. The target value denotes the R\&D expenditures for each country $i$ at year $t$, and the input vector ($\mathbf{x}$) consists of four main components. We categorize these components into the following vectors: autoregressive (AR) terms of the target ($\mathbf{Y}_{t-\tau, i} \in \mathbb{R}^{2 \tau}$), search volume data, \textit{i.e.} Google trends data ($\mathbf{S}_{t, j, i} \in \mathbb{R}^ {(1 + \tau) \cdot k_s}$), macroeconomic variables ($\mathbf{Z}_{t-\tau, i} \in \mathbb{R}^{\tau \cdot k_z}$), and the general categorical features ($\mathbf{C}_{j, i} \in \mathbb{R}^ {k_c}$), as defined further below. We can re-write the prediction model as:  

\begin{equation}
\label{eq:baseline}
y_{t, i} = f^{(m)}\left(\mathbf{Y}_{t-\tau, i},\ \mathbf{S}_{t, j, i}^{(m)},\ \mathbf{Z}_{t-\tau, i},\ \mathbf{C}_{j, i}^{(m)} ; \ \boldsymbol{\theta}^{(m)}\right) + \ \varepsilon_{t, j, i}^{(m)}.
\end{equation}

This model accounts for variations that occur on a monthly basis thanks to the inclusion of monthly-level features in the predictor set, as indicated by superscript $(m)$. Note also the presence of the superscript on the model $f(.)$ itself, highlighting the fact that the model contains features with a different sampling frequency than the target value. 
Finally, $\boldsymbol{\theta}$ represents the corresponding vector of model parameters.

We now discuss each of the four components of the model. The vector $\mathbf{Y}_{t-\tau, i}$ includes $\tau > 0$ lagged values of the target to account for potential serial correlation in $y_{t, i}$, without adding any forward-looking (look-ahead) bias:\footnote{After having tested various lags, the empirical analysis will consider $\tau=3$. Note that even with less than $\tau=3$ and only considering the final state ($\tau=1$), the out-of-sample performance of the model is not significantly different, showing only minor improvement from $\tau=1$ to $\tau=3$ (not reported). However, due to the greater comprehensiveness of considering more lags and in order to evaluate properly the ability of neural networks to handle high-dimensional data, we present the empirical results for $\tau=3$.}

\begin{equation*}
\mathbf{Y}_{t-\tau, i} = \left[\begin{array}{lll}
y_{t-1, i} \ \ldots \ y_{t-\tau, i} \quad y'_{t-1, i} \ \ldots \ y'_{t-\tau, i}
\end{array}\right]^{\top}.
\end{equation*}

In addition to the AR terms $y_{t, i}$, we incorporate a binary variable associated with each AR term $y'_{t, i}$, that indicates missing values on the $y_{t, i}$ variable. In case of missing value for any of the AR terms, we estimate it using linear interpolation (see Section \ref{sec:data_rd}).

The second vector of predictors, $\mathbf{S}_{t, j, i}^{(m)}$, consists of monthly data obtained from Google trends. The data contain $k_{s}$ number of different topics associated with a selection of search terms, as we elaborate in Section \ref{sec:data}.\footnote{Google trends provides \textit{related topics} by associating some standardized groups to the search terms. The topics are more reliable than search terms because they include not only the exact search terms, but also their misspellings and acronyms, and they are harmonized over all languages \cite{GoogleTrends2023}. As an example, the search term `R\&D expenditure' relates to the topics \textit{`Research and development', `Innovation', `Technology', `Patents', etc.}} Note the subscript $j$, which indicates values corresponding to the $(12-j)^{th}$ month of the year, with $j = 0, ..., 11$.\footnote{Considering the scalars $s_{k_{s}, t, j, i}$ with $j$ corresponding to the $(12-j)^{th}$ month of the year $t$ for country $i$, is equivalent to $s_{k_{s}, t-j/12, i}$. It would be interpreted as follows: December $s_{k_{s}, t-0/12, i}$, November $s_{k_{s}, t-1/12, i}$, ... , June $s_{k_{s}, t-0.5, i}$, ... , and January $s_{k_{s}, t-11/12, i}$.} Formally, we define  $\mathbf{S}_{t, j, i}^{(m)}$ as follows:

\begin{equation*}
\mathbf{S}_{t, j, i}^{(m)} = \left[\begin{array}{lll}
\mathbf{\bar{S}}_{t-\tau, i} \quad \mathbf{\bar{{S}}}_{t, j, i}^{\text{YTD}^{(m)}}
\end{array}\right]^{\top}.
\end{equation*}

The first sub-vector, $\mathbf{\bar{S}}_{t-\tau, i}$, contains the lagged monthly Google trends values averaged on a yearly frequency for the past $\tau$ years and for $k_{s}$ number of predictors representing various topics associated with the selected search terms. We define:

\begin{equation*}
\mathbf{\bar{S}}_{t-\tau, i} = \left[\begin{array}{lll}
\mathbf{\bar{s}}_{t-1, i} \ \ldots \ \mathbf{\bar{s}}_{t-\tau, i}
\end{array}\right],
\end{equation*}

\begin{equation*}
\mathbf{\bar{s}}_{t-\tau, i} = \left[\begin{array}{lll}
\bar{s}_{1, t-\tau, i} \ \ldots \ \bar{s}_{k_{s}, t-\tau, i}
\end{array}\right],
\end{equation*}

\begin{equation*}
\bar{s}_{k_{s}, t-\tau, i} = \frac{1}{12} \sum_{j=0}^{11} {s}_{k_{s}, t-\tau, j, i}^{(m)}.
\end{equation*}

By aggregating the Google trends data into annual average for each topic $\bar{s}_{k_{s}, t-\tau, i}$ and constituting $\mathbf{\bar{s}}_{t-\tau, i}$ for all topics, we reduce the monthly variability and emphasize long-term trends. Also, by constructing the time-lagged vector $\mathbf{\bar{S}}_{t-\tau, i}$ for $\tau$ lagged values of these averages, we enable the model to capture temporal dynamics across several years and test their associations with the future outcomes.

The second sub-vector of $\mathbf{S}_{t, j, i}^{(m)}$, which is $\mathbf{\bar{S}}_{t, j, i}^{\text{YTD}^{(m)}}$, captures the year-to-date (YTD) Google trends values averaged over the past months of the current year (excluding the current month):

\begin{equation*}
\mathbf{\bar{S}}_{t, j, i}^{\text{YTD}^{(m)}} = \left[\begin{array}{lll}
\bar{s}_{1, t, j, i}^{\text{YTD}^{(m)}} \ \ldots \ \bar{s}_{k_{s}, t, j, i}^{\text{YTD}^{(m)}}
\end{array}\right],
\end{equation*}

\begin{equation*}
\bar{s}_{k_{s}, t, j, i}^{\text{YTD}^{(m)}} = 
\begin{cases} 
\frac{1}{11-j} \sum\limits_{j=j+1}^{11} s_{k_{s}, t, j, i}^{(m)} \quad \forall j \in [0,10], \\[3ex]
0 \hspace{4em} \forall j = 11.
\end{cases}
\end{equation*}

In short, while the subvector $\mathbf{\bar{S}}_{t-\tau, i}$ captures the information from previous year(s), the subvector $\mathbf{\bar{S}}_{t, j, i}^{\text{YTD}^{(m)}}$ captures the information available in the current year $(t)$ up to the previous month that is being estimated. Indeed, this definition implies that for January $(j=11)$, we have $\bar{s}_{k_{s}, t, j, i}^{\text{YTD}^{(m)}}=0$.

The third vector of predictors, $\mathbf{Z}_{t-\tau, i}$, contains the lagged values of $k_{z}$ different macroeconomic variables. We formally define it as: 
\begin{equation*}
\mathbf{Z}_{t-\tau, i} = \left[\begin{array}{lll}
\mathbf{z}_{t-1, i} \ \ldots \ \mathbf{z}_{t-\tau, i}
\end{array}\right]^{\top},
\end{equation*}

\begin{equation*}
\mathbf{z}_{t-\tau, i} = \left[\begin{array}{llll}
z_{1, t-\tau, i} \ \ldots \ z_{k_{z}, t-\tau, i}
\end{array}\right].
\end{equation*}

Finally, the fourth vector of predictors, $\mathbf{C}_{j, i}^{(m)}$, contains features associated with countries and months, which we will discuss further in Section \ref{sec:mlp_setup}:

\begin{equation*}
\mathbf{C}_{j, i}^{(m)} = \left[\begin{array}{llll}
c_{1, i} & c_{2, i} & m_{1}^{(m)} \ \ldots \ m_{12}^{(m)}
\end{array}\right]^{\top}.
\end{equation*}

As a baseline, expressing the model in equation (\ref{eq:baseline}) considering a classical regression setup, in which the input and output data satisfy a linear relation, yields:

\begin{equation*}
y_{t, i} = \mathbf{S}_{t, j, i}^{{(m)}^{\top}} \ \boldsymbol{\omega}_{S}^{(m)} + \mathbf{Y}_{t-\tau, i}^{\top} \ \boldsymbol{\omega}_{AR}^{(m)} + \mathbf{Z}_{t-\tau, i}^{\top} \ \boldsymbol{\omega}_{Z}^{(m)} + \mathbf{C}_{j, i}^{{(m)}^{\top}} \ \boldsymbol{\omega}_{C}^{(m)} + \varepsilon_{t,i}^{(m)},
\end{equation*}

and the vector of model parameters is defined as:

\begin{equation*}
\boldsymbol{\theta}^{(m)} = \left[\begin{array}{llll}
{\boldsymbol{\omega}_{AR}^{(m)}}^{{\top}} \quad {\boldsymbol{\omega}_{X}^{(m)}}^{{\top}} \quad {\boldsymbol{\omega}_{Z}^{(m)}}^{{\top}} \quad {\boldsymbol{\omega}_{C}^{(m)}}^{{\top}} 
\end{array}\right]^{\top}.
\end{equation*}

We depart from this linear regression model by developing a neural network architecture from the multilayer perceptron (MLP) class \citep{goodfellow2016deep}. In this approach, the function $f(.)$ is represented by a neural network that will learn weights and biases through the course of training for a suitable representation of the data. This problem falls under the class of supervised learning. We observe some data (train set) $S_{\text{train}} = \{\mathbf{x}_n, y_n\}_{n=1}^N \in \mathcal{X} \times \mathcal{Y}$ and given a new $\mathbf{x}$, we aim to predict its label $y$, corresponding to R\&D expenditures.

Our approach is informed by the expressive power of neural networks---having superior feature learning capabilities compared to traditional regression models. It aligns with \citet{radhakrishnan2022feature}'s insights on the importance of feature learning models compared to non-feature learning ones. Also, according to the double-descent risk curve introduced by \citet{belkin2019reconciling}, in which they incorporate both classical and modern regimes, rich models such as neural networks that are over-parameterized and considered to be over-fitted, still exhibit high accuracy on out-of-sample data. The tendency of neural networks for better generalization and reduced risk of overfitting compared to under-parameterized classes of model (low-capacity function classes) further motivated our choice.
 
We adopt a series of configurations for modeling $f(\cdot)$ by varying the input space---and noting that all configurations contain features associated with country and month, $\mathbf{C}_{j, i}^{(m)}$. We start by considering two `conventional' input spaces. The configuration with the minimal dimensional input space includes only autoregressive terms, referred to as \textit{LagRD} model. Then, restricting the set of predictors to macroeconomic variables and autoregressive terms, leads to a configuration labeled as the \textit{Macros} model. Next, we consider an input space that contains exclusively Google trends data and we explore two configurations. Model \textit{AGT} includes the annual Google trends data from previous years, so that $\mathbf{S}_{t, i} = [\mathbf{\bar{S}}_{t-\tau, i}]$. Model \textit{MGT} integrates the yearly data with the monthly Google trends data from the most recent months, that is $\mathbf{S}_{t, i} = [\mathbf{\bar{S}}_{t-\tau, i} \quad \mathbf{\bar{S}}_{t, i}^{\text{YTD}}]$. We then expand the input space by including autoregressive terms of R\&D expenditures as additional predictors, hereafter referred to as the \textit{AGTwRD} and \textit{MGTwRD} models, respectively. Finally, we consider the broadest possible input space by combining macroeconomic variables, historical target values, and Google trends data at both yearly and monthly intervals, hereafter called \textit{AllVar}. This model is the general configuration introduced in equation (\ref{eq:baseline}). By varying the input space, we will be able to assess the improvements in predictive accuracy that the various pieces of data offer. Table \ref{tab:model-configurations} summarizes the seven configurations that we will explore.

\begin{table}[!htb]
\centering
\caption{Summary of Model Configurations}
\label{tab:model-configurations}
\begin{tabular}{l l}
\toprule
Configuration & Prediction function given an input space \\
\midrule
\textit{LagRD} & $f(\mathbf{x}) := f^{(m)} \left(\mathbf{Y}_{t-\tau, i},\ \mathbf{C}_{j, i}^{(m)}\right)
$ \\

\textit{Macros} & $f(\mathbf{x}) := f^{(m)} \left(\mathbf{Y}_{t-\tau, i},\ \mathbf{Z}_{t-\tau, i},\ \mathbf{C}_{j, i}^{(m)}\right)
$ \\

\textit{AGT} & $f(\mathbf{x}) := f^{(m)} \left(\mathbf{S}_{t, j, i}(\mathbf{\bar{S}}_{t-\tau, i}),\ \mathbf{C}_{j, i}^{(m)}
\right)$ \\

\textit{MGT} & $f(\mathbf{x}) := f^{(m)} \left(\mathbf{S}_{t, j, i}^{(m)} (\bar{\mathbf{S}}_{t-\tau, i},\ \mathbf{\bar{S}}_{t, j, i}^{\text{YTD}^{(m)}}),\ \mathbf{C}_{j, i}^{(m)}
\right)$ \\

\textit{AGTwRD} & $f(\mathbf{x}) := f^{(m)} \left(\mathbf{Y}_{t-\tau, i},\ \mathbf{S}_{t, j, i}(\mathbf{\bar{S}}_{t-\tau, i}),\ \mathbf{C}_{j, i}^{(m)}
\right)$ \\

\textit{MGTwRD} & $f(\mathbf{x}) := f^{(m)} \left(\mathbf{Y}_{t-\tau, i},\ \mathbf{S}_{t, j, i}^{(m)}(\mathbf{\bar{S}}_{t-\tau, i},\ \mathbf{\bar{S}}_{t, j, i}^{\text{YTD}^{(m)}}),\ \mathbf{C}_{j, i}^{(m)}
\right)$ \\

\textit{AllVar} & $f(\mathbf{x}) := f^{(m)} \left(\mathbf{Y}_{t-\tau, i},\ \mathbf{S}_{t, j, i}^{(m)}(\mathbf{\bar{S}}_{t-\tau, i},\ \mathbf{\bar{S}}_{t, j, i}^{\text{YTD}^{(m)}}),\ \mathbf{Z}_{t-\tau, i},\ \mathbf{C}_{j, i}^{(m)}
\right)$ \\
\toprule

\end{tabular}
\end{table}

The next section elaborates on the MLP set-up to model the function $f(.)$.

%%%%%%%%%%%%%%%%%%%%%%%%%%%%%%%%%%%%%%%%%%%%%%%%%%%%%%%%%%%%%%%%%%%%%%%%%%%%%

\subsection{MLP set-up}
\label{sec:mlp_setup}
As mentioned in Section \ref{sec:theoretical_model}, unlike non-feature learning algorithms that require explicit programming for feature extraction, neural networks autonomously learn and utilize features inherent in the data \citep{radhakrishnan2022feature}. Also, neural networks have the ability to interpolate data and generalize effectively, while being over-parameterized and considered as high-capacity function classes \citep{belkin2021fit}. In this subsection, we present the neural networks set-up from the MLP class developed for our context.

Mathematically, an MLP architecture can be expressed as a series of function compositions that map an input vector to an output prediction. We can define \( f: \mathbb{R}^d \rightarrow \mathbb{R} \) as a fully connected network with $L$ hidden layers for $L>1$, weight matrices $\{\mathbf{W}^{(l)}\}_{l=1}^{L+1}$, bias vectors $\{b^{(l)}\}_{l=1}^{L+1}$, and activation function $\phi$, of the form:

\begin{equation*}
f(\mathbf{x}) = {\mathfrak{h}(\mathbf{x})}^{\top} \mathbf{W}^{(L+1)} + b^{(L+1)};
\end{equation*}
\begin{equation*}
\mathbf{x}^{(\ell)} = h^{(\ell)}\left(\mathbf{x}^{(\ell-1)}\right) := \phi\left( {\mathbf{W}^{(\ell)}}^{\top} \mathbf{x}^{(\ell-1)} + b^{(\ell)} \right) \quad \text{for  } \ \ell \in \{2, \dots, L\};
\end{equation*}
\begin{equation*}
{\mathfrak{h}(\mathbf{x})} := \mathbf{x}^{(L)} =  h^{(L)}\left(\mathbf{x}^{(L-1)}\right) = \phi\left( {\mathbf{W}^{(L)}}^{\top} \mathbf{x}^{(L-1)} + b^{(L)} \right).
\end{equation*}

with $\mathbf{x}^{(0)} = \mathbf{x}$ for $\ell=1$, and $\mathfrak{h}: \mathbb{R}^d \rightarrow \mathbb{R}^K$. The function $\mathfrak{h}(.)$ represents the feature extractor part of the architecture, with activation function $\phi$ that performs an element-wise non-linear transformation on its input. 

The model learns the weight matrices and bias vectors with the objective of minimizing the training loss (also known as empirical risk minimization). The implemented architecture is MLP with a rectified linear unit (ReLU) activation function and $L=3$, which is a feed-forward type comprising three fully connected hidden layers consisting of 200, 20, and 20 neurons at each layer, respectively.\footnote{The ReLU activation function $\phi(x)$ is defined as $\phi(x) = (x)_+ = \max\{0, x\}$.} These dense layers are followed by an output layer with a single neuron, such that we reference the network as `NN4.' We assessed the model's performance using both the pyramid strategy, as recommended by \citet{masters1993practical}, and the flatter strategy, \textit{i.e.} same size for all layers. \citet{lecun2002efficient} points out that using the same number of neurons for all dense layers works at least as well as a pyramid-like setup or an upside-down pyramid. Also, they add that in most cases, the performance of an architecture with an overcomplete first hidden layer is better than an undercomplete one \citep{lecun2002efficient}.\footnote{An `overcomplete' first hidden layer stands for cases in which the number of neurons in the first hidden layer is larger than the input vector.} Lastly, they mention that the set-up choice ultimately depends on the data. Hence, settling on the present set-up is informed by having tested several set-ups for our context. The selected architecture, in fact, has an overcomplete first hidden layer, which aligns with \citet{lecun2002efficient}'s findings.

In order to diminish the sensitivity of the model's output to the random initial parametrization, we construct an ensemble of ten such neural networks. In this setup, each neural network is initialized with random parameters, and we take the average of all predictions. This approach stabilizes the model's performance by combining predictions from multiple neural networks \citep{woloszko2020tracking}.

Furthermore, as mentioned in Section \ref{sec:theoretical_model}, the vector of predictors $\mathbf{C}_{j, i}^{(m)}$ contains features associated with months and countries. As for months, we implement `one-hot encoding,' treating each month separately as an independent feature. In regard to countries, first we use `LabelEncoder' to encode the country names to transform them into numerical values, and then we map them to vectors---as an `embedding layer.' More precisely, given that we consider an embedding dimension of size two for the countries, the embedding layer maps each unique integer (representing a country) to a two-dimensional continuous vector (\textit{i.e.}, a tensor of rank one). Having countries in vector format allows the neural network to understand latent relationships between countries in terms of the prediction task, which outperforms one-hot encoding methods \citep{guo2016entity}. As training progresses, countries with similar behaviors or characteristics might get vectors that are closer together, indicating their similarity in terms of the target variable the model is predicting \citep{guo2016entity}. The embedding vector of each country is learned over the training process and constitutes the weight matrix associated with the embedding layer. 

Moreover, in the context of model specificity, \citet{woloszko2020tracking} elaborates on the trade-offs between country-specific and cross-country modeling. While the former provides nuanced, country-centric insights, it requires a richer variable set. In contrast, the latter aggregates data across countries, thereby increasing the sample size and improving estimation robustness. This trade-off exemplifies the Bias-Variance trade-off. Although pooling data across countries might introduce certain biases, it significantly diminishes the estimator's variance. Hence, we consider a cross-country set-up in which we pool all countries together.

In addition, we incorporate batch normalization into the set-up, which is applied post-ReLU activation for each hidden layer. It acts as a stabilizer for neuron activation, ensuring smoother learning, and might fasten the convergence process \citep{ioffe2015batch,santurkar2018does}.

The model utilizes the AdamW optimizer, which complements the popular Adam optimizer with weight decay. This choice is informed by empirical evidence by \citet{AdamW-loshchilov2017decoupled} suggesting superior generalization capabilities.

Finally, we also implement scenarios with `early stopping' according to the validation set. To check and mitigate the risks of overfitting, we interrupt the training regimen for each MLP in the ensemble if the validation loss stagnates or worsens over a particular number of iterations, defined by the `patience' parameter. Using `early stopping' in scenarios ensures that the results are not driven by overfitting. In support of open scientific research, all codes and datasets used in this study are available in a dedicated GitHub repository.\footnote{The dedicated GitHub repository is available at \href{https://github.com/AtiinA1/Nowcasting_RD_Expenditures.git}{https://github.com/AtiinA1/Nowcasting\_RD\_Expenditures.git}.}

%%%%%%%%%%%%%%%%%%%%%%%%%%%%%%%%%%%%%%%%%%%%%%%%%%%
%% DATA
%%%%%%%%%%%%%%%%%%%%%%%%%%%%%%%%%%%%%%%%%%%%%%%%%%%
\section{Data} \label{sec:data}
This section provides an overview of the data underpinning the empirical application. The target variable is R\&D expenditure, formally known as Gross Domestic Expenditures on Research and Development (GERD). It is published on a yearly basis by the OECD. As explained in Section \ref{sec:model_annual}, we generate nowcasting values for yearly R\&D using mixed-frequency data, consisting of macro variables and Google trends data. In this study, we focus particularly on training and calibrating the model for eight selected countries out of the top twenty innovative countries, according to WIPO’s Global Innovation Index (GII) \citep{WIPO2023}. The selected countries in our sample set are as follows: Switzerland, the United States, the United Kingdom, Germany, South Korea, China, Japan, and Canada.\footnote{Selected countries in our sample are arbitrary selection from different continents, based on the top twenty innovative countries according to WIPO’s GII.}

\subsection{Gross domestic expenditures on research and development (GERD)} \label{sec:data_rd}
GERD measures the total expenditures (both current and capital) on R\&D activities conducted by all organizations within a nation's territory, including companies, research institutes, universities, and government laboratories \citep{oecdrd2023}. It includes R\&D funded from external sources but excludes domestic funds designated for R\&D activities outside the domestic economy. These data are primarily gathered through surveys conducted by national authorities and then reported to the OECD \citep{oecdrdmeasure2015}. National authorities follow the international norms set forth by the Frascati Manual \citep{oecdrdmeasure2015}. The R\&D series are expressed in billions of 2015 USD PPPs to ensure comparability over time and across countries.

The GERD series have data gaps for certain countries, such as Switzerland. We deal with data gaps by estimating the missing values using linear interpolation, thereby guaranteeing the integrity and continuity of the time series data. We also introduce a binary indicator variable $y'_{t, i}$ that takes value 1 when the original R\&D figure is missing and 0 otherwise. However, note that we use the interpolated figures only for the lagged predictor variables and not for the target variable we aim to predict. We exclude target variables with missing values from the learning phase to ensure the reliability of our predictive model.

\subsection{Google trends}
\label{sec:GT}
Google trends, developed by Google, measures search intensities for any given queried keyword(s) by geographical area and over any selected time period starting in 2004 \citep{GoogleTrends2023}. The tool provides normalized data on the relative search volume of queries. Each data point in Google trends is normalized by the total search volume of the geography and time range considered, on a scale between 0 and 100. The search volume index (SVI) for a given search term or topic $k_{s}$ is denoted as ${s}_{k_{s}, t, j, i}^{(m)}$ and defined as follows:

\begin{equation*}
    {s}_{k_{s}, t, j, i}^{(m)} = \frac{\text{SV}_{k_{s}, t, j, i}^{(m)}}{\text{TSV}_{t, j, i}^{(m)}} * c_{i, k_{s}} \in [0, 100]
\end{equation*}

in which $\text{SV}_{k_{s}, t, j, i}^{(m)}$ and $\text{TSV}_{t, j, i}^{(m)}$ stand for volume of searches for $k_{s}$ in country $i$ at a given time (year $t$ and month $j$), and total number of searches in country $i$ at the selected time range, \textit{i.e.} from 2004 to the given time (year $t$ and month $j$), respectively. The SVI is normalized and also indexed on a scale of 0 to 100 by the constant $ c_{i, k_{s}}$ for a given time series ${s}_{k_{s}, t, j, i}^{(m)}$.

We use Google trends data to predict R\&D expenditures dynamics with a fine level of granularity. To keep the relevant search terms that might correlate with R\&D activities, we start by reconstructing the ecosystem of stakeholders associated with R\&D activities. This ecosystem comprises entities that either directly perform R\&D or play a supporting role in the broader R\&D landscape. We identify the various stakeholders and the specific search terms by studying the `systems of innovation' literature \citep{edquist2013innosys, lundvall2010innosys}. 
Table \ref{tab:gt_ecosystem} presents the ecosystem developed for this study, containing the finalized selection of stakeholders and the corresponding search terms that we identified.\footnote{The initial ecosystem developed for this study encompassed a broader range of search terms, detailed in the appendix, Table \ref{tab:gt_ecosystem_comp}.}

\begin{table} [!htb]
	\centering
  	\caption{Stakeholders and their respective search terms for R\&D expenditure.}
	\label{tab:gt_ecosystem}
	\begin{tabular}{ll}
		\toprule
		Stakeholder & Search Terms \\
		\midrule
		Businesses & R\&D Expenditure, Product Development \\
		Consulting Firms & Innovation Management, Innovation Strategy \\
		Government Agencies & Government Grants, Research Funding \\
		Innovation Hubs & Startup Incubation, Technology Park \\
		Patent Attorneys & Patent Attorney, Patent Registration \\
		R\&D Employees & R\&D Jobs \\
		Research Institutions & Collaboration with Industry, Research Grant \\
		Tax Authorities & R\&D Tax Credit \\
		Venture Capitalists (VCs) & VC Investment, Startup Funding \\
		\bottomrule
    \\
	\end{tabular}
\end{table}

After identifying specific search terms, our method involves extracting the related topics from Google trends. While the primary use of Google trends is often to search for individual search terms, it is not limited to that. In order to have the big picture, the data are also grouped into topics and we can retrieve the topic(s) associated with a given search term. The use of topics allows us to address the challenges posed by language-specific search terms and to mitigate any potential ambiguity inherent in relying only on individual search term searches \citep{woloszko2020tracking}. Consequently, the use of topics ensures that our analysis remains comparable across countries with diverse linguistic backgrounds, as these identifiers represent standardized, language-neutral markers of user interest. Nevertheless, the selection of relevant topics in Google trends, given the absence of a fixed topic list, demands careful exploration. We build upon the specific search terms (Table \ref{tab:gt_ecosystem}) as anchors to extract associated topic identifiers from Google trends. In the end, we capture 57 different Google trends topics, such that, $k_s = 57$ in the setting of this problem.

One feature of Google trends is that search indices are derived from a sample of total search volume, for computational tractability reasons. Indices that represent low-volume searches may exhibit significant sampling variance \citep{woloszko2020tracking}. Accordingly, in order to minimize sampling variance associated with Google's methods, we retrieve five samples per query and average search volume index across the samples for each topic to obtain a more reliable time series \citep{woloszko2020tracking, medeiros2021proper, woloszko2023nowcasting}.

\subsection{Macroeconomic variables}
The set of predictors also contains six `traditional' macroeconomic variables capturing elements of countries' wealth, economic conditions, and competitiveness ($k_z = 6$). These variables, which correlate with R\&D expenditures, include:

\begin{itemize}
    \item Gross domestic product per capita: Reflects the wealth and economic strength of an economy.
    \item Unemployment rate: Indicates the health of the labor market and broader economic conditions.
    \item Population: Used as a scale variable, represents the size of an economy.
    \item Inflation rate: Represents the rate of price growth and indicates the state of the economy.
    \item Export and import volumes: Chosen as indicators of a nation's competitiveness and its engagement with the global economy.
\end{itemize}

We sourced these variables from the International Monetary Fund (IMF)'s World Economic Outlook Database, April 2023 release \citep{IMF2023WEO}. We filled missing values with the mean of the respective variable for that country, calculated over all available years. 

%%%%%%%%%%%%%%%%%%%%%%%%%%%%%%%%%%%%%%%%%%%%%%%%%%%
%% Results
%%%%%%%%%%%%%%%%%%%%%%%%%%%%%%%%%%%%%%%%%%%%%%%%%%%
\section{Results}
\label{sec:results}

\subsection{Out-of-sample predictions} 
As explained earlier, we present the performance of the MLP setup over different configurations of the input space. By varying the input space, we aim to understand the value of the Google trends data compared to traditional economic variables. 

\begin{figure}[!htb]
    \centering
    \caption{Comparison of RMSE and MAPE Values for Different Configurations.}    
    \begin{subfigure}{0.45\textwidth}
         {\includegraphics[width=\linewidth]{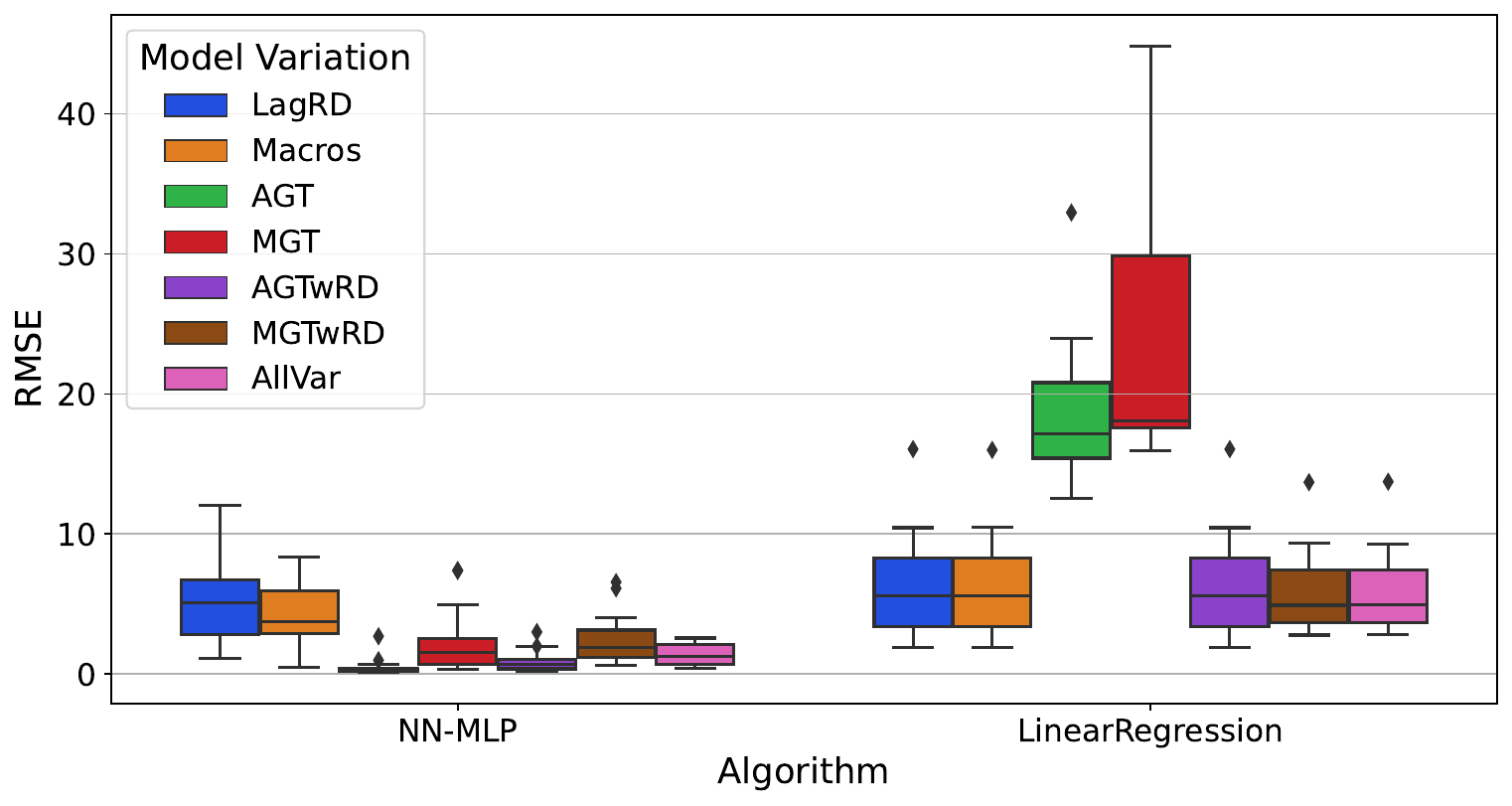}}
        \caption{RMSE (Values in Billions of USD)}
        \label{fig:rmse_linearscale}
    \end{subfigure}
    \hfill    
    \begin{subfigure}{0.45\textwidth}
         {\includegraphics[width=\linewidth]{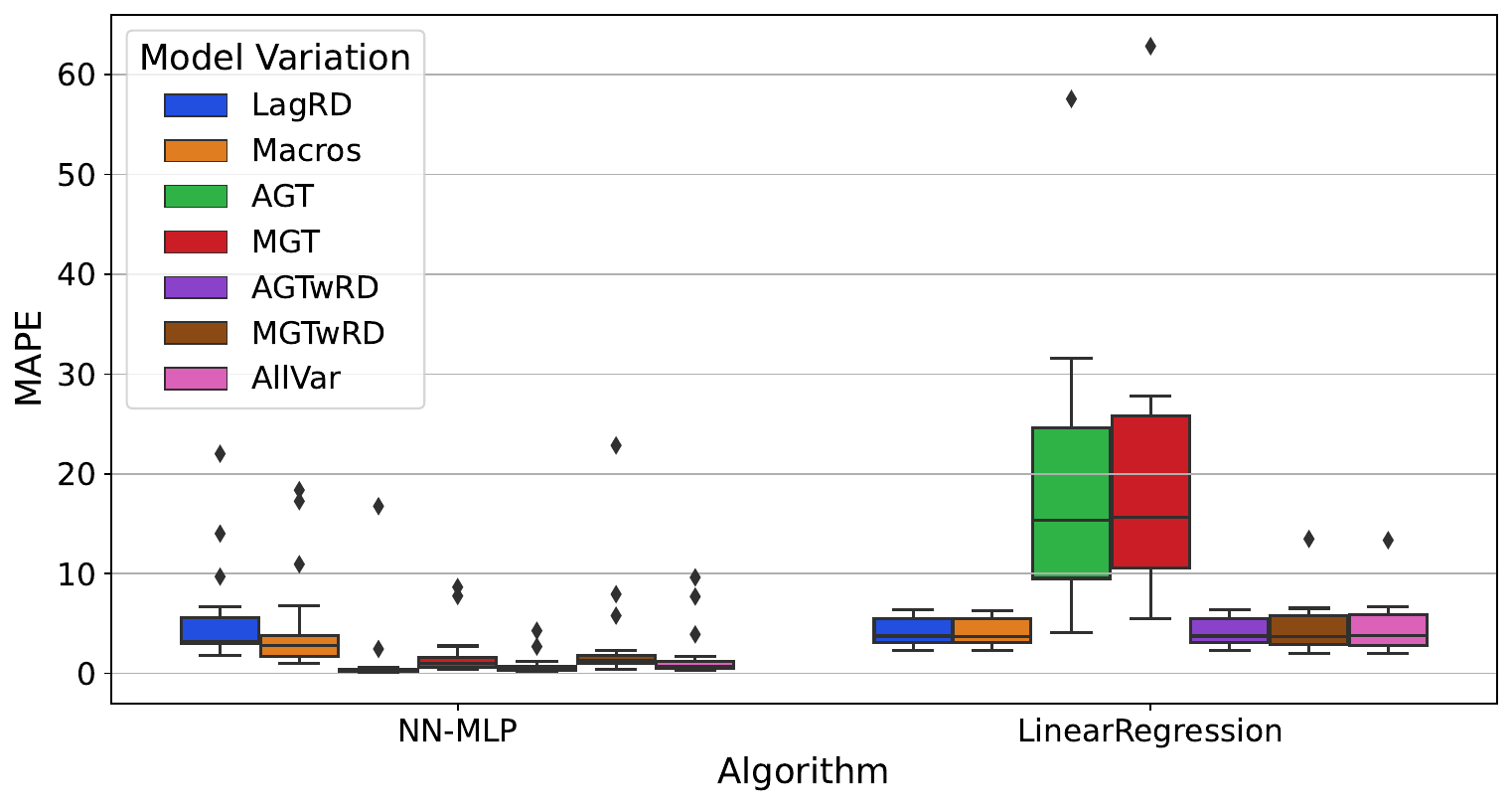}} 
        \caption{MAPE (Values in Percentages)}
        \label{fig:mape_linearscale}
    \end{subfigure}    
    \label{fig:rmse_mape_comparison}
\end{figure}

Figures \ref{fig:rmse_linearscale} and \ref{fig:mape_linearscale} respectively present box plots of the Root Mean Square Errors (RMSE) and Mean Absolute Percentage Errors (MAPE) of the MLP predictions for each of the seven input spaces considered. The figures also present RMSE and MAPE for corresponding OLS regression models as benchmarks. Two remarks are in order. 

First, the neural network predictions generally outperform the linear regression results, having lower MAPE, with the notable exception of the parsimonious \textit{LagRD} model, which exhibits similar performances. This exception can be explained by the low dimensionality of the input space (\textit{i.e.}, smaller input space cardinality), resulting in a low signal-to-noise ratio and underfitting of the model. Focusing for a moment on the linear regression model, the scale of metrics values for the \textit{AGT} and \textit{MGT} configurations have the highest median values and the broadest interquartile range. All other configurations exhibit very similar median values and interquartile ranges. The common point between all these configurations is the presence of AR terms (lagged target values), which have the highest coefficients and dominantly drive the prediction outcome. Turning now to the neural network models, it is apparent that they perform better than the linear regression models. They perform particularly well with a large number of predictors, reflecting their ability to capture complex, non-linear relationships between the input features and the target variable---relationships that linear models are inherently unable to represent.

Second, the model exploiting the annual Google trends data as input (\textit{AGT}) using a neural network exhibits the best predictive accuracy on average. The RMSE and MAPE are the lowest of all configurations---even lower than the models that also integrate lagged R\&D data. The intuition behind this performance is that there are too few data points available on a yearly basis to train the model properly and at the same time too much degree-of-freedom associated with those data points available due to the cardinality of the input space, compared to other configurations involving the Google trends data. This intuition also explains the higher variance in the MAPE for \textit{AGT} compared to \textit{AGTwRD}. Using as analogy a classification problem with data points in the input space, it means that we have too many hyperplanes to classify them. While we may achieve perfect classification on our training data, the performance can vary significantly on diverse test sets, indicating high variance. Therefore, relying solely on the performance of \textit{AGT} can be misleading, as it signals overfitting. It is more reliable and informative to consider \textit{AGT} with respect to all configurations.

Overall, the observations of Figure \ref{fig:rmse_mape_comparison} highlight the potential of neural networks and high-frequency data to capture complex dynamics and generate insightful forecasts. This can be interpreted from the outperformance of the configurations with search volume data included, compared to those without any high-frequency data in neural network set-up.

\begin{figure}[!htb]
    \centering
    \caption{True vs. Predicted R\&D Expenditures in USD (bn) for Different Countries in \textit{AllVar} Configuration.}    
     {\includegraphics[width=0.6\linewidth]{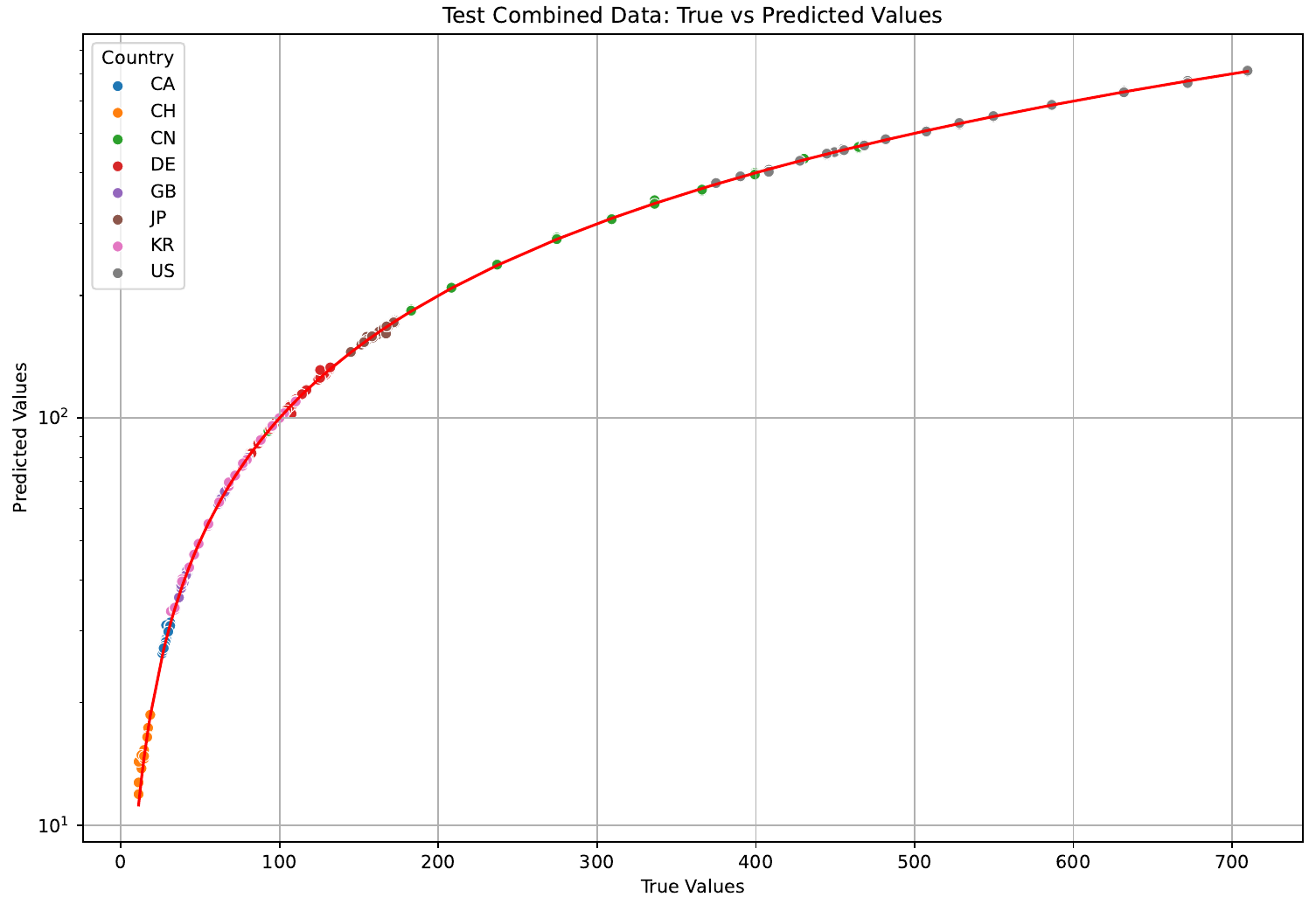}}
    \label{fig:annual_comparison_allvar_log}
\end{figure}

Figure \ref{fig:annual_comparison_allvar_log} depicts the out-of-sample performance of the \textit{AllVar} neural-network prediction model. We focus on this model because it exploits all the available information. The figure suggests that the model's predictions are fairly consistent across countries. Appendix \ref{app:oos} reports the out-of-sample performance for the individual countries. Zooming into countries suggests the presence of some outlier predictions, which could arise from country-specific factors, and the lack of representativeness of the Google trends data for those countries, either in the early days of Google trends or in general. 

Regarding Switzerland, the R\&D expenditures data are reported biennially since 2015 and prior to that, only once every four years. This irregular frequency and sparse data availability can hinder the model's capability to accurately predict for Switzerland. Increasing the frequency and consistency of data points could potentially improve the model's performance. In the case of China, the influence of government censorship, commonly referred to as the Great Firewall, has significantly restricted access to Google. This limitation affects the representativeness and reliability of Google trends data, as a substantial portion of internet searches can be routed through local search engines, \textit{e.g.} Baidu \citep{StatCounter2024China}. For countries like Japan and South Korea, Google competes with other search engines that are more integrated with the respective languages and cultural contexts, \textit{e.g.} Yahoo! in Japan \citep{StatCounter2024Japan} and Naver in South Korea \citep{StatCounter2024Korea}. This competition affects the volume and characteristics of data available from Google trends in these countries and how comprehensive the view of search behaviors are.
Lastly, a common factor affecting all models' performance for all countries is the varying rate of Google's adoption over time, which directly influences the values obtained from Google trends. If these factors can be effectively addressed, it may enhance the model's accuracy even more, leading to more reliable and robust outcomes across different countries.

\subsection{Global interpretability of model: SHAP}
In response to the tension between accuracy and interpretability inherent in complex ML models, \citet{lundberg2017unified} offers a unified framework for interpreting feature importance, so-called SHAP (SHapley Additive exPlanations). 

SHAP values have become a standard tool to obtain both local and global interpretability \citep{lundberg2017unified, woloszko2020tracking}. They enable local interpretability by showing how a model makes a decision for an individual prediction. They also enable global interpretability by explaining the model’s behavior across the entire sample. In this regard, we implement the recent model-agnostic approximation method, \textit{Kernel SHAP}, introduced by \citet{lundberg2017unified}. This method applies to any ML model, including deep learning models like ours. We use $k$-means to sample our data based on $k$ clusters, for which we considered five representative points per country and in total $k=40$. The chosen method ensures that the sample is representative of the larger dataset.

\begin{figure}[!htb]
    \centering
    \caption{SHAP Summary Plot for \textit{AllVar} Configuration.}    
     {\includegraphics[width=0.45\linewidth]{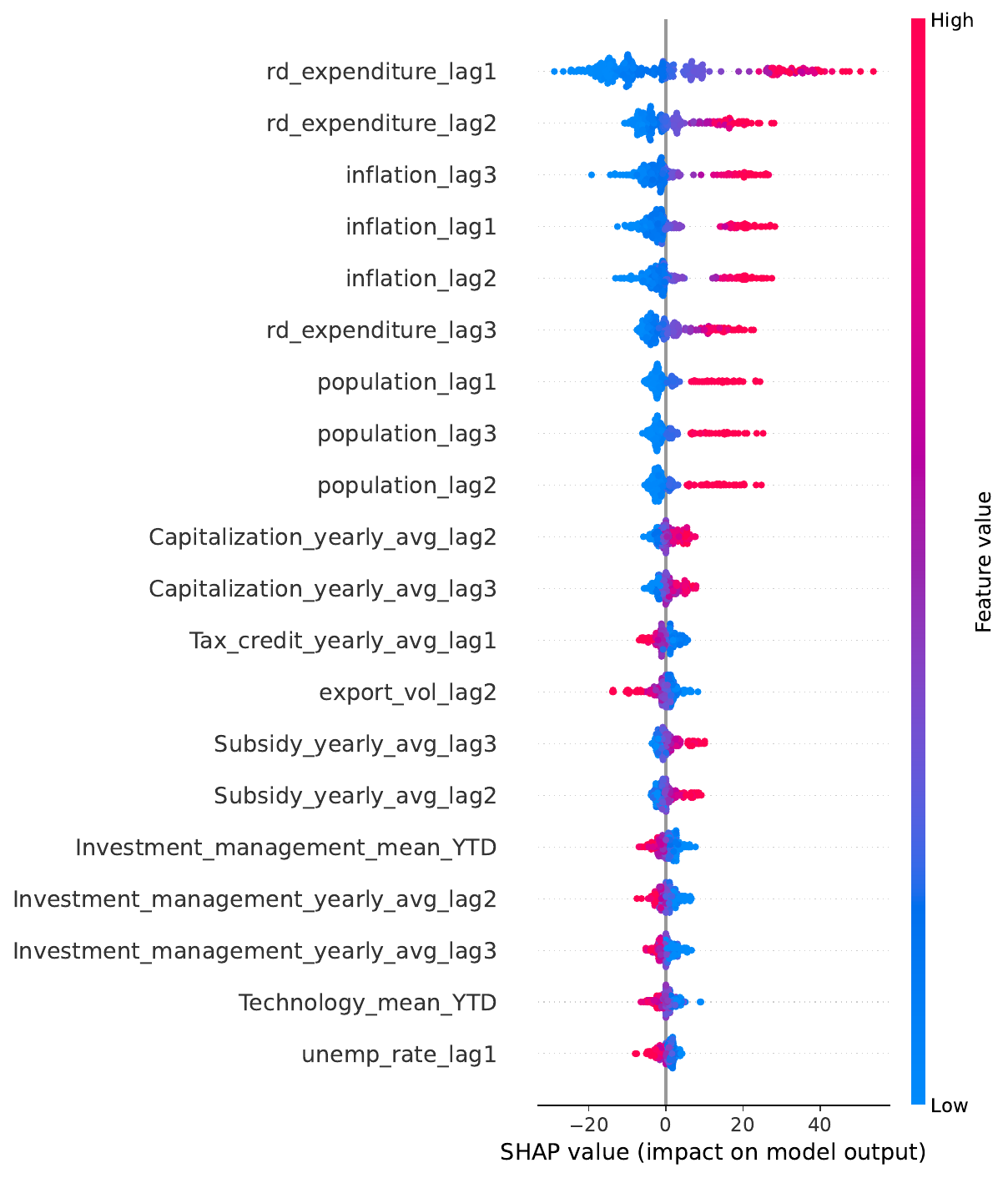}}
    \label{fig:shap_summary_plt_allvar}
\end{figure}

The SHAP summary plot in Figure \ref{fig:shap_summary_plt_allvar} illustrates the top 20 most important features and their contributions to the prediction. It is evident that the contributions are not confined to a few dominant features; rather, a wide array of features, particularly Google trends features, collectively contribute to high-quality predictions. This result is a manifestation of the `Illusion of Sparsity,' which posits that while a small number of variables may be identified as primary predictors in a prediction model, a broad range of features could still collectively exhibit significant influence \citep{giannone2021economic}. It highlights the fact that we are dealing with a \textit{dense}-modelling problem, in which the target value is recovered by many features with small contributions \citep{giannone2021economic, woloszko2023nowcasting}. Also, it  validates the choice of neural networks, as part of \textit{dense} class of estimators, to address non-linearities between input features and the output \citep{woloszko2023nowcasting}. 
In addition, Figure \ref{fig:shap_summary_plt_allvar} reveal that the model effectively learns the importance of historical data on R\&D expenditures, as captured in autoregressive terms, and also integrates signals from other macroeconomic variables.

%%%%%%%%%%%%%%%%%%%%%%%%%%%%%%%%%%%%%%%%%%%%%%%%%%%
%% Temporal Disaggregation
%%%%%%%%%%%%%%%%%%%%%%%%%%%%%%%%%%%%%%%%%%%%%%%%%%%
\section{Interpolating R\&D expenditures at a higher frequency}
\label{sec:model_bem}

We have established so far that Google trends data are well suited to the nowcasting of yearly R\&D expenditures. Another advantage of the data lies in the fact that they are available at a high frequency, making it possible to nowcast monthly R\&D expenditures. We take a step in this direction by interpolating R\&D expenditures, \textit{i.e.} applying temporal disaggregation, which involves using high-frequency indicators to construct high-frequency renditions of low-frequency information \citep{mosley2022sparse}.\footnote{The task of nowcasting monthly R\&D expenditures differs from the task of interpolating yearly series into monthly figures in an important way. In the latter task, we apply temporal disaggregation on the yearly amount of R\&D expenditures, by factoring in the records of full year monthly Google trends data.} 

We interpolate the series using three methods. We start with the classical regression-based temporal disaggregation method introduced by \citet{chowlin1971best}, followed by a recent extension for high-dimensional settings proposed by \citet{mosley2022sparse}. We then propose a full neural-network-based approach. 

In order to construct the unobserved monthly time series ${y}_{t, j, i}^{(m)}$, \citet{chowlin1971best} and \citet{mosley2022sparse} assume the following regression model at the monthly frequency \citep[Equation 1]{mosley2022sparse}:

\begin{equation}
\label{eq:reg_temp_disagg_m}
    {y}_{t,j, i} = \mathbf{x}_{t,j,i}^{{(m)}^{\top}} \boldsymbol{\theta} + \mathbf{u}_{t,j,i}^{(m)}.
\end{equation}

The residual vector $\mathbf{u}_{t,j, i}^{(m)}$ is mean-zero, has covariance matrix $\mathbf{V}_{m}$, and follows a first-order autoregressive process $\mathbf{u}_{t,j, i}^{(m)}=\rho \ \mathbf{u}_{t,j-1, i}^{(m)} + \varepsilon_{t,j, i}^{(m)}$ with $\varepsilon_{t,j, i}^{(m)} \sim \mathcal{N}(0, \sigma^2)$ and $|\rho| < 1$. Due to ${y}_{t,j, i}$ being unobserved, the counterpart of equation (\ref{eq:reg_temp_disagg_m}) is written as 
\citep[Equation 3]{mosley2022sparse}

\begin{equation}
\label{eq:reg_temp_disagg_y}
    {y}_{t, i} = \mathbf{x}_{t,i}^{\top} \boldsymbol{\theta} + \mathbf{u}_{t,i}.
\end{equation}

We obtain it by multiplying equation (\ref{eq:reg_temp_disagg_m}) by an aggregation matrix $\mathbf{A} = \mathbf{I}_n \otimes \mathbf{1}_{12}$. In our context, the twelve monthly figures in a year must sum to their corresponding yearly figure, without loss of generality \citep{mosley2022sparse}. Similarly, we transform the covariance matrix $\mathbf{V}_m$ to $\mathbf{V}_a$ using the aggregation matrix and its transpose.

Given the yearly figures being observed, we can solve equation (\ref{eq:reg_temp_disagg_y}) using standard techniques, for which \citet{chowlin1971best} establishes a generalized least squares (GLS) estimator for $\boldsymbol{\theta}$. This estimator can be derived from the Chow-Lin cost function term in equation (\ref{eq:reg_temp_disagg_estimator}) and it depends on the unknown parameter $\rho$, which needs to be estimated \citep[Equation 15]{chowlin1971best}. In an extension suited to high-dimensional data, \citet{mosley2022sparse} proposes the estimator as the form in equation (\ref{eq:reg_temp_disagg_estimator}) \citep[Equation 7]{mosley2022sparse}. They incorporate a penalty function with the classic Chow-Lin cost function to regularize and encode the sparsity assumption in high-dimensional settings. As for the regulizer term, they consider LASSO (${l}_1$) penalty $P_{\lambda}(\boldsymbol{\theta}) = \lambda \|\boldsymbol{\theta}\|_1 := \lambda_{\rho} \sum_{j=1}^{d} |\boldsymbol{\theta}_j|$, and refer to this method as ${l}_1$-spTD. The index $\lambda$ is a non-negative regularization parameter that controls the degree of shrinkage.

\begin{equation}
\label{eq:reg_temp_disagg_estimator}
    \hat{\boldsymbol{\theta}} = \hat{\boldsymbol{\theta} }_\rho = \arg \min_{\boldsymbol{\theta}  \in \mathbb{R}^d} \left\{ \underbrace{\left\| \mathbf{V}_a^{-1/2} ({y}_{t, i} - \mathbf{x}_{t,i}^{\top} \boldsymbol{\theta}) \right\|_2^2}_{\text{Chow-Lin cost function}} + \underbrace{P_\lambda(\boldsymbol{\theta})}_{\text{Regularizer}} \right\}.
\end{equation} 

We implement the classical regression-based temporal disaggregation method proposed by \citet{chowlin1971best}, in particular the \textit{`chow-lin-maxlog'} method using the `tempdisagg' \textit{R} package by \citet{sax2013temporal}. As for the sparse temporal disaggregation method by \citet{mosley2022sparse}, we use the `DisaggregateTS' \textit{R} package by \cite{DisaggregateTS2022}.

While these regression-based estimates are well-accepted temporal disaggregation methods, they may fail to capture the inherent characteristics of the data. Accordingly, we propose a new method that leverages our neural-network model.

In broad terms, we corrupt the input and record its corresponding output, allowing us to derive a neural network-driven elasticity value for each feature of the \textit{AGT} configuration. Considering the elasticities for all input features, in addition to the proportion of each input feature at the monthly level with respect to its aggregated value at the yearly level, we distribute the yearly figures of R\&D expenditures to monthly figures, labeled $\hat{y}_{t, j, i}$.\footnote{Appendix \ref{app:interpolate_appendix_corruptedinput} proposes an alternative method based on the corrupted input approach. While this method can be very intuitive from computer science perspective, it does not perform as well as the proposed temporal disaggregation method.}

More formally, we calculate the \textit{empirical predicted expected elasticity} $\mathbb{E}[\hat{\eta}_{s_{k_{s}, i}}]$, for a particular input feature, \textit{i.e.} topic $s_{k_{s}}$, and a given country $i$, from the prediction model on the observations (train set) as in equation (\ref{eq:nn_elasticity}). \noindent $\mathbb{E}_{S_{\text{train}}}$ denotes the expectation over the samples in the training set, while $\mathbb{E}_{\delta \sim \Delta}$ represents the expectation over the perturbations $\delta$, which are sampled from the distribution $\Delta$. In our setting, we apply perturbations to the input features by sampling from a normal distribution with a mean of 0.01 and a standard deviation of 0.005. 

\begin{equation}
\label{eq:nn_elasticity}
\bar{\eta}_{\bar{s}_{k_{s}, i}} = \mathbb{E}_{S_{\text{train}}}\left[ \hat{\eta}_{\bar{s}_{k_{s}, i}} \right] = \mathbb{E}_{S_{\text{train}}}\left[\mathbb{E}_{\delta \sim \Delta}\left[ {\eta}_{\bar{s}_{k_{s}, t-\tau, i}} \right]\right] = \mathbb{E}_{S_{\text{train}}}\left[\mathbb{E}_{\delta \sim \Delta}\left[ \frac{\frac{{y}_{t, i}^{\delta} - {y}_{t, i}}{{y}_{t, i}}}{\frac{\bar{s}_{k_{s}, t-\tau, i}^{\delta} - \bar{s}_{k_{s}, t-\tau, i}}{\bar{s}_{k_{s}, t-\tau, i}}} \right]\right].
\end{equation}

Letting $p_{s_{k_{s}}, t, j, i}^{(m)}$ denote the proportion of Google trends value for a topic associated with search terms $k_s$ on a monthly frequency, with respect to its aggregated value on a yearly frequency of year $t$, for month $j$, and country $i$, we have:
\begin{equation*}
p_{s_{k_{s}}, t, j, i}^{(m)} = \frac{s_{k_{s}, t, j, i}^{(m)}}{\sum_{j=0}^{11} s_{k_{s}, t, j, i}^{(m)}}.
\end{equation*}

Given the estimator for the expected elasticity value $\hat{\eta}_{\bar{s}_{k_{s}, i}}$, the adjusted proportion of the Google trends value of year $t$ for month $j$ and country $i$ can be defined as:
\begin{equation*}
\tilde{p}_{s_{k_{s}}, t, j, i}^{(m)} = \bar{\eta}_{\bar{s}_{k_{s}, i}} \times p_{s_{k_{s}}, t, j, i}^{(m)}.
\end{equation*}

As a result, estimated monthly R\&D expenditures of year $t$ for month $j$ and country $i$ is given by:\footnote{In case of unavailability of the original data point $y_{t, i}$, we can estimate it using the neural network-based nowcasting model $\hat{y}_{t, i}=f(\mathbf{x})$.}
\begin{equation*}
\hat{y}_{t, j, i}^{(m)} = y_{t, i} \times \frac{\sum_{k_{s}} \tilde{p}_{s_{k_{s}}, t, j, i}^{(m)}}{\sum_{j} \sum_{k_{s}} \tilde{p}_{s_{k_{s}}, t, j, i}^{(m)}} \quad \text{s.t.} \quad y_{t, i} = \sum_{j=0}^{11} \hat{y}_{t, j, i}^{(m)}.
\end{equation*}

\begin{figure}[!htb]
    \centering
    \caption{Comparison of Estimated Monthly R\&D Expenditures for the United States.}    
    \begin{subfigure}{0.48\textwidth}
         {\includegraphics[width=\linewidth]{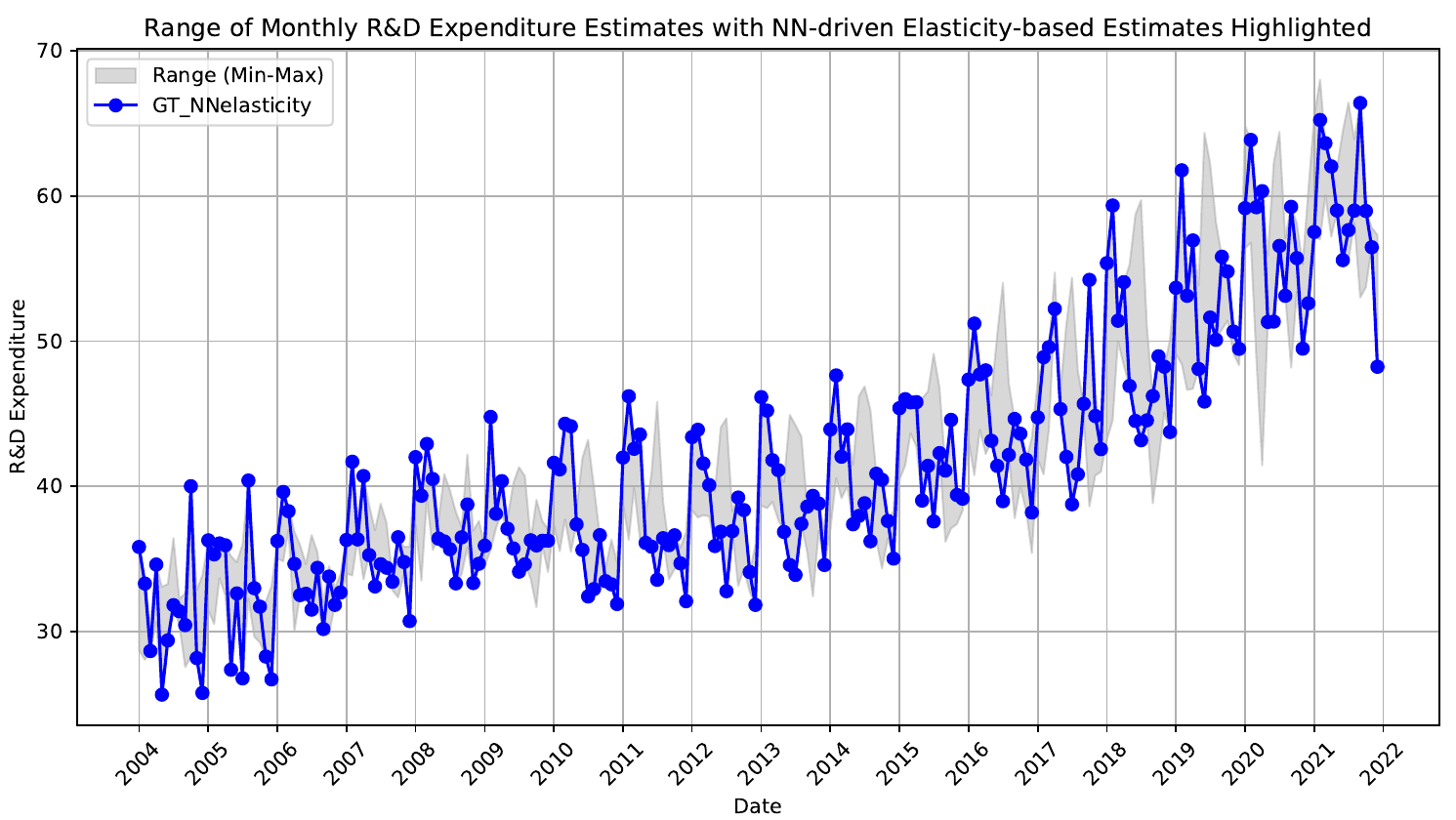}}
        \caption{NN-driven Elasticity-based Estimates}
        \label{fig:tempdis_gt_nn_elasticity}
    \end{subfigure}
    \hfill    
    \begin{subfigure}{0.48\textwidth}
         {\includegraphics[width=\linewidth]{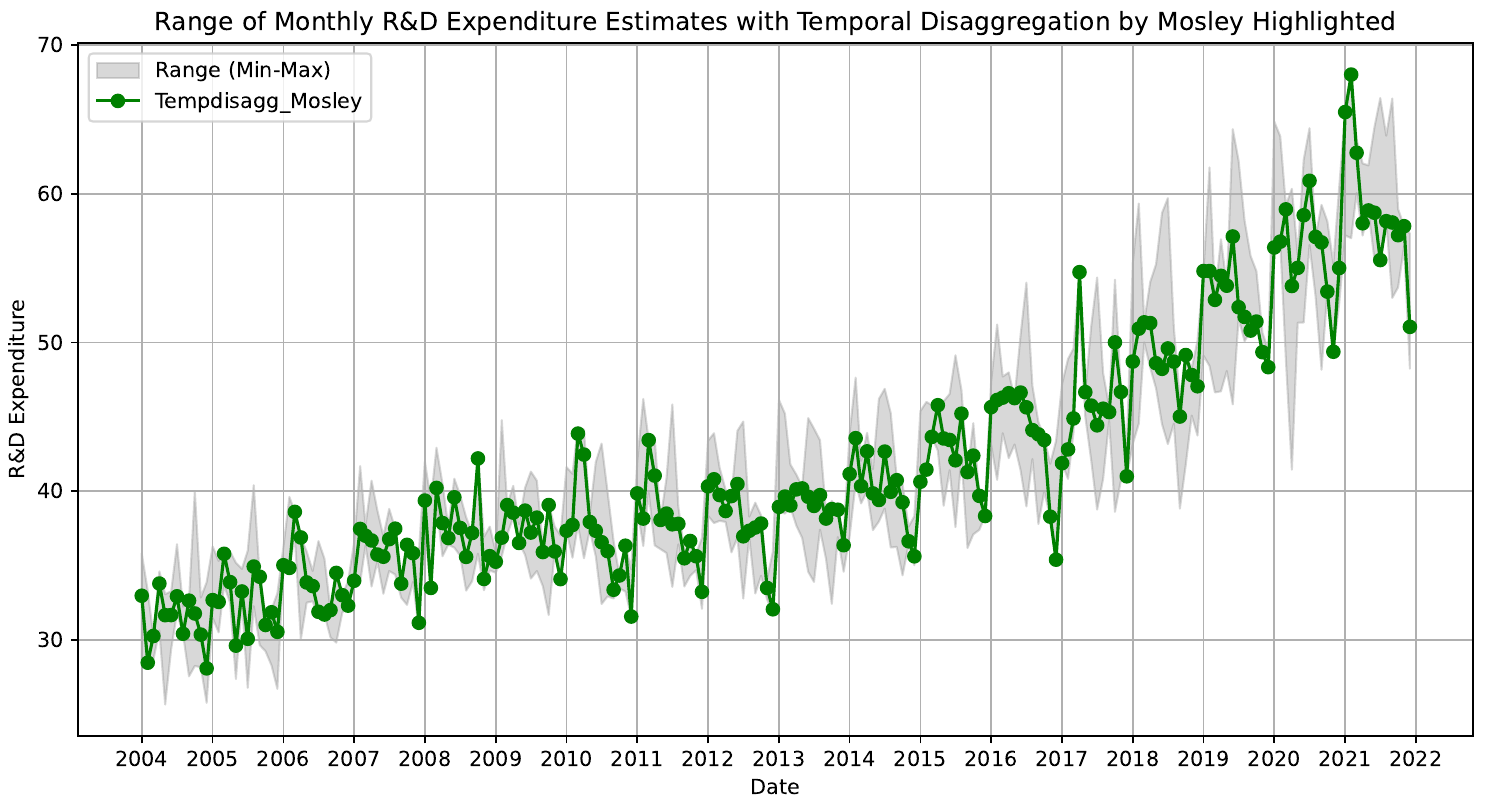}} 
        \caption{\cite{mosley2022sparse}-based Sparse Temporal Disaggregation}
        \label{fig:tempdis_mosley}
    \end{subfigure}
    \hfill    
    \begin{subfigure}{0.48\textwidth}
         {\includegraphics[width=\linewidth]{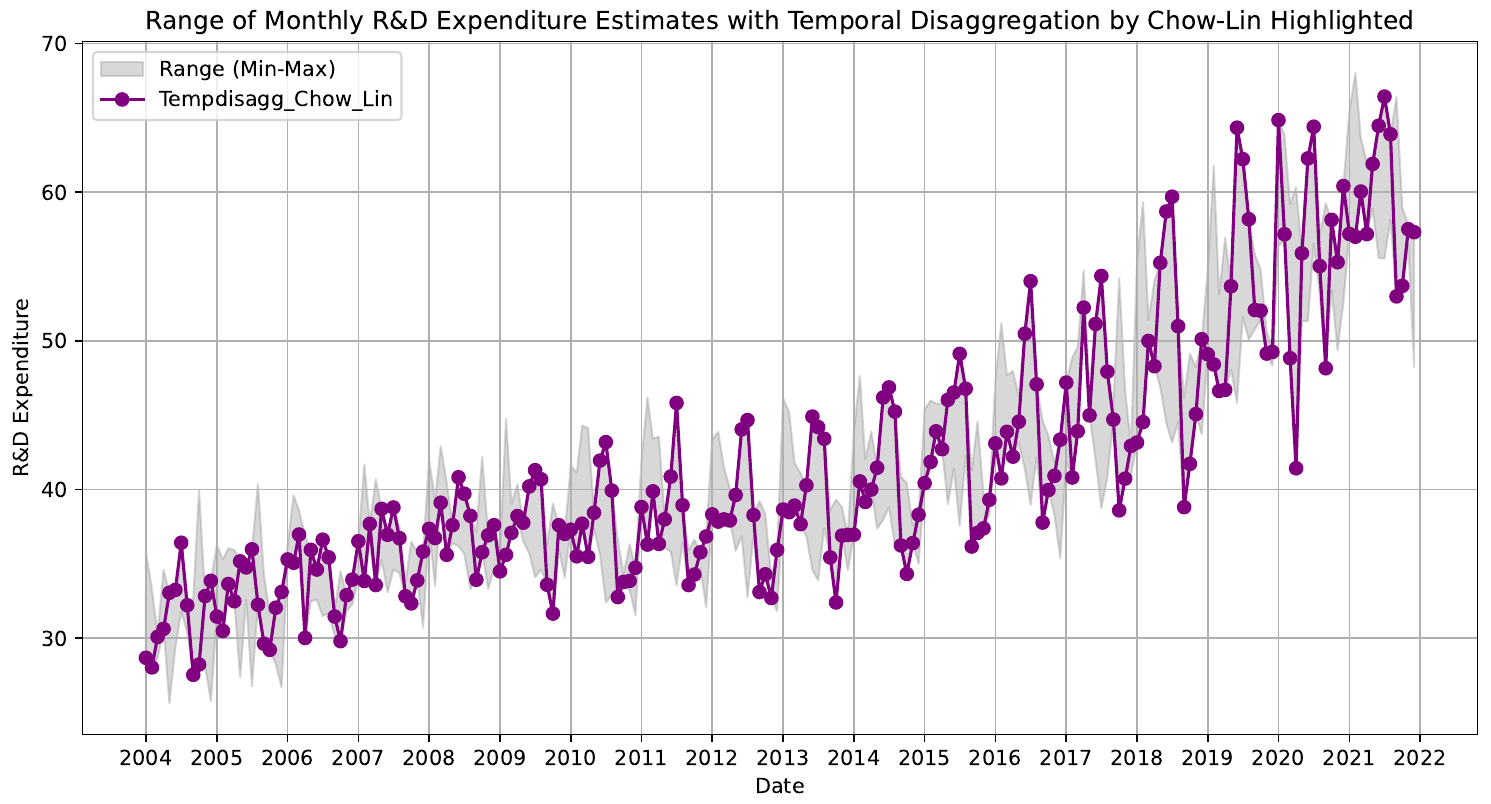}} 
        \caption{\citet{chowlin1971best}-based Temporal Disaggregation}
        \label{fig:tempdis_sax}
    \end{subfigure}    
    \label{fig:different_estimators}
\end{figure}

Figure \ref{fig:different_estimators} depicts different estimated monthly R\&D expenditures over time for the United States. The neural-network-driven elasticity-based estimates in Figure \ref{fig:tempdis_gt_nn_elasticity} strongly correlate with the sparse temporal disaggregation estimates~\citep{mosley2022sparse} shown in Figure \ref{fig:tempdis_mosley}, with correlation coefficient $r(N-2) = r(214)=0.93, \ p<0.001$ in level, and $r(214)=0.66, \ p<0.001$ on the growth rates. While these two estimators exploit full information by allowing for high-dimensional indicator matrices, the classical temporal disaggregation method of \citet{chowlin1971best} requires a dimensionality-reduction step to limit the set of predictors to a handful of indicators. In this regard, we leverage the insights generated by the neural networks model, in particular, the top six important features in terms of their contribution to the prediction based on the shapley values in the \textit{AGT} configuration. The correlation between the resulting temporal disaggregation shown in Figure \ref{fig:tempdis_sax} and the NN-driven elasticity-based estimates is $r(214)=0.73, \ p<0.001$ in level, and $r(214)=-0.17, \ p=0.02$ on the growth rates which is not statistically significantly different from 0.\footnote{The correlation between the temporal disaggregation shown in Figure \ref{fig:tempdis_sax} and Figure \ref{fig:tempdis_mosley} is $r(214)=0.87, \ p<0.001$ in level, and $r(214)=0.25, \ p=0.0011$ on the growth rates.}

\begin{figure}[!htb]
    \centering
    \caption{Estimated Monthly R\&D Expenditures vs. Employees in Scientific R\&D Services.}    
     {\includegraphics[width=0.55\linewidth]{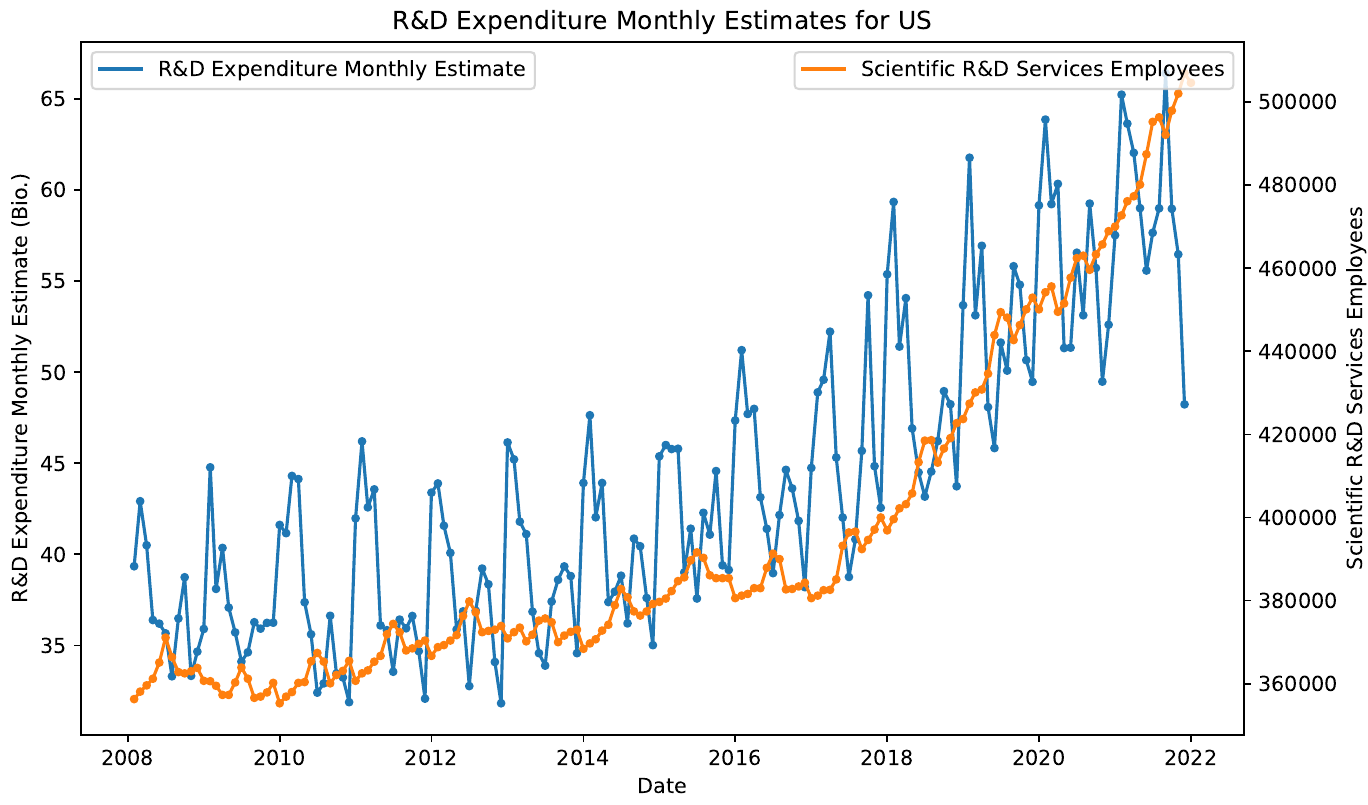}}
    \label{fig:tempdis_employ}
\end{figure}

In a second validation exercise, we correlate the monthly series to another series that we expect to relate strongly to R\&D expenditures. We use monthly figures for employment in scientific R\&D services in the United States from Data USA \citep{DataUSA2023SRDS}, as shown in Figure \ref{fig:tempdis_employ}. A priori, it is unclear how monthly R\&D expenditures should correlate with monthly R\&D employment. The time lag between R\&D expenditures and employment is ambiguous because R\&D expenditures include wages as well as consumables and equipment costs. An increase in employment at time $t = 0$ could result in a subsequent rise in R\&D expenditures at time $t + x$, as new employees begin their research, thereby increasing future non-employee expenditures for equipment and consumables. Conversely, the purchase of expensive equipment might occur before the hiring of engineers and scientists needed to operate it. Consequently, the time lags between these variables are not straightforward. Furthermore, the correlation could even be negative or null depending on the degree of substitution between human capital and physical capital.

We compute the correlation between estimated R\&D expenditures (in growth rates) and scientific R\&D services employees (in growth rates) for different lags and different disagregation methods, as shown in Table \ref{tab:correlation_employees}. We report the correlation coefficients for lags statistically significantly different from zero, specifically with a p-value less than $0.01$. A positive lag means a backward shift in the R\&D expenditures growth rate, \textit{i.e.}, capturing how the R\&D expenditures growth rate from a previous period (month) correlates to the current R\&D services employees growth rate. Conversely, a negative lag value indicates a forward shift, comparing the future R\&D expenditures growth rate with the current employees' growth rates.

We observe significant correlations at various lags across all three methods, indicating a relationship between R\&D expenditure and employee growth. The estimates obtained using \citet{mosley2022sparse} seem to be captured by the neural-network-based estimates, with roughly similar correlation coefficients at lags -5, -4, and 4. The estimates based on \citet{chowlin1971best} exhibit a strong positive correlation coefficient at lag $0$ and negative at lags $-3$ and $3$. The correlation coefficient at lag $5$ is similar to the neural-network-based estimates.

It is worth emphasizing that all methods are based on the predicted features of our neural network model. \citet{chowlin1971best} exploits only a handful of the features and exhibit correlation coefficients that are markedly different from the other two methods, which exploit the full set of features. We have no way of establishing which method delivers the most `valid' results. Lacking ground truth, we cannot rule out or favor a method. The fact that all methods significantly correlate with the R\&D series indicates that they all capture meaningful aspects of the R\&D dynamics.\footnote{Data on patent filings are available at high frequency, but do not relate to current R\&D expenditures since the capture the output of the R\&D process. The gap between R\&D expenditures and patent filings is a least one year, as documented by \cite{gaetan_depinnov}. Furthermore, patent statistics are published 18 months after filing, making it a poor candidate for nowcasting.}

\begin{table}[h!]
\centering
\caption{Correlation between Monthly R\&D Expenditure and R\&D Services Employees Growths}
\label{tab:correlation_employees}
\begin{tabular}{l r r r}
\toprule
\textbf{Monthly R\&D Expenditure Estimates Growth} & \textbf{Lag} & \textbf{Correlation} & \textbf{P-Value} \\
\midrule
\multirow{6}{*}{Neural Network-driven elasticity-based estimates} & 0  & -0.42 & <0.0001 \\ 
 & -5 & -0.39 & <0.0001 \\ 
 & 5  & 0.29  & 0.0002 \\ 
 & -2 & 0.27  & 0.0004 \\ 
 & 4  & 0.21  & 0.0080 \\ 
 & -4 & -0.20 & 0.0097 \\ 
\midrule
\multirow{3}{*}{\citet{mosley2022sparse}-based estimates} & -5 & -0.32 & <0.0001 \\ 
 & -4 & -0.25 & 0.0010 \\ 
 & 4  & 0.22  & 0.0049 \\ 
\midrule
\multirow{6}{*}{\citet{chowlin1971best}-based estimates} & 0  & 0.46  & <0.0001 \\ 
 & -2 & -0.37 & <0.0001 \\ 
 & 1  & 0.31  & <0.0001 \\ 
 & 3  & -0.25 & 0.0010 \\ 
 & -3 & -0.22 & 0.0050 \\ 
 & 5  & 0.20  & 0.0090 \\ 
\bottomrule
\end{tabular}
\end{table}

%%%%%%%%%%%%%%%%%%%%%%%%%%%%%%%%%%%%%%%%%%%%%%%%%%%
%% Conclusion
%%%%%%%%%%%%%%%%%%%%%%%%%%%%%%%%%%%%%%%%%%%%%%%%%%%
\section{Conclusion and future work} 
\label{sec:conclusion}
In this study, we propose a nowcasting model to predict annual gross domestic expenditures on R\&D. We develop an MLP set-up and leverage a large set of covariates, including macroeconomic variables and Internet search volume data. Our choice of a neural network prediction model is based on these models' superior ability in feature learning and their tendency for better generalization and, accordingly, a reduced risk of overfitting. The empirical evidence from our research highlights the existence of non-linear mapping between the covariates and our target. Indeed, the neural network models not only outperform traditional linear regression models but also demonstrate their ability to capture the complex interplay of features. This finding also validates that traditional linear models often fail to perform well when considering high-dimensional settings.

Besides predicting annual R\&D expenditures, we explore the feasibility of offering higher-frequency R\&D  series. We leverage the neural-network-based annual nowcasting model and integrate additional steps to produce a distribution of monthly R\&D expenditures. More specifically, by perturbing the input and analyzing its corresponding output, we derive neural-network-driven elasticity values for each feature, that combined with the monthly proportion of each feature relative to its annual aggregate, enable the distribution of yearly R\&D expenditures to monthly figures. The results from this extension indicate the model's potential to provide more frequent and detailed insights on R\&D. It offers a step in the direction of a monthly nowcasting model. However, lack of ground truth data prevented us from recommending one temporal disaggregation method over the others. More studies are needed to evaluate the robustness, accuracy, and responsiveness of the model, particularly regarding its ability to adapt and respond to economic shocks. 

Finally, we hope that policymakers will adopt our model to produce R\&D statistics and to inform their decisions. A valuable feature of the framework is that one can adapt it at different levels, \textit{e.g.} regional level, to offer more granular insights. Furthermore, the method can be adapted to capture a different set of topics that go beyond technological innovation reflected by R\&D expenditures. This method's potential to track, say, social innovation \citep{mulgan2007social} or open innovation \citep{chesbrough2003open} is a particularly exciting extension.

%%%%%%%%%%%%%%%%%%%%%%%%%%%%%%%%%%%%%%%%%%%%%%%%%%%
%% References
%%%%%%%%%%%%%%%%%%%%%%%%%%%%%%%%%%%%%%%%%%%%%%%%%%%
% \bibliographystyle{unsrtnat}
% \bibliography{references}  %%% Uncomment this line and comment out the ``thebibliography'' section below to use the external .bib file (using bibtex) .
\bibliographystyle{apalike}
\bibliography{references}

\newpage
%%%%%%%%%%%%%%%%%%%%%%%%%%%%%%%%%%%%%%%%%%%%%%%%%%%
%% Appendix
%%%%%%%%%%%%%%%%%%%%%%%%%%%%%%%%%%%%%%%%%%%%%%%%%%%
\input{annex.tex}

\end{document}

%% file: annex.tex
\begin{appendix}
\renewcommand{\thesection}{\Alph{section}} % Changes numbering to letters
\counterwithin{figure}{section}
\counterwithin{table}{section}

\section{Google search terms}
\label{app:gt}

The initial ecosystem developed for this study encompassed a broader range of search terms, detailed in table \ref{tab:gt_ecosystem_comp}.
\begin{table} [!htb]
\caption{Initial search terms to build a network of stakeholders.}
\centering
\begin{tabularx}{\textwidth}{lX}
\toprule
\textbf{Stakeholder} & \textbf{Keywords/Topics/Categories} \\
\midrule
Firms/Companies & R\&D Expenditure, Product Development, Technology Innovation, Patent Application, Tech Research, Pharma Research, New Drug Application, Research Grants, Intellectual Property. \\
Venture Capitalists (VCs) & Startup Funding, Technology Startups, Pharma Startups, VC Investment in R\&D, Innovation Investment, Return on Investment, Exit Strategy, Seed Funding, Angel Investment. \\
Banks and Financial Institutions & Business Loans, R\&D Loans, Investment Banking, Corporate Finance, Financial Risk, Credit Assessment, Interest Rate, Loan Application, Credit Score. \\
Universities and Research Institutions & Academic Research, Collaboration with Industry, Research Funding, University Patents, Postgraduate Studies, Doctoral Research, Research Publication, Research Grant. \\
Government agencies & R\&D Policy, Research Funding, Government Grants, Innovation Policy, Public-Private Partnership, Tax Incentives for R\&D, Technology Transfer, Patent Law, Economic Development. \\
R\&D Employees & Research Methods, Data Analysis, Patent Filing, Lab Equipment, Scientific Journal, Professional Development, Research Ethics, Project Management, Collaboration Tools. \\
Tax Authorities & R\&D Tax Credit, Tax Incentives, Tax Deduction, Tax Filing, Corporate Tax, Tax Law, Tax Consultancy. \\
Consulting Firms & Business Strategy, Market Analysis, Risk Assessment, Business Growth, Innovation Strategy, Portfolio Management, Project Planning, Financial Modeling. \\
Innovation Hubs/Incubators & Startup Incubation, Innovation Hub, Technology Park, Business Accelerator, Entrepreneurship, Mentorship, Networking, Business Pitch, Startup Ecosystem. \\
Patent Attorneys & Patent registration, Intellectual property rights, Patent law, Technology patents, Patent disputes. \\
Tax Consultants/Accountants & R\&D tax credits, Corporate tax, Business expenses, Tax deductions, Tax advice for R\&D, Accounting for R\&D. \\
\bottomrule
\label{tab:gt_ecosystem_comp}
\end{tabularx}
\end{table} 
%%%%%%%%%%%%%%%%%%%%%%%%%%%%%%%%%%%%%%%%%%%%%%%%%%%%%%%%%%%%%%%%%%%%%%%%%%%%%%%%%%%%
\newpage
\section{Out-of-sample performance and global interpretability of different configurations}

In this section, we show the out-of-sample performance of the prediction model, for all the configuration individually.
\subsection{\textit{AGT} Configuration}

\begin{figure}[!htb]
    \centering
    \caption{Analysis for \textit{AGT} Configuration.}             
    \begin{subfigure}[c]{0.45\linewidth}
        \centering
        \includegraphics[width=\linewidth]{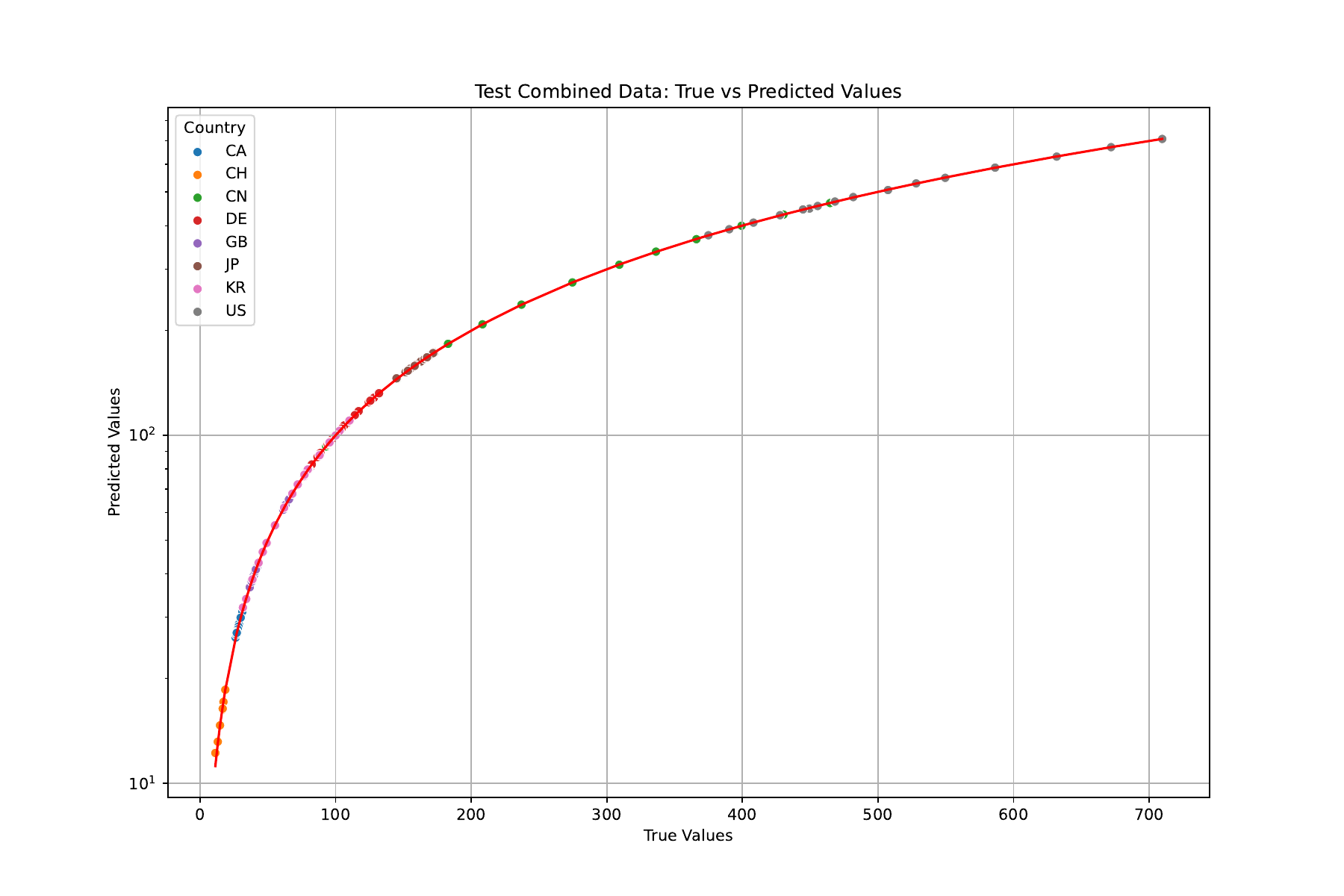}
        \caption{True vs. Predicted R\&D Expenditures in USD (bn) for Different Countries.}
        \label{fig:annual_comparison_agt_log}
    \end{subfigure}
    \begin{subfigure}[c]{0.4\linewidth}
        \centering
        \includegraphics[width=\linewidth]{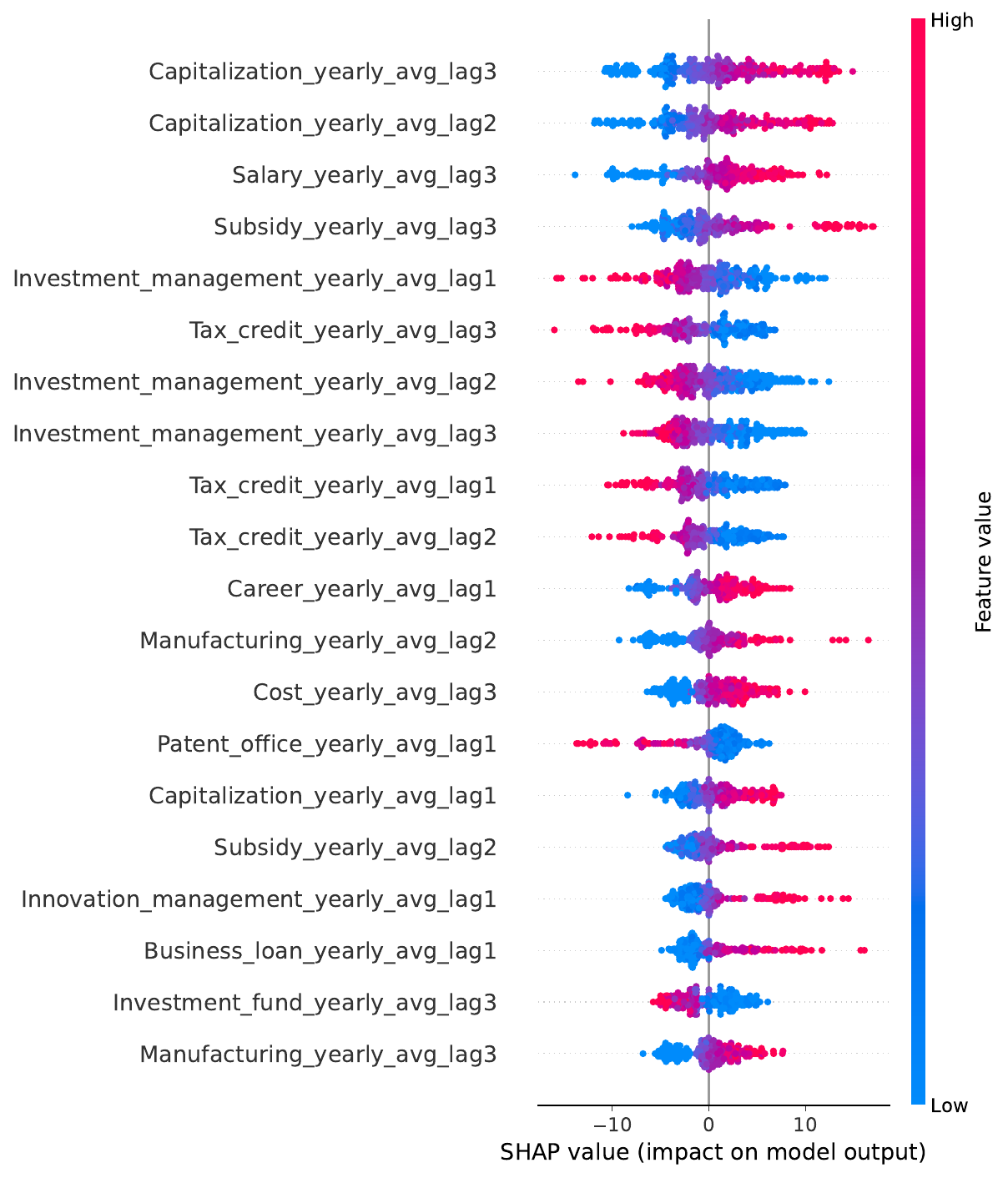}
        \caption{SHAP summary plot.}
        \label{fig:shap_summary_plt_agt}
    \end{subfigure}
\end{figure}
\subsection{\textit{AGTwRD} Configuration}

\begin{figure}[!htb]
    \centering
    \caption{Analysis for \textit{AGTwRD} Configuration.}             
    \begin{subfigure}[c]{0.45\linewidth}
        \centering
        \includegraphics[width=\linewidth]{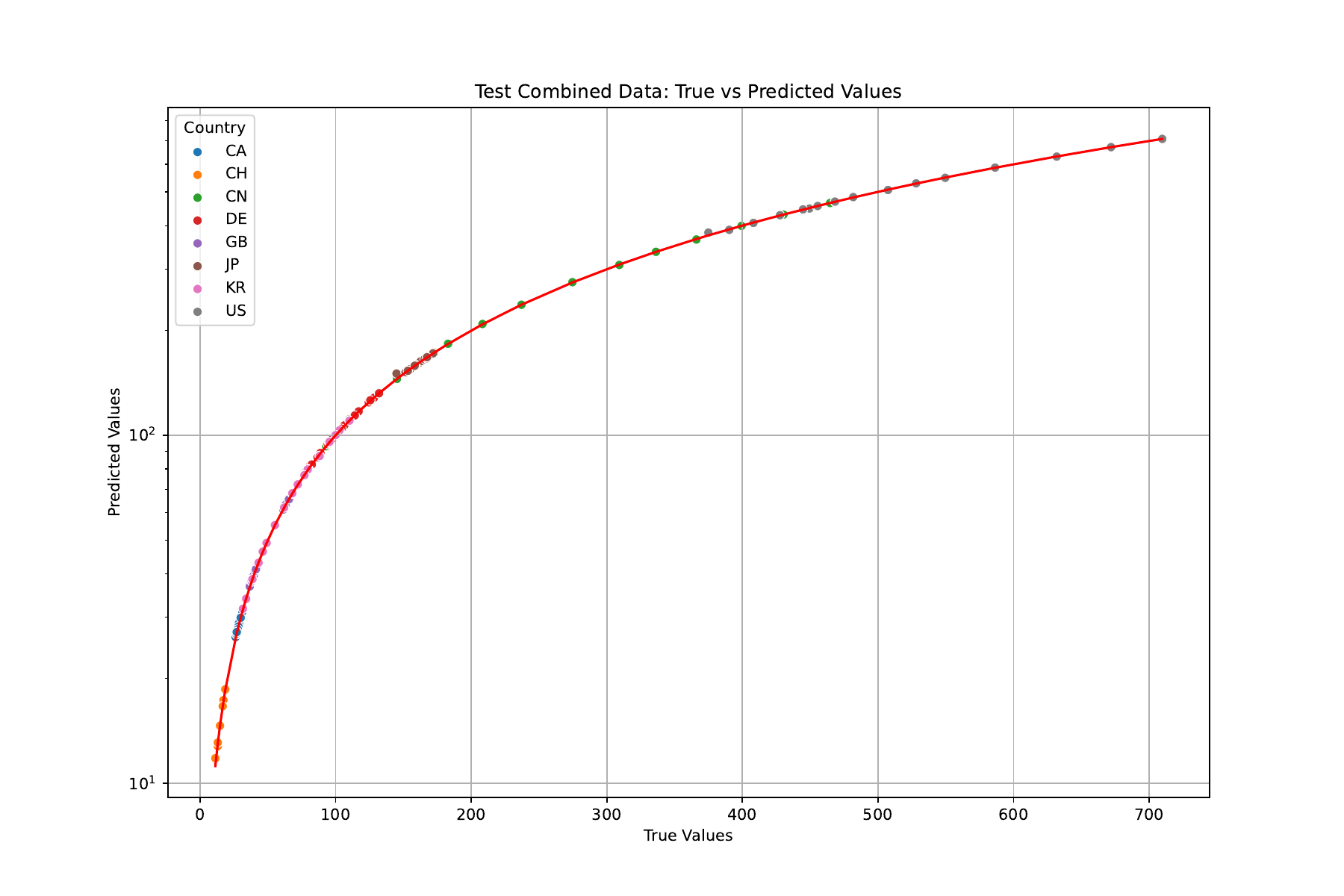}
        \caption{True vs. Predicted R\&D Expenditures in USD (bn) for Different Countries.}
        \label{fig:annual_comparison_agtwrd_log}
    \end{subfigure}
    \begin{subfigure}[c]{0.4\linewidth}
        \centering
        \includegraphics[width=\linewidth]{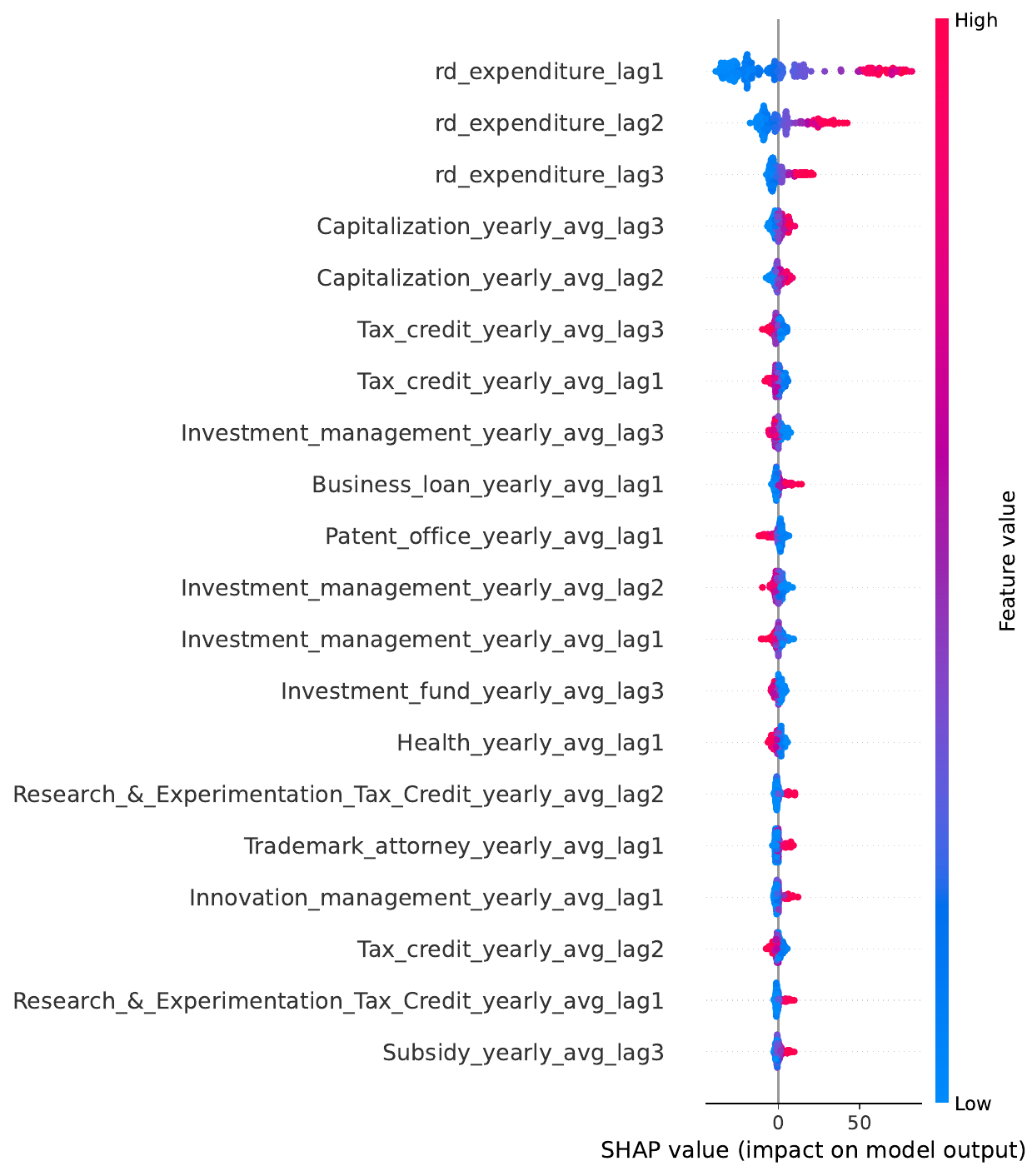}
        \caption{SHAP summary plot.}
        \label{fig:shap_summary_plt_agtwrd}
    \end{subfigure}
\end{figure}

\newpage
\subsection{\textit{MGT} Configuration}

\begin{figure}[!htb]
    \centering
    \caption{Analysis for \textit{MGT} Configuration.}         
    \begin{subfigure}[c]{0.45\linewidth}
        \centering
        \includegraphics[width=\linewidth]{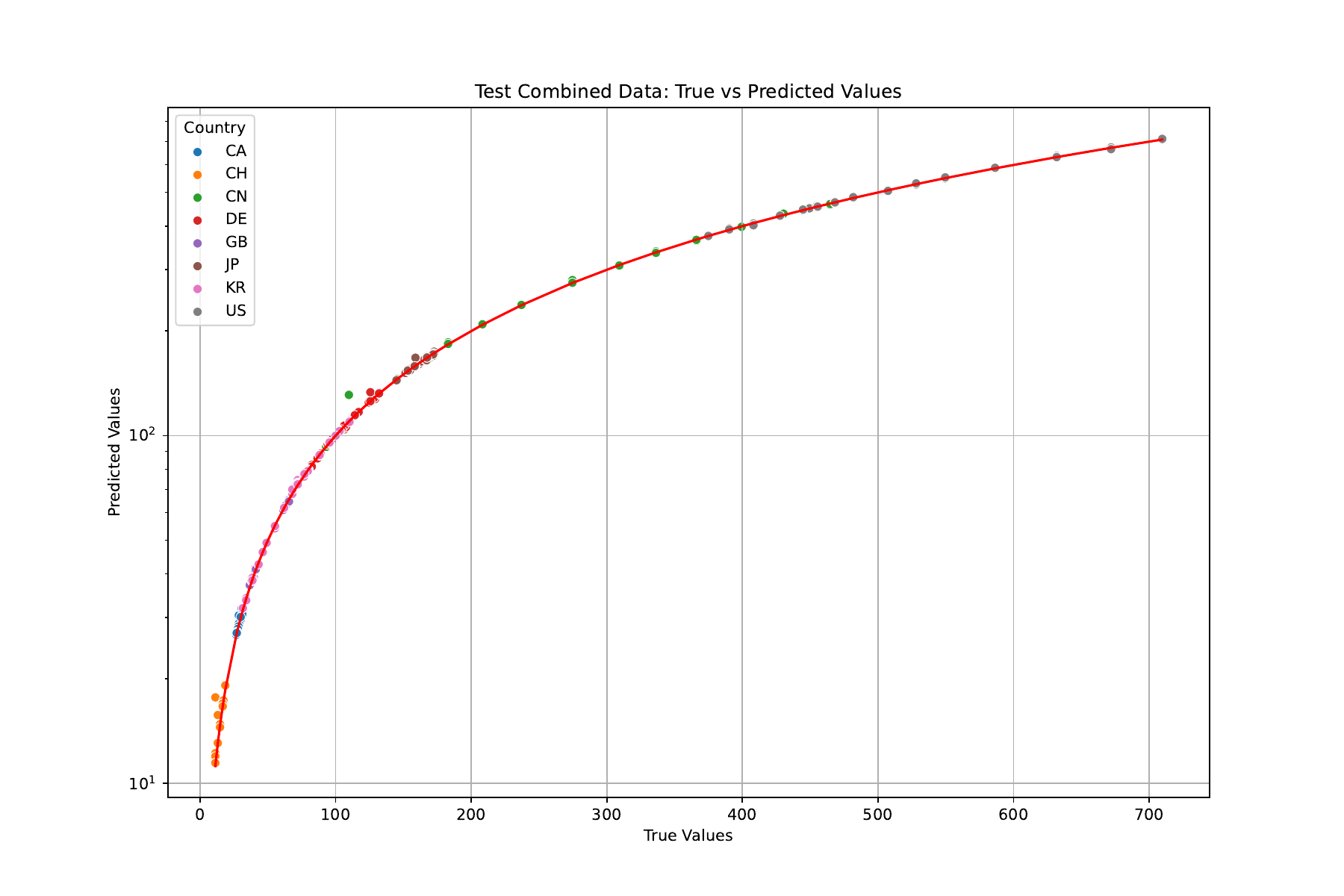}
        \caption{True vs. Predicted R\&D Expenditures in USD (bn) for Different Countries.}
        \label{fig:annual_comparison_mgt_log}
    \end{subfigure}
    \begin{subfigure}[c]{0.4\linewidth}
        \centering
        \includegraphics[width=\linewidth]{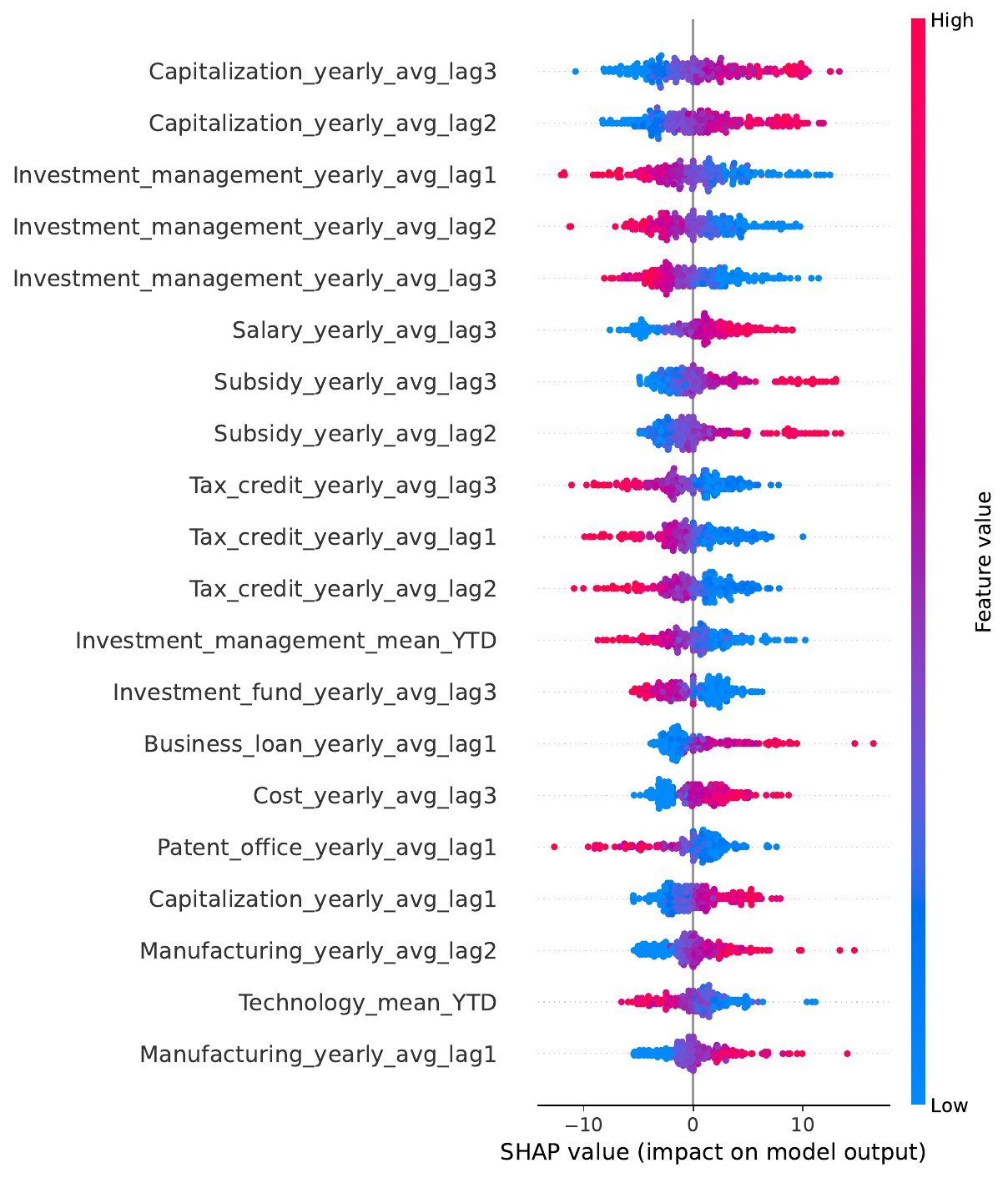}
        \caption{SHAP summary plot.}
        \label{fig:shap_summary_plt_mgt}
    \end{subfigure}
\end{figure}

\subsection{\textit{MGTwRD} Configuration}

\begin{figure}[!htb]
    \centering
    \caption{Analysis for \textit{MGTwRD} Configuration.}     
    \begin{subfigure}[c]{0.45\linewidth}
        \centering
        \includegraphics[width=\linewidth]{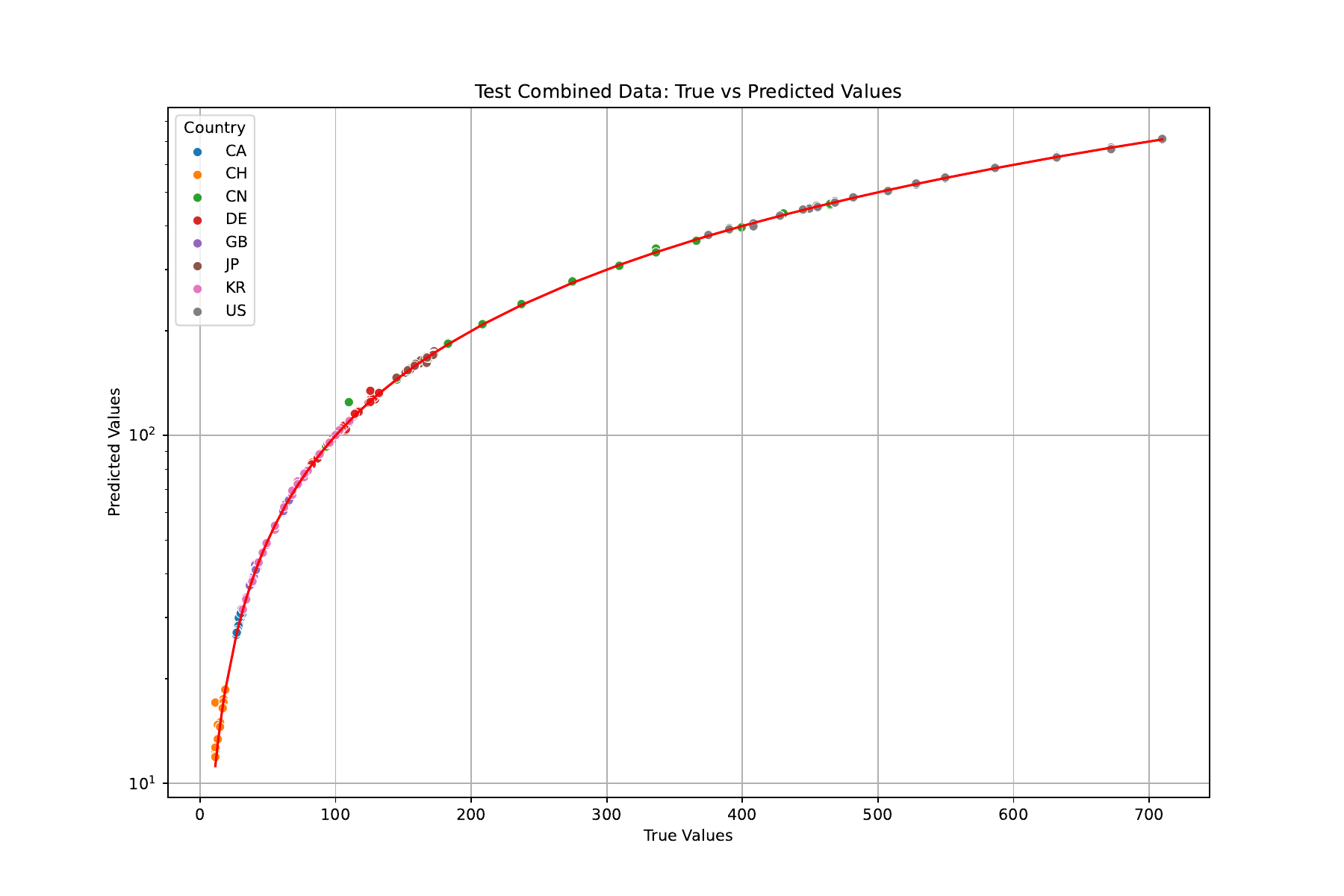}
        \caption{True vs. Predicted R\&D Expenditures in USD (bn) for Different Countries.}
        \label{fig:annual_comparison_mgtwrd_log}
    \end{subfigure}
    \begin{subfigure}[c]{0.4\linewidth}
        \centering
        \includegraphics[width=\linewidth]{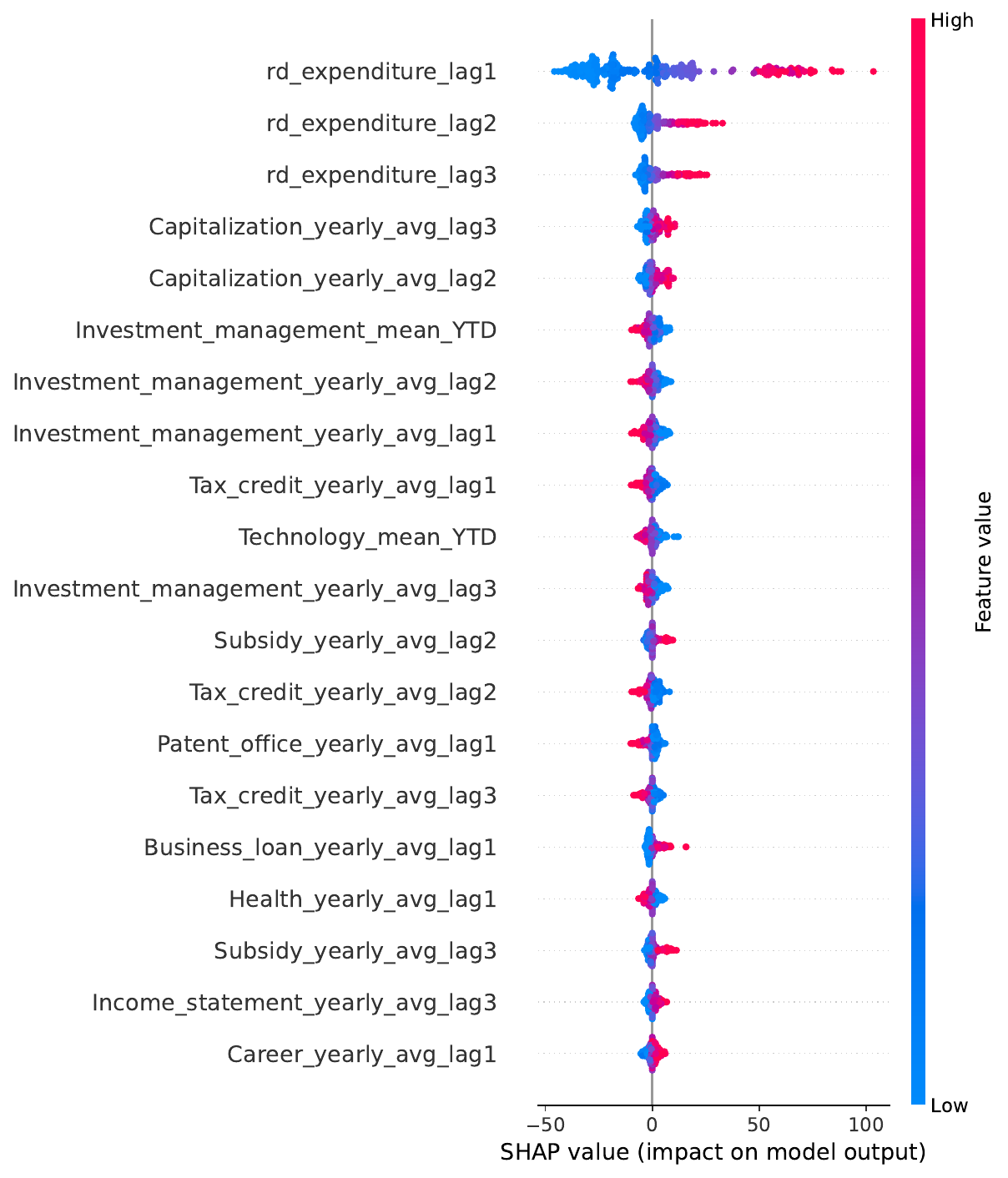}
        \caption{SHAP summary plot.}
        \label{fig:shap_summary_plt_mgtwrd}
    \end{subfigure}
\end{figure}

\subsection{\textit{Macros} Configuration}

\begin{figure}[H]
    \centering
    \caption{Analysis for \textit{Macros} Configuration.}     
    \begin{subfigure}[c]{0.45\linewidth}
        \centering
        \includegraphics[width=\linewidth]{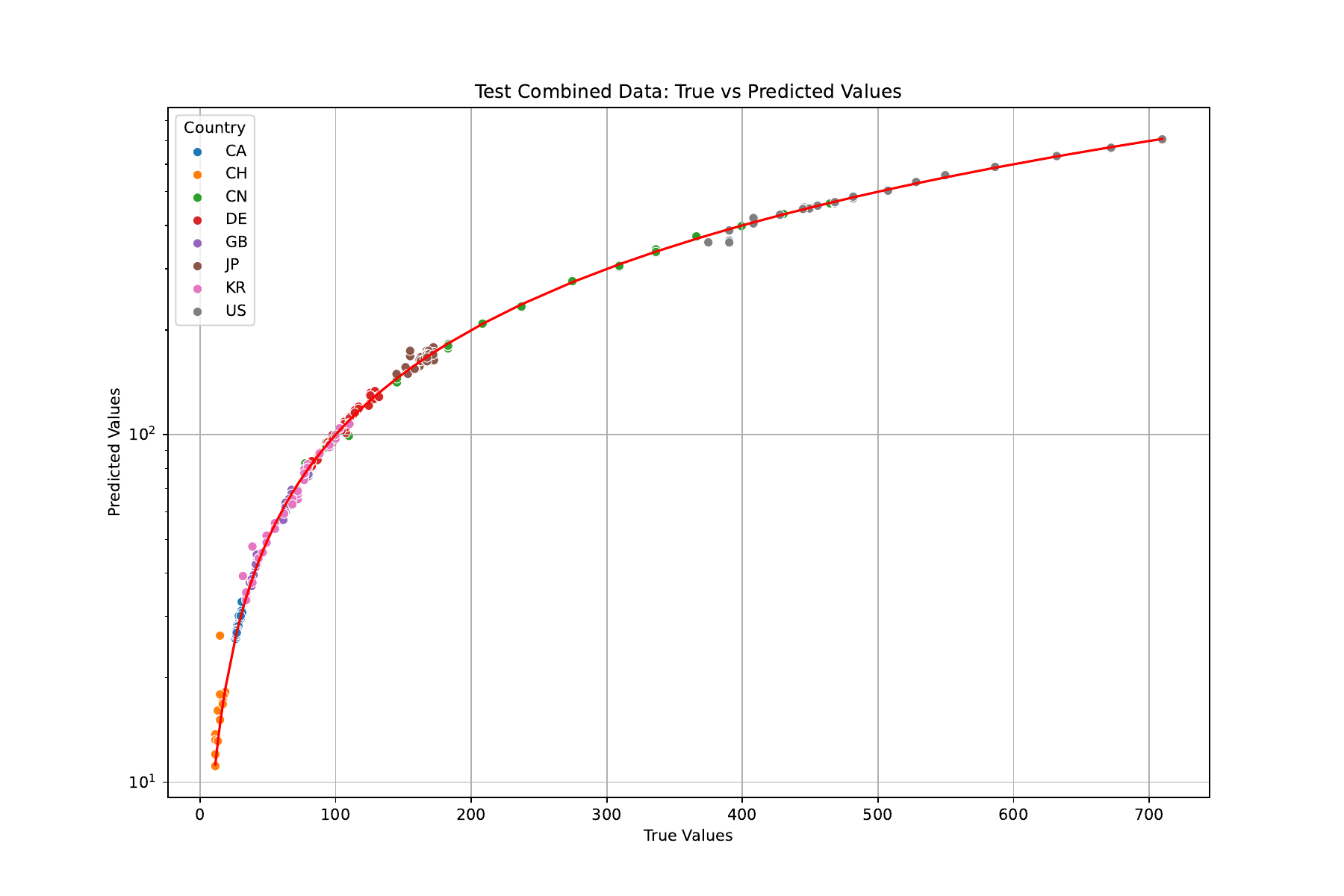}
        \caption{True vs. Predicted R\&D Expenditures in USD (bn) for Different Countries.}
        \label{fig:annual_comparison_macros_log}
    \end{subfigure}
    \begin{subfigure}[c]{0.4\linewidth}
        \centering
        \includegraphics[width=\linewidth]{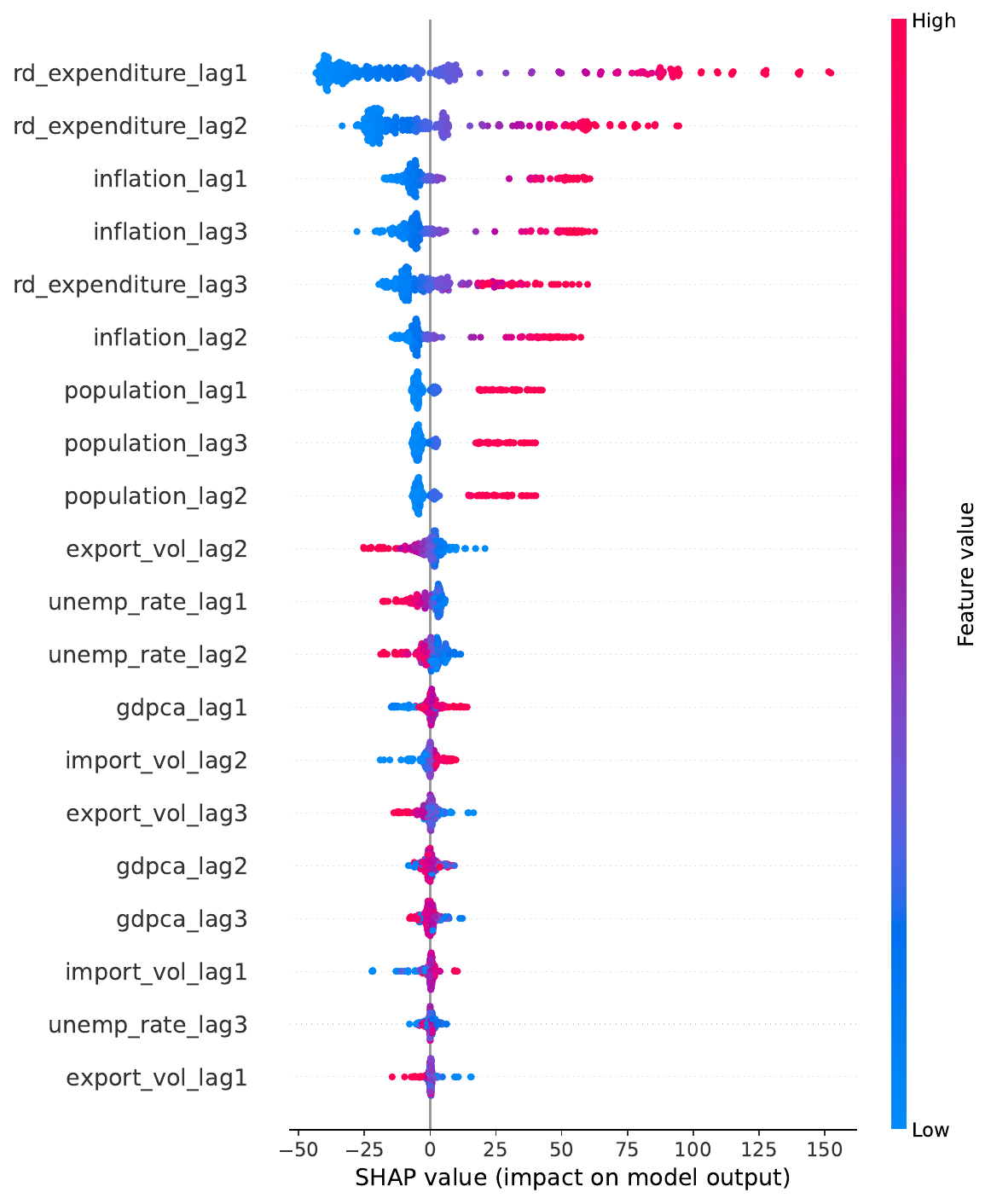}
        \caption{SHAP summary plot.}
        \label{fig:shap_summary_plt_macros}
    \end{subfigure}
\end{figure}

\subsection{\textit{LagRD} Configuration}

\begin{figure}[!htb]
    \centering
    \caption{Analysis for \textit{LagRD} Configuration.}    
    \begin{subfigure}[c]{0.45\linewidth}
        \centering
        \includegraphics[width=\linewidth]{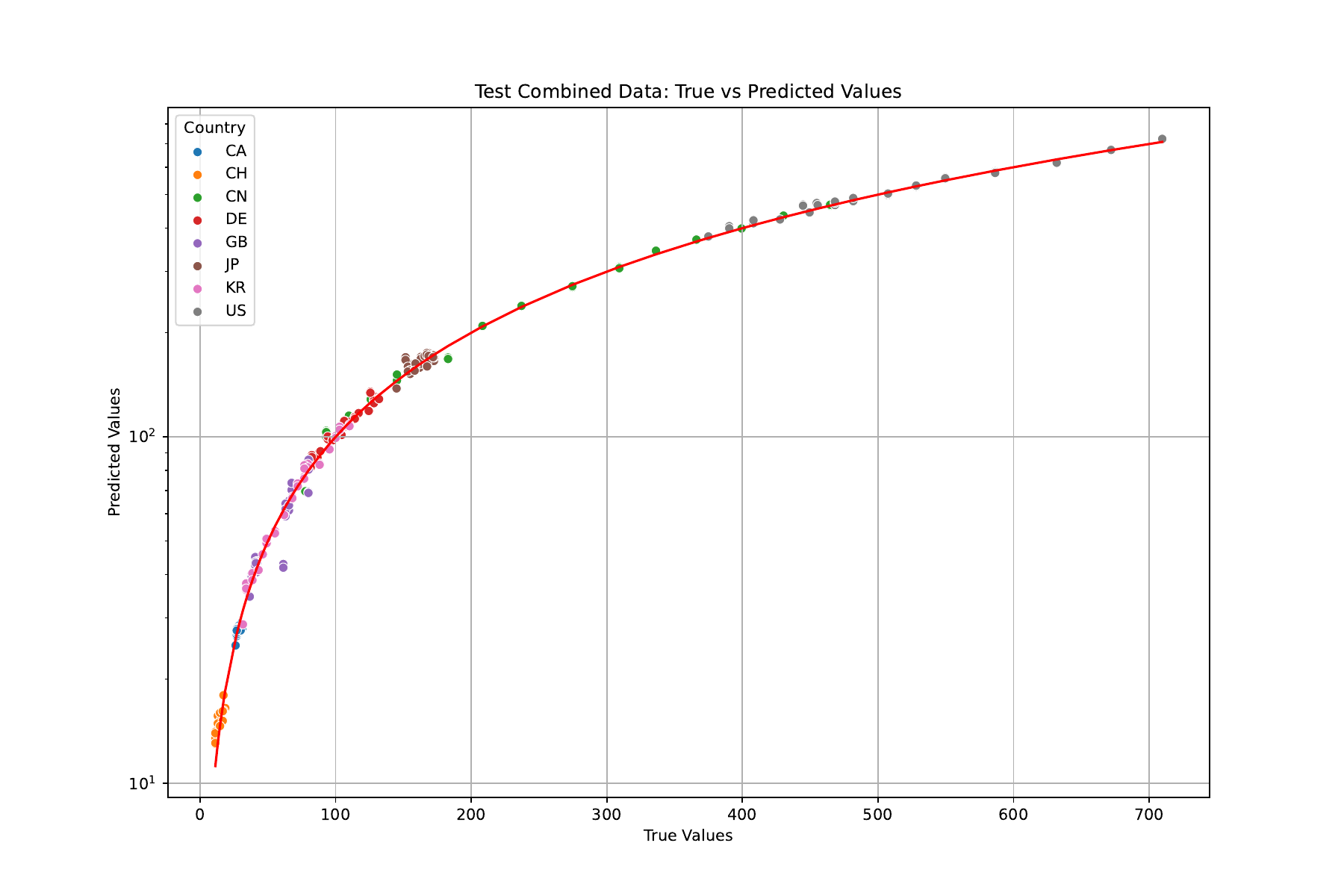}
        \caption{True vs. Predicted R\&D Expenditures in USD (bn) for Different Countries.}
        \label{fig:annual_comparison_lagrd_log}
    \end{subfigure}
    \begin{subfigure}[c]{0.4\linewidth}
        \centering
        \includegraphics[width=\linewidth]{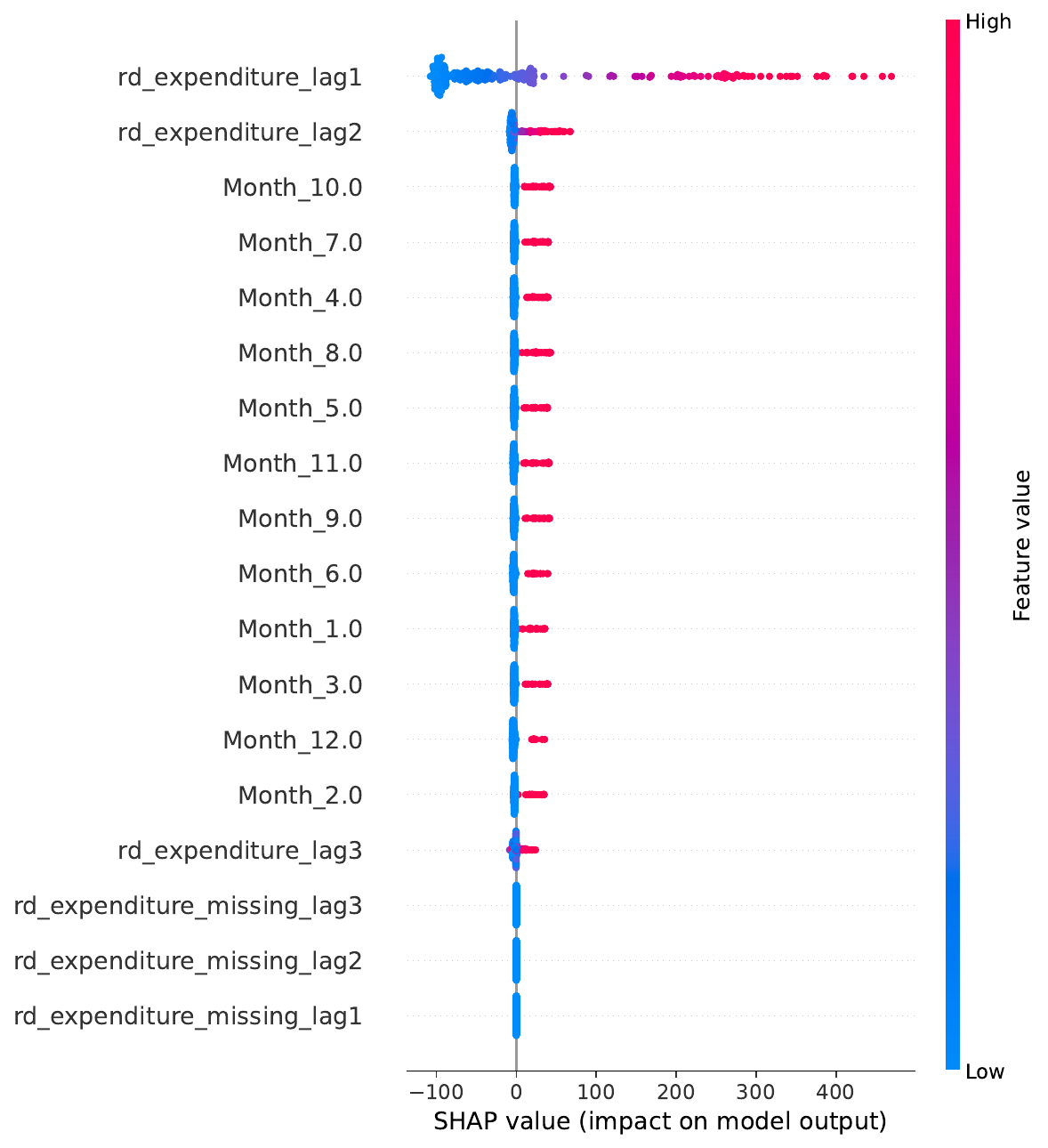}
        \caption{SHAP summary plot.}
        \label{fig:shap_summary_plt_lagrd}
    \end{subfigure}
\end{figure}

%%%%%%%%%%%%%%%%%%%%%%%%%%%%%%%%%%%%%%%%%%%%%%%%%%%%%%%%%%%%%%%%%%%%%%%%%%%%%%%%%%%%
\newpage
\section{Country-level out-of-sample performances}
\label{app:oos}

\subsection{\textit{AllVar} Configuration}
\begin{figure}[!htb]
    \centering
    \caption{Yearly R\&D expenditures for selected countries in Europe: True Values vs. Estimates.}
    \label{fig:predict_rd_expenditure_2_allvar}    
    \begin{subfigure}{0.45\textwidth}
        \centering
         {\includegraphics[width=\textwidth]{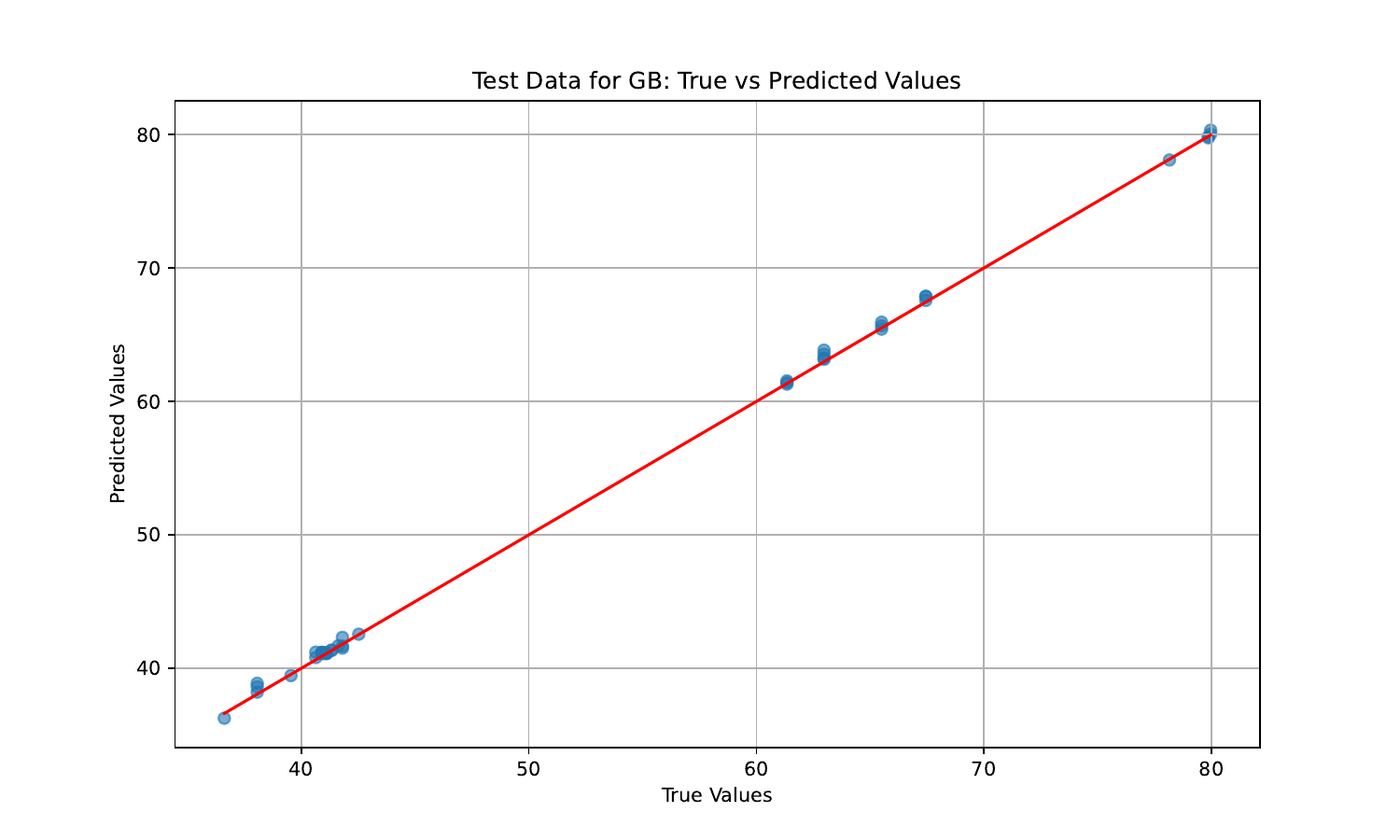}}
        \caption{United Kingdom}
        \label{fig:gb_predict_allvar}
    \end{subfigure}\hfill
    \begin{subfigure}{0.45\textwidth}
        \centering
         {\includegraphics[width=\textwidth]{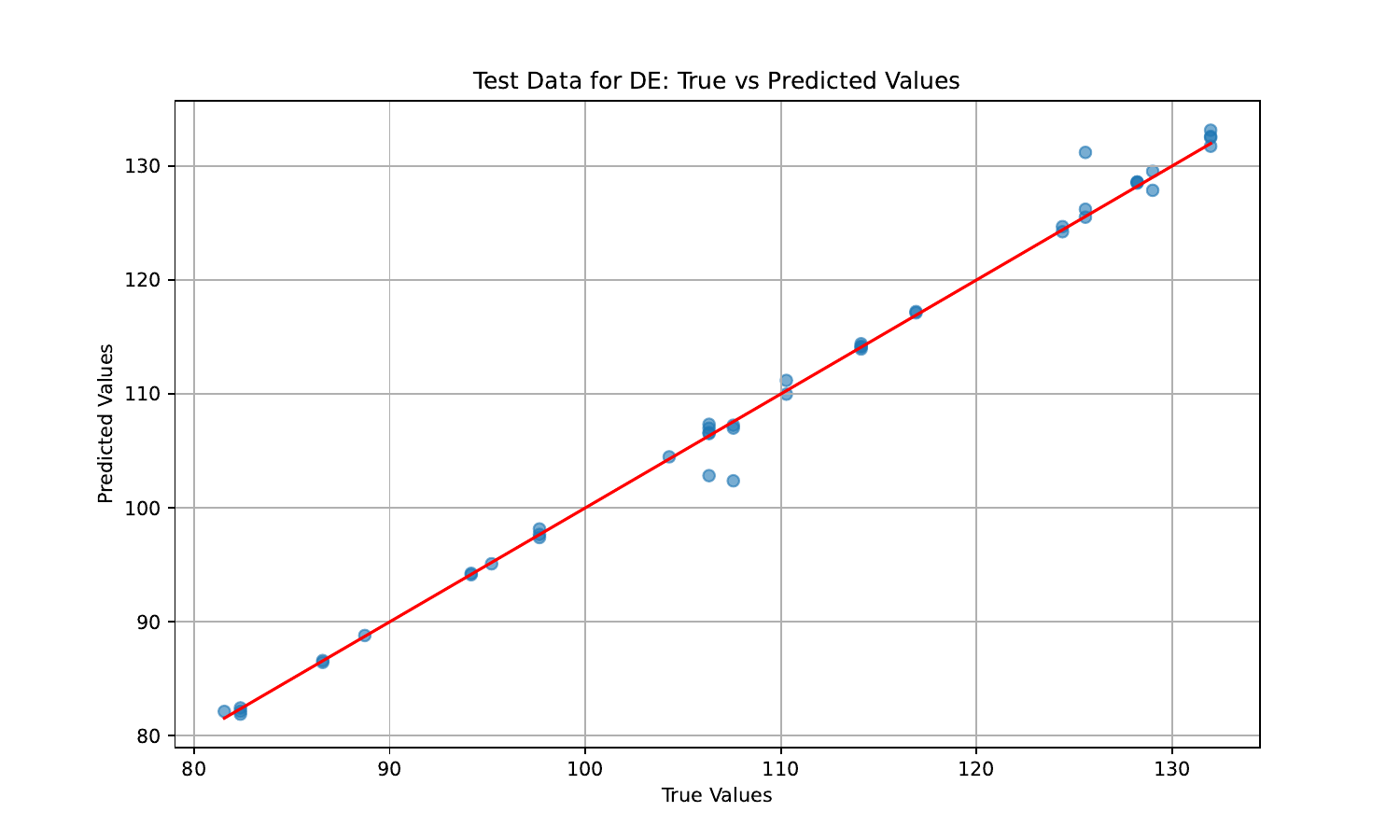}}
        \caption{Germany}
        \label{fig:de_predict_allvar}
    \end{subfigure}
\end{figure}

\begin{figure}[!htb]
    \centering
    \caption{Yearly R\&D expenditures for selected countries in North America and East Asia: True Values vs. Estimates.}
    \label{fig:predict_rd_expenditure_3_allvar}    
    \begin{subfigure}{0.45\textwidth}
        \centering
         {\includegraphics[width=\textwidth]{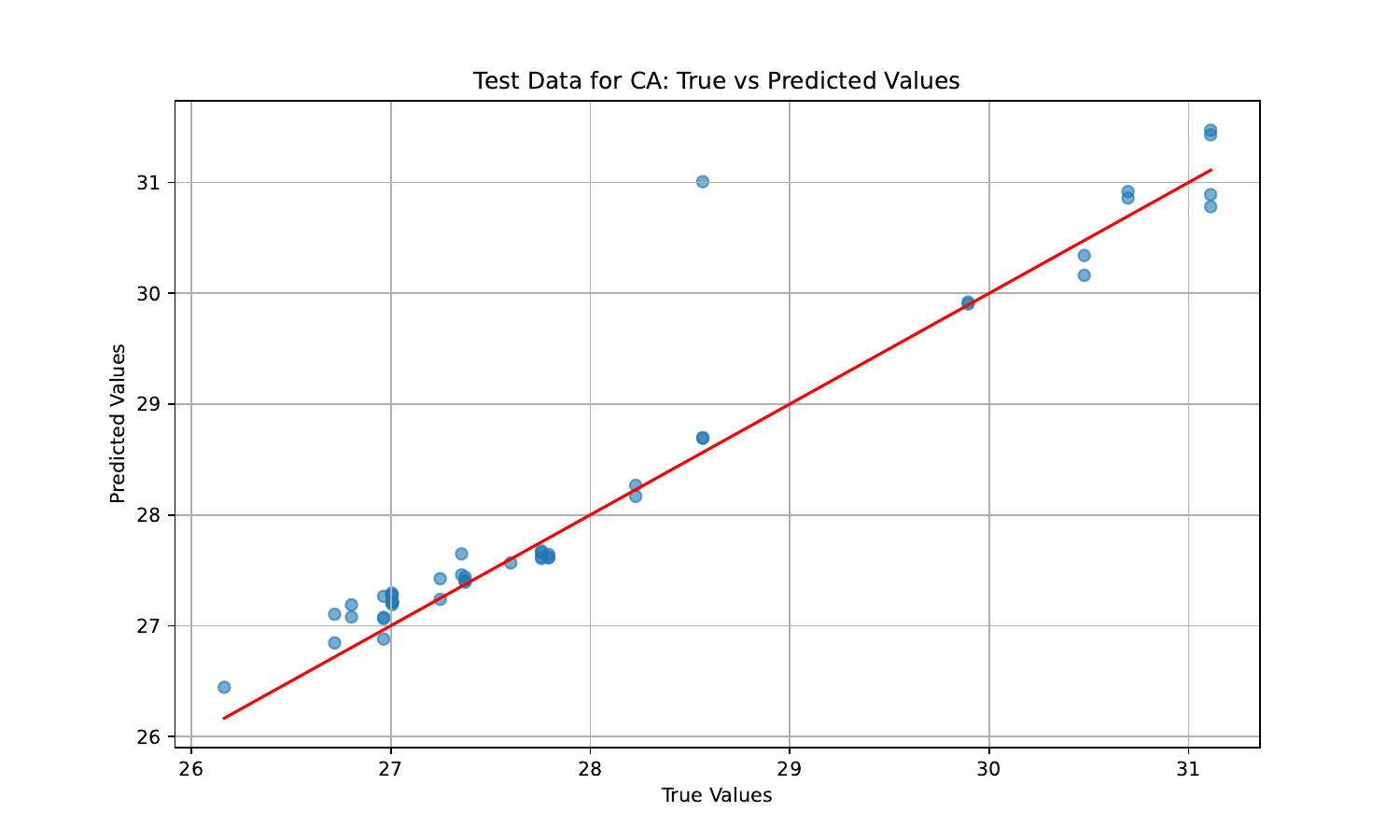}}
        \caption{Canada}
        \label{fig:ca_predict_allvar}
    \end{subfigure}\hfill
    \begin{subfigure}{0.45\textwidth}
        \centering
         {\includegraphics[width=\textwidth]{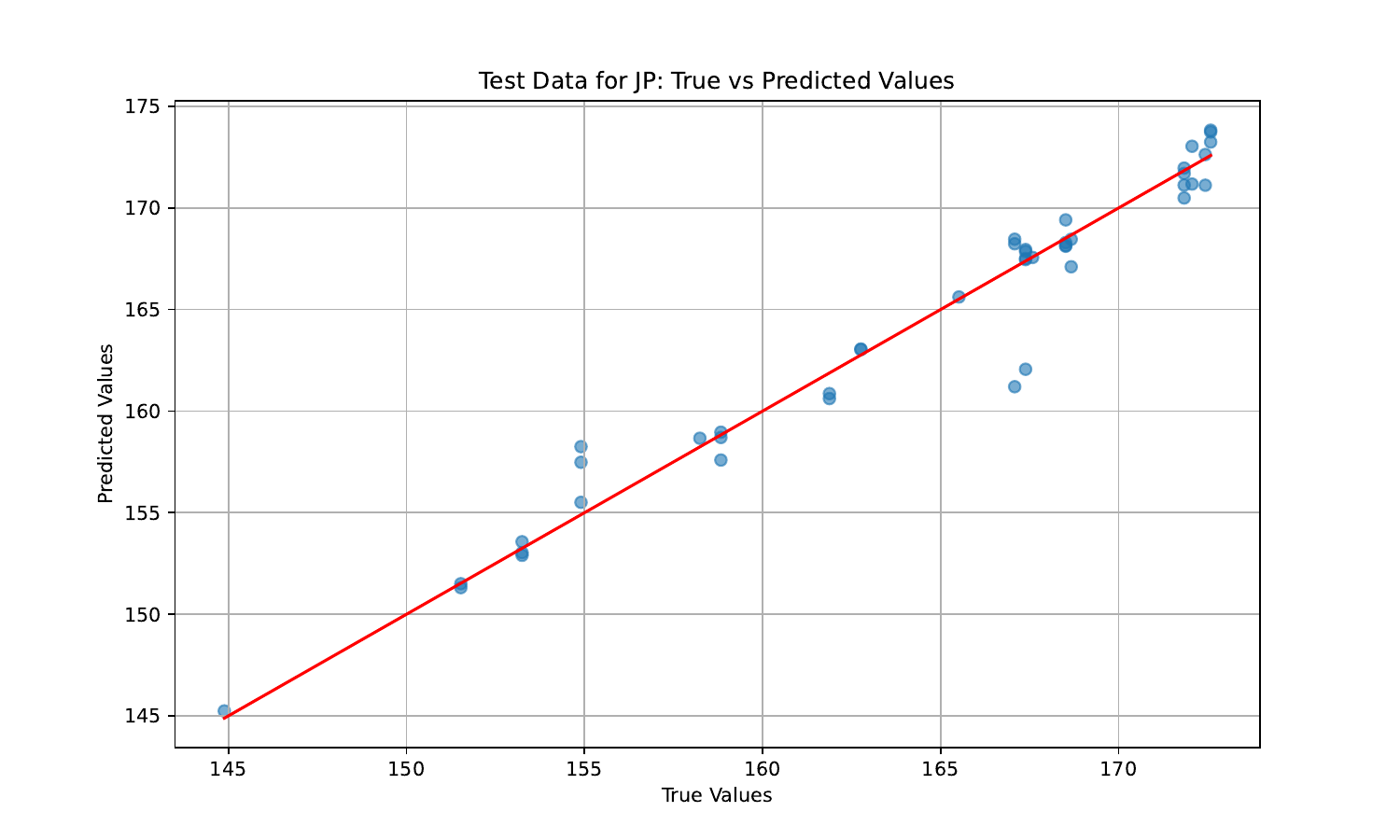}}
        \caption{Japan}
        \label{fig:jp_predict_allvar}
    \end{subfigure}
\end{figure}

\begin{figure}[H]
    \centering
    \caption{Yearly R\&D expenditures for selected countries in East Asia and Europe: True Values vs. Estimates.}
    \label{fig:predict_rd_expenditure_4_allvar}    
    \begin{subfigure}{0.45\textwidth}
        \centering
         {\includegraphics[width=\textwidth]{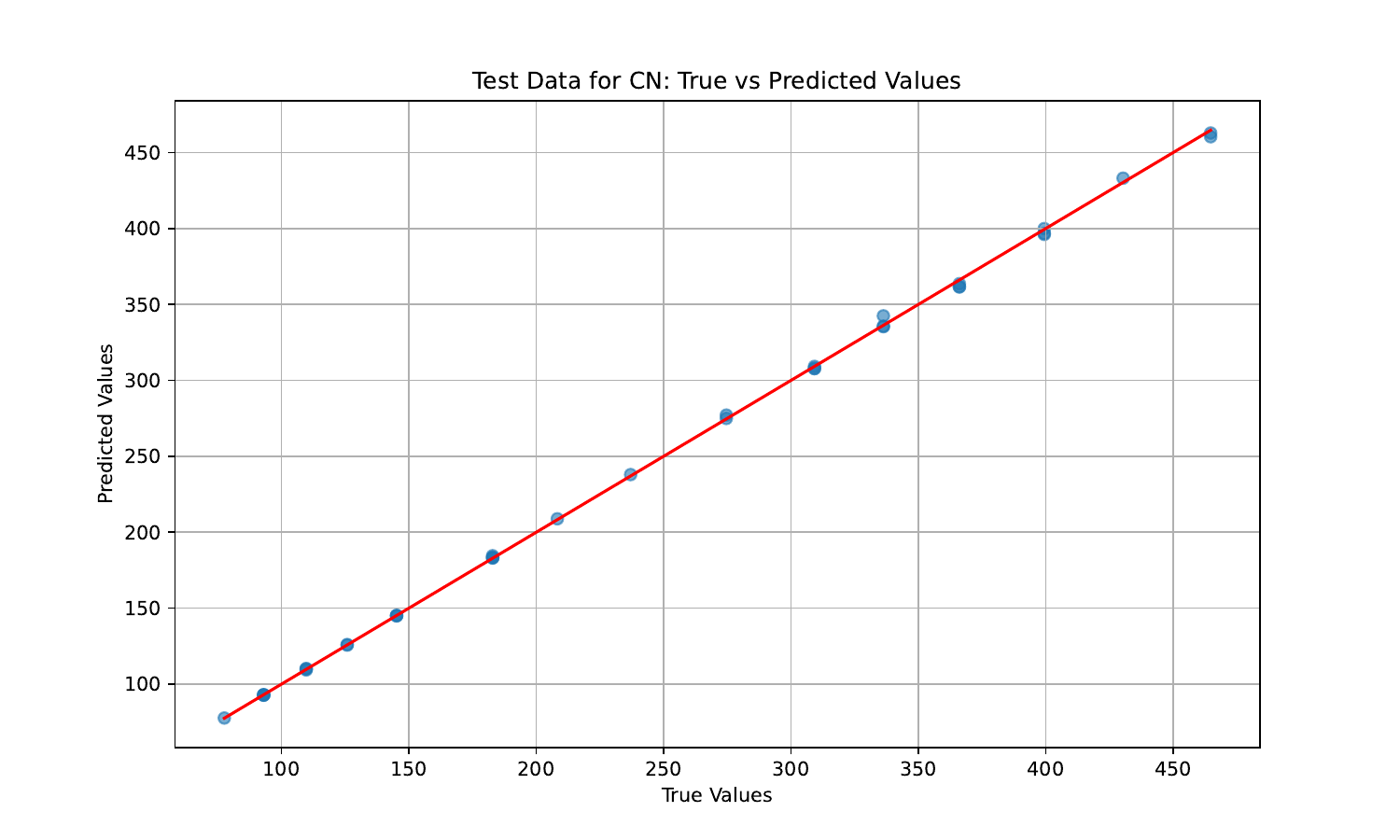}}
        \caption{China}
        \label{fig:cn_predict_allvar}
    \end{subfigure}\hfill
    \begin{subfigure}{0.45\textwidth}
        \centering
         {\includegraphics[width=\textwidth]{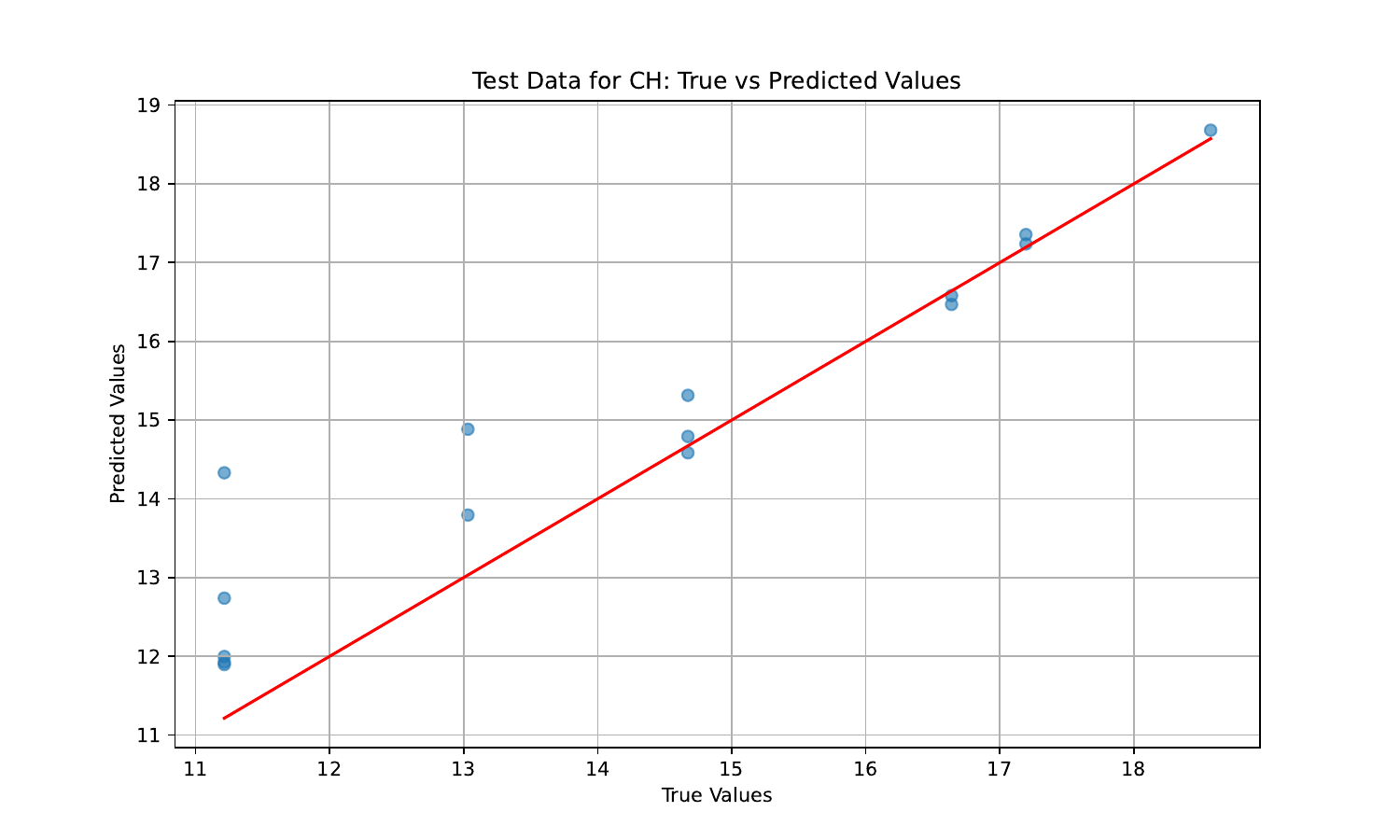}}
        \caption{Switzerland}
        \label{fig:ch_predict_allvar}
    \end{subfigure}
\end{figure}

In the following figures \ref{fig:predict_rd_expenditure_1_allvar}, \ref{fig:predict_rd_expenditure_2_allvar}, \ref{fig:predict_rd_expenditure_3_allvar} and \ref{fig:predict_rd_expenditure_4_allvar}, we illustrate the out-of-sample performance of the prediction model, for each country, in \textit{AllVar} configuration, as the most comprehensive configuration in terms of input vector.
\begin{figure}[!htb]
    \centering
    \caption{Yearly R\&D expenditures for selected countries in North America and Asia: True Values vs. Estimates.}
    \label{fig:predict_rd_expenditure_1_allvar}    
    \begin{subfigure}{0.45\textwidth}
        \centering
         {\includegraphics[width=\textwidth]{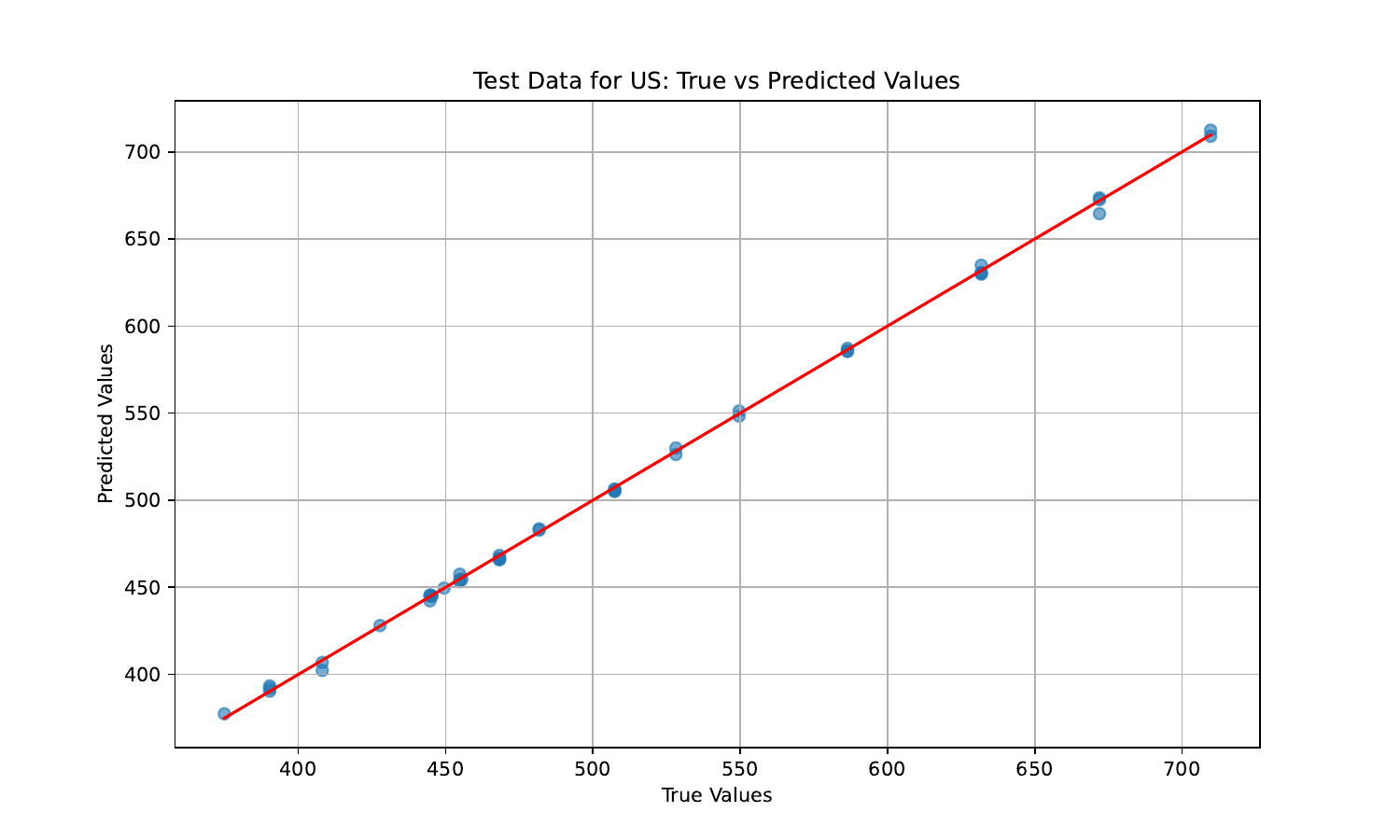}}
        \caption{United States}
        \label{fig:us_predict_allvar}
    \end{subfigure}\hfill
    \begin{subfigure}{0.45\textwidth}
        \centering
         {\includegraphics[width=\textwidth]{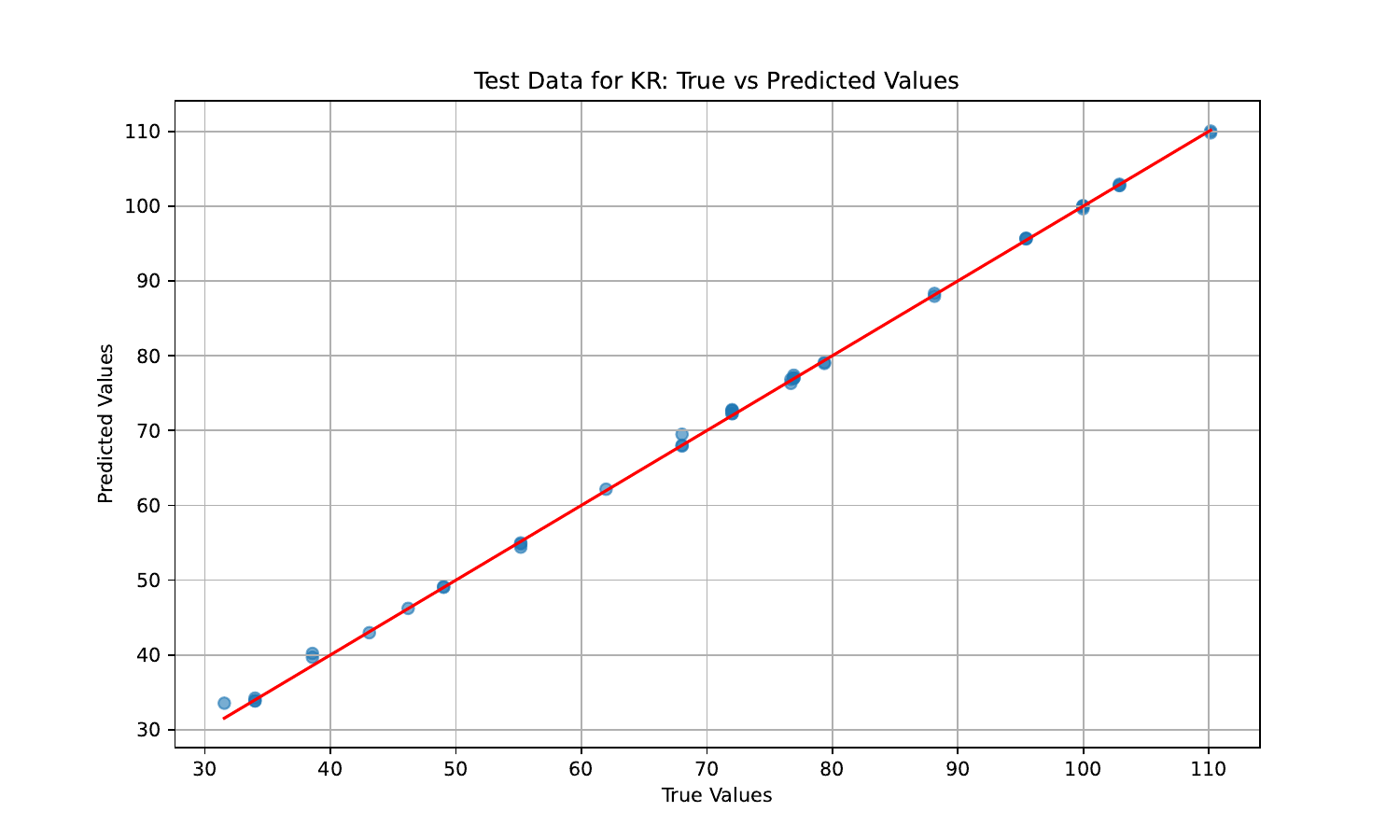}}
        \caption{Korea}
        \label{fig:kr_predict_allvar}
    \end{subfigure}
\end{figure}

\subsection{\textit{AGT} Configuration}
\begin{figure}[H]
    \centering
    \caption{Yearly R\&D expenditures for selected countries in North America and Asia: True Values vs. Estimates.}
    \label{fig:predict_rd_expenditure_1}    
    \begin{subfigure}{0.45\textwidth}
        \centering
         {\includegraphics[width=\textwidth]{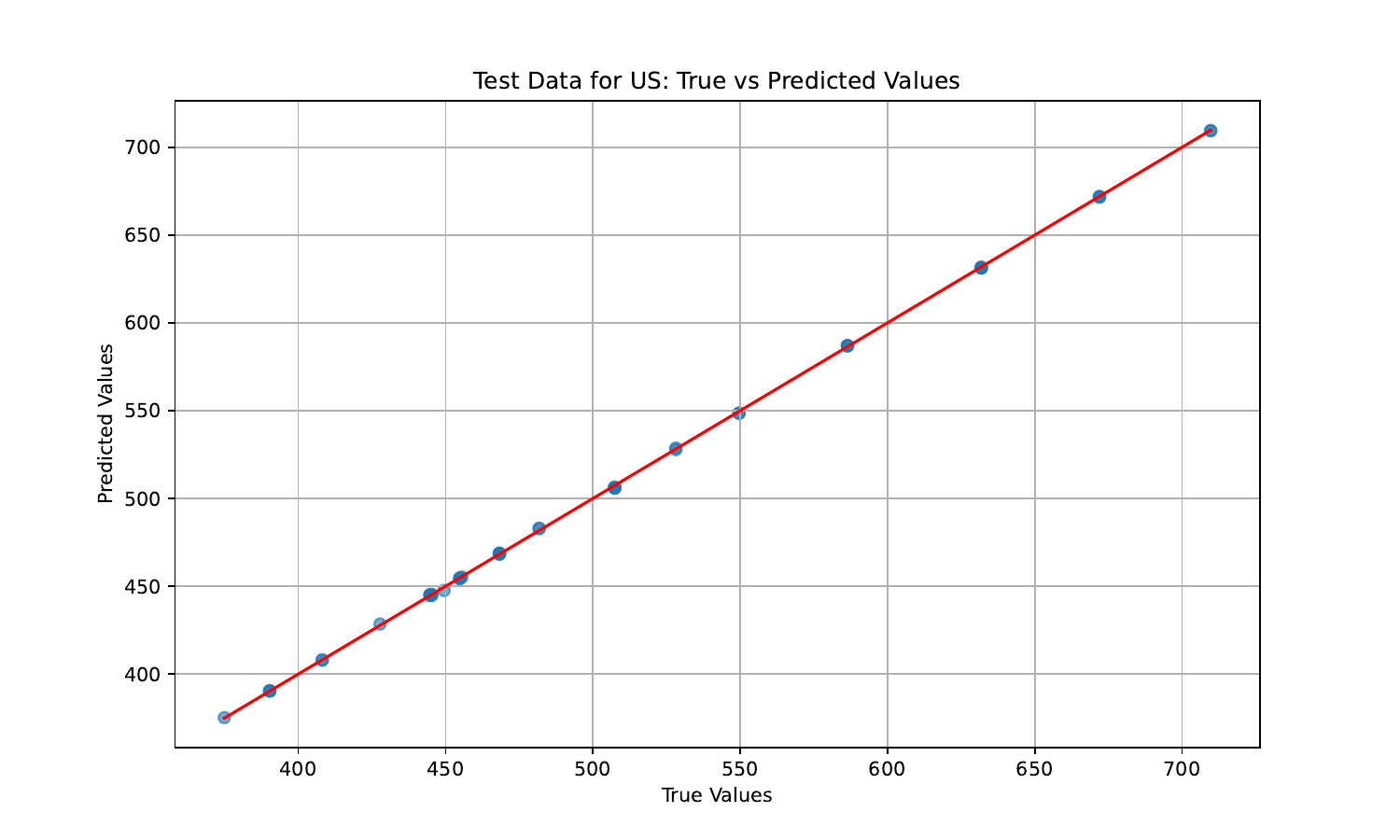}}
        \caption{United States}
        \label{fig:us_predict}
    \end{subfigure}\hfill
    \begin{subfigure}{0.45\textwidth}
        \centering
         {\includegraphics[width=\textwidth]{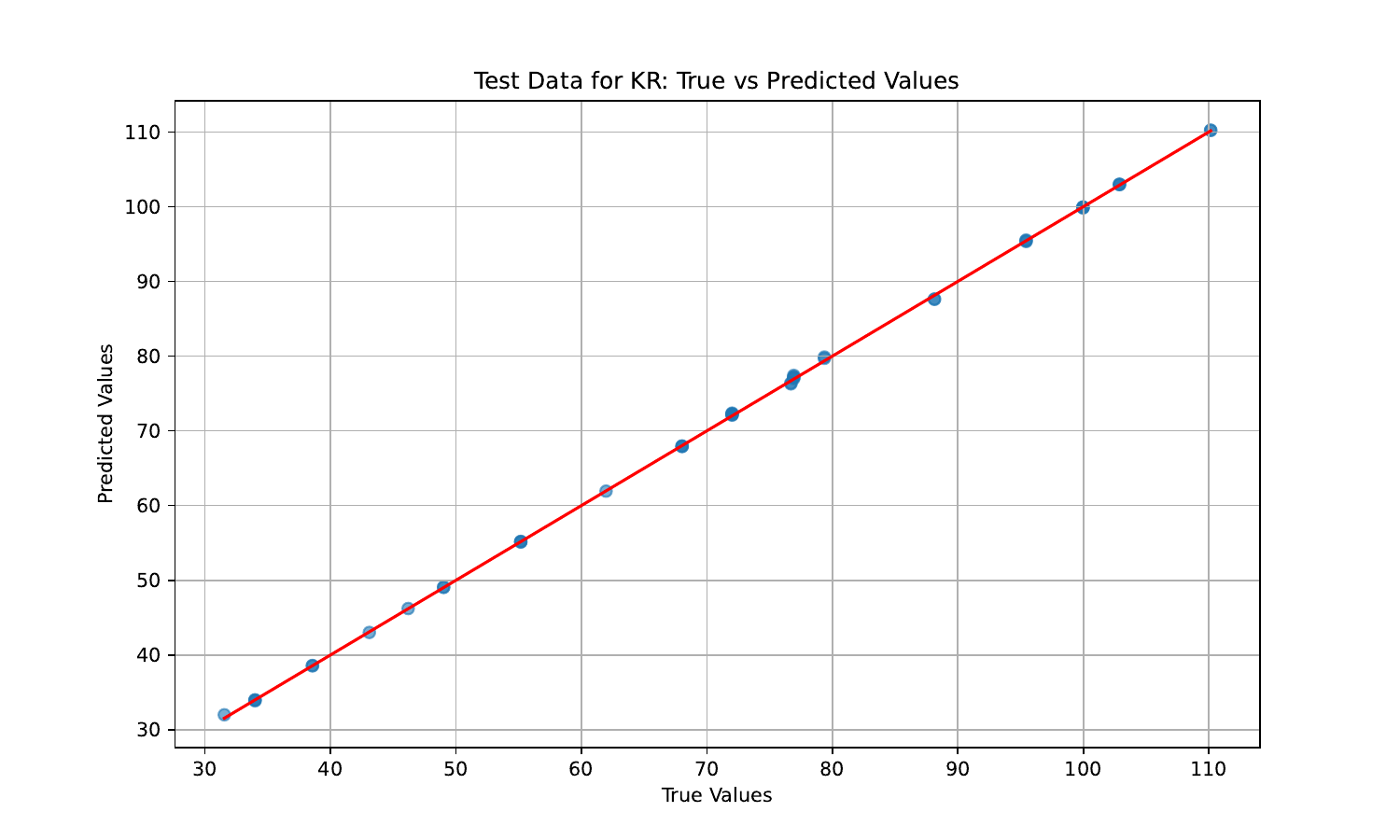}}
        \caption{Korea}
        \label{fig:kr_predict}
    \end{subfigure}
\end{figure}

\begin{figure}[H]
    \centering
    \caption{Yearly R\&D expenditures for selected countries in Europe: True Values vs. Estimates.}
    \label{fig:predict_rd_expenditure_2}    
    \begin{subfigure}{0.45\textwidth}
        \centering
         {\includegraphics[width=\textwidth]{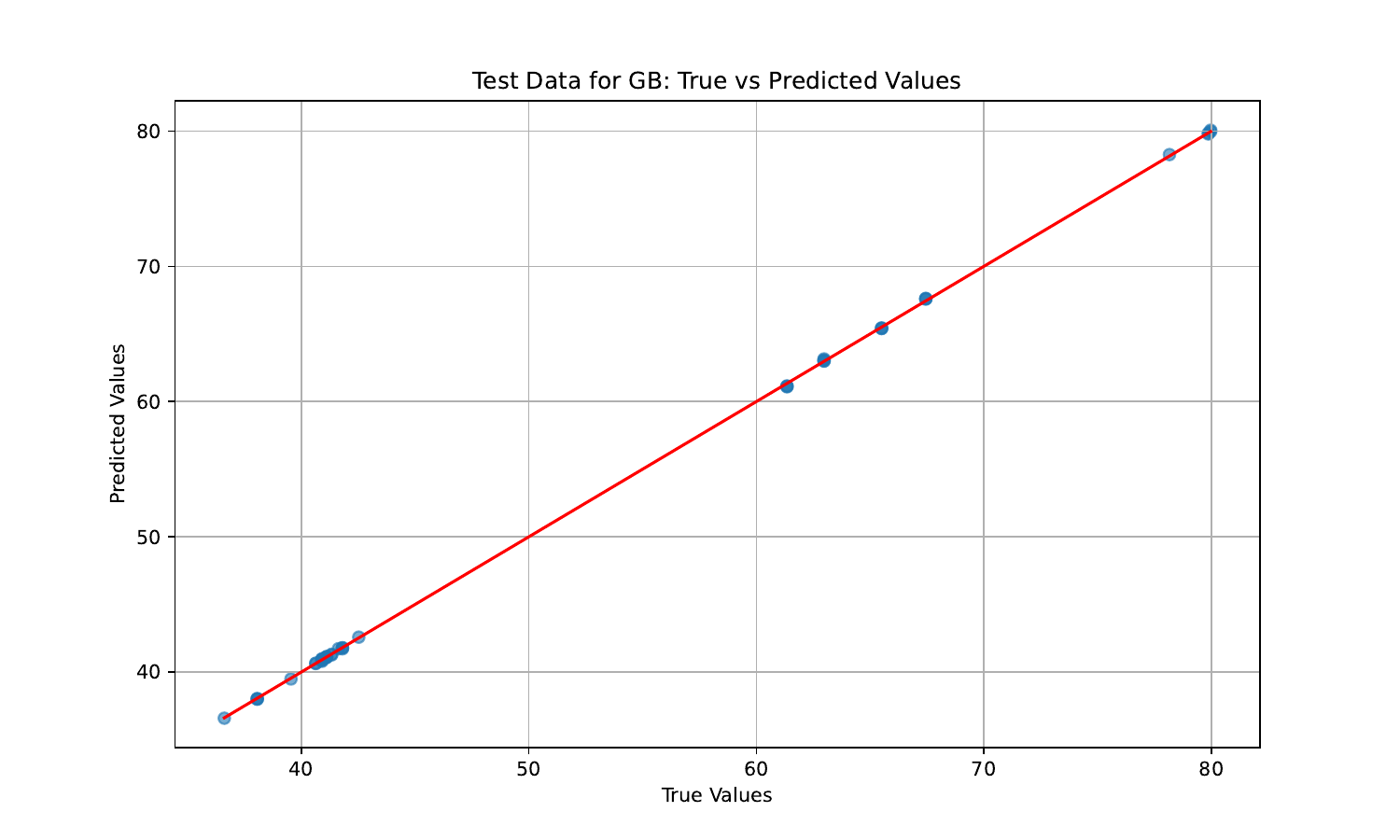}}
        \caption{United Kingdom}
        \label{fig:gb_predict}
    \end{subfigure}\hfill
    \begin{subfigure}{0.45\textwidth}
        \centering
         {\includegraphics[width=\textwidth]{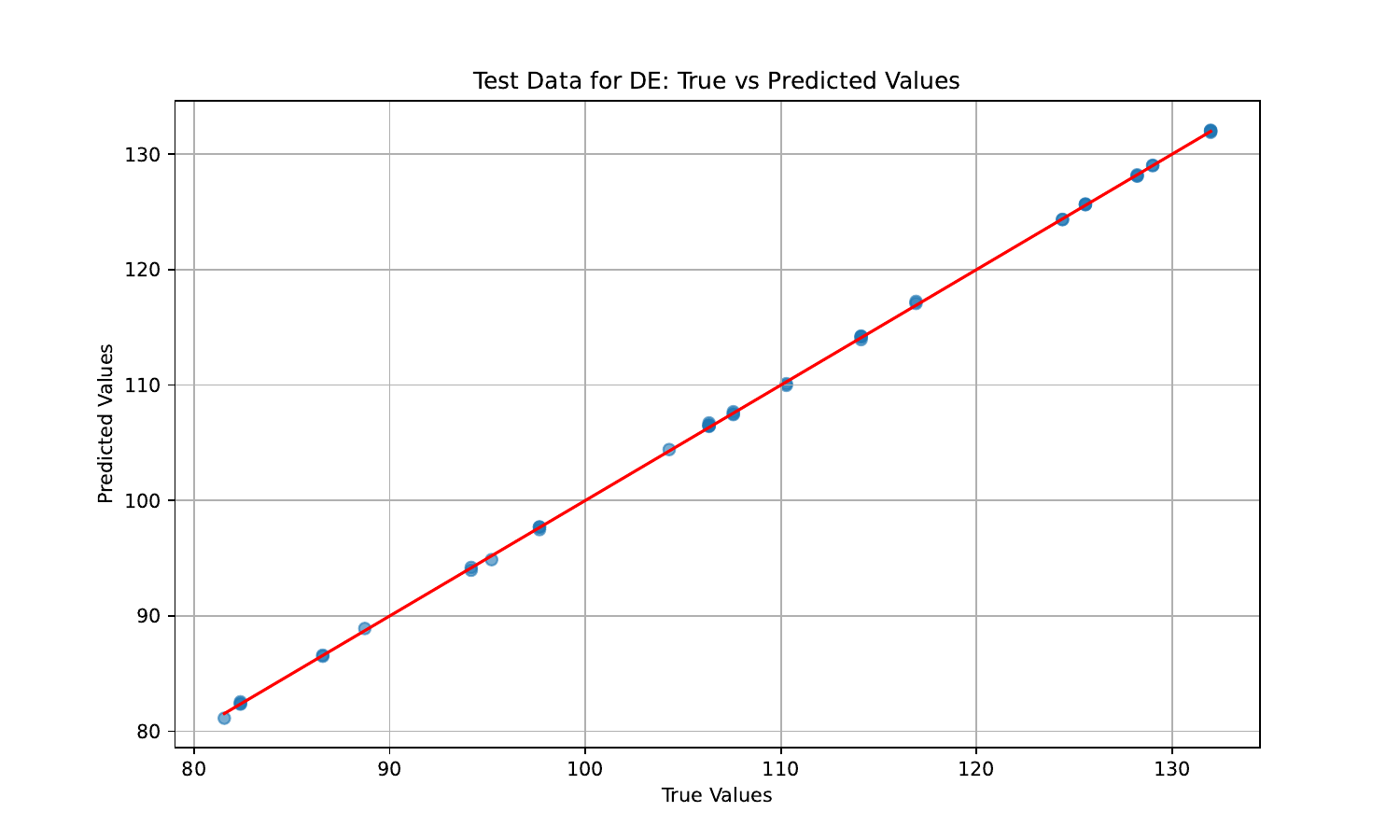}}
        \caption{Germany}
        \label{fig:de_predict}
    \end{subfigure}
\end{figure}

\begin{figure}[H]
    \centering
    \caption{Yearly R\&D expenditures for selected countries in North America and East Asia: True Values vs. Estimates.}
    \label{fig:predict_rd_expenditure_3}    
    \begin{subfigure}{0.45\textwidth}
        \centering
         {\includegraphics[width=\textwidth]{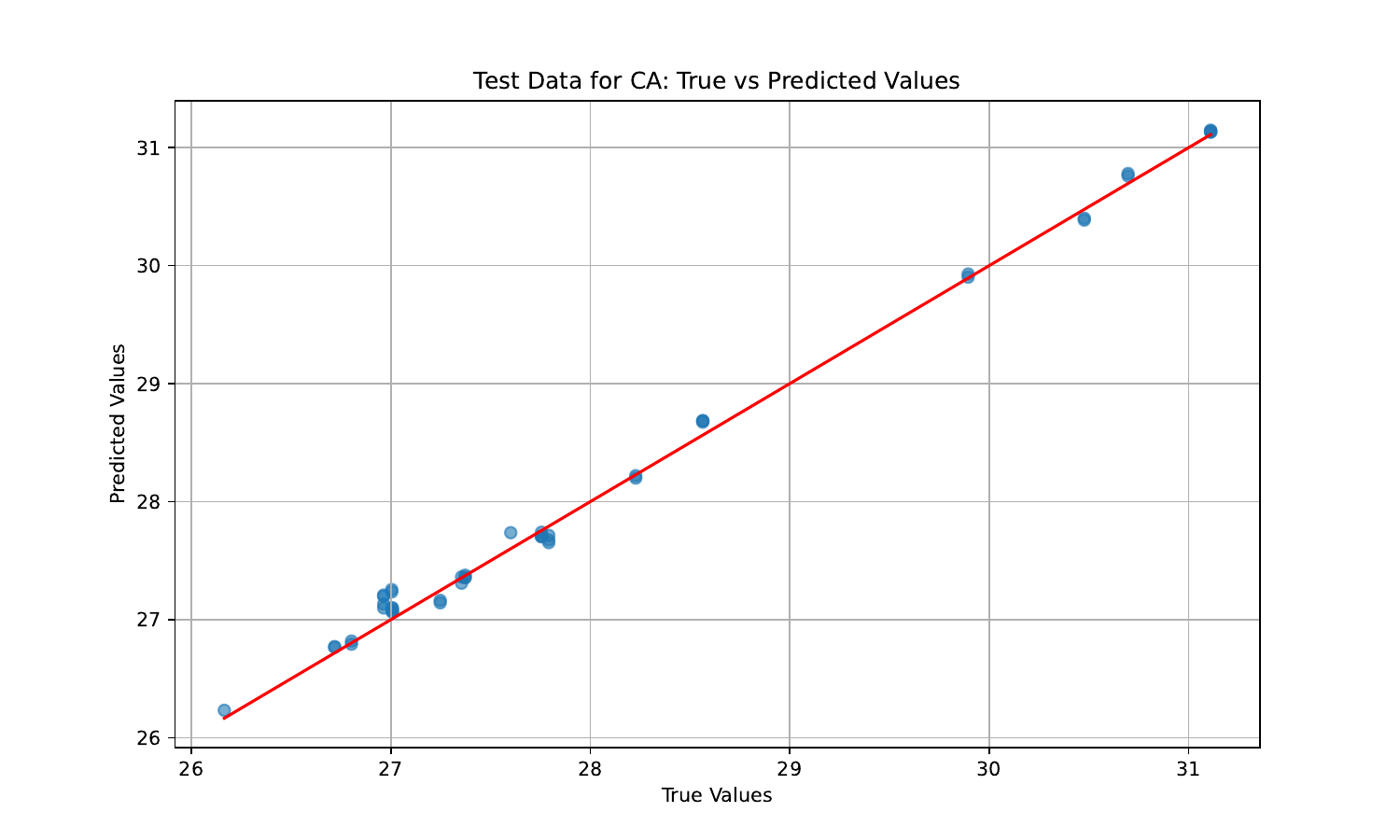}}
        \caption{Canada}
        \label{fig:ca_predict}
    \end{subfigure}\hfill
    \begin{subfigure}{0.45\textwidth}
        \centering
         {\includegraphics[width=\textwidth]{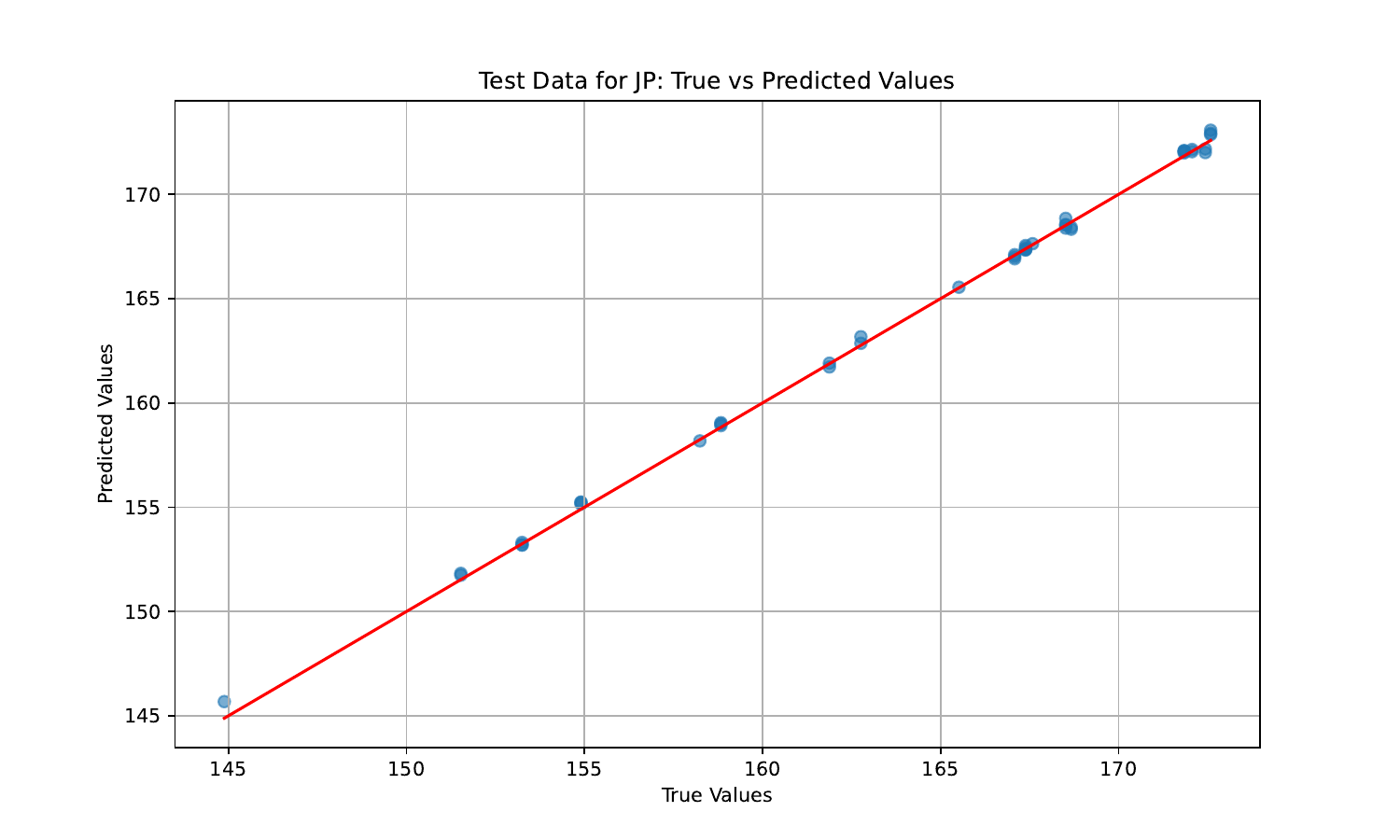}}
        \caption{Japan}
        \label{fig:jp_predict}
    \end{subfigure}
\end{figure}

\begin{figure}[H]
    \centering
    \caption{Yearly R\&D expenditures for selected countries in East Asia and Europe: True Values vs. Estimates.}
    \label{fig:predict_rd_expenditure_4}    
    \begin{subfigure}{0.45\textwidth}
        \centering
         {\includegraphics[width=\textwidth]{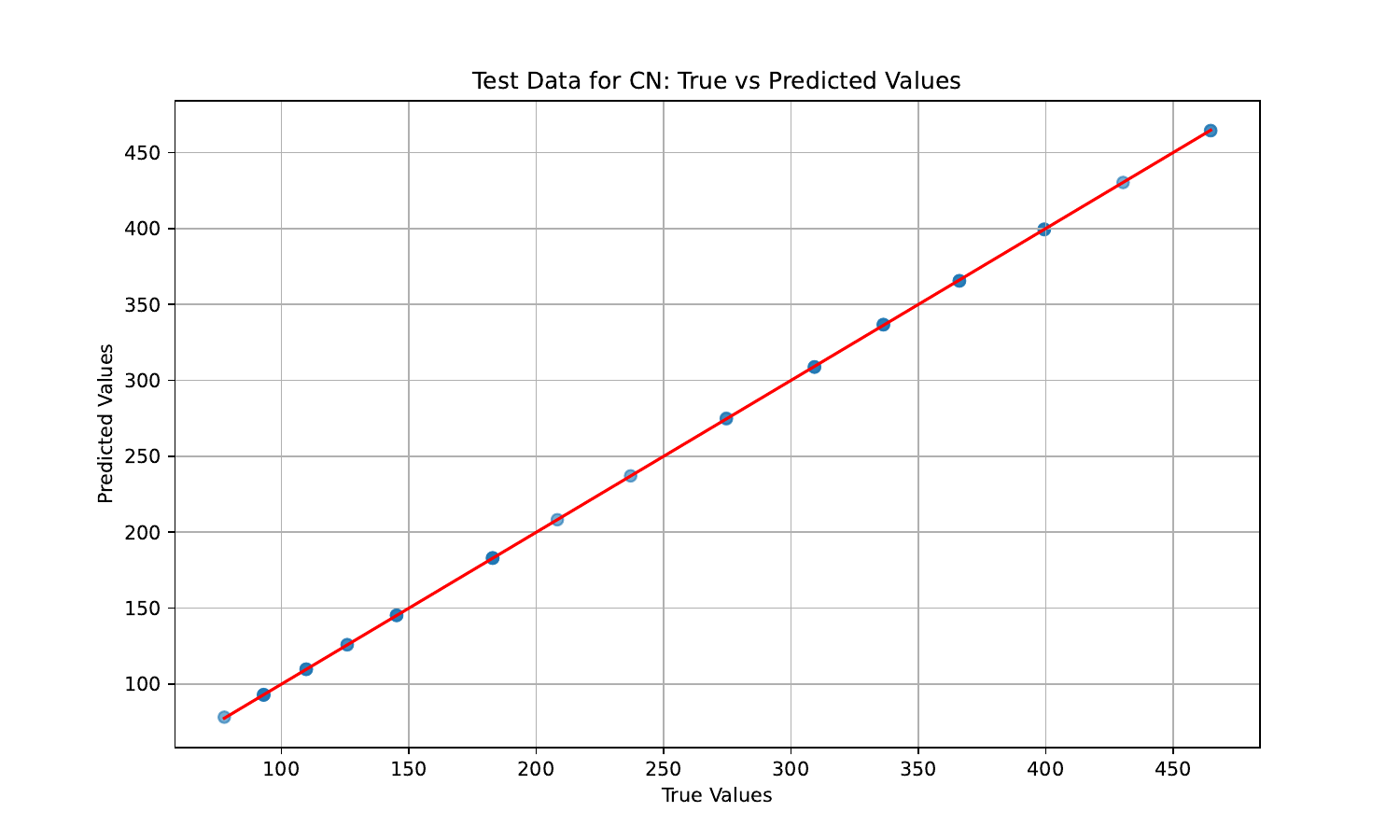}}
        \caption{China}
        \label{fig:cn_predict}
    \end{subfigure}\hfill
    \begin{subfigure}{0.45\textwidth}
        \centering
         {\includegraphics[width=\textwidth]{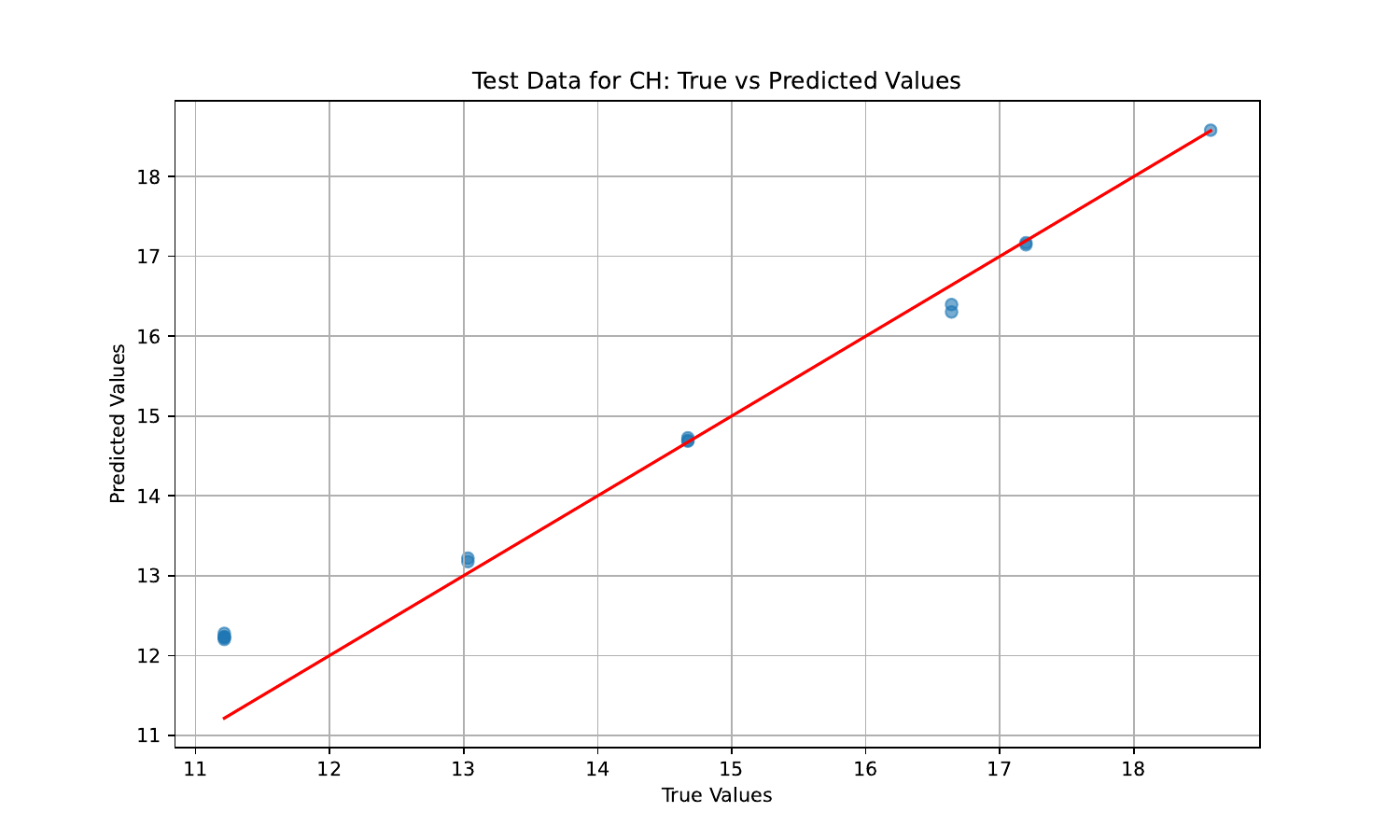}}
        \caption{Switzerland}
        \label{fig:ch_predict}
    \end{subfigure}
\end{figure}

Similarly, in the next batch of figures \ref{fig:predict_rd_expenditure_1}, \ref{fig:predict_rd_expenditure_2}, \ref{fig:predict_rd_expenditure_3} and \ref{fig:predict_rd_expenditure_4}, we present the out-of-sample performance of the prediction model, for each country, in \textit{AGT} configuration, as the outperformer.

%%%%%%%%%%%%%%%%%%%%%%%%%%%%%%%%%%%%%%%%%%%%%%%%%%%%%%%%%%%%%%%%%%%%%%%%
%%%%%%%%%%%%%%%%%%%%%%%%%%%%%%%%%%%%%%%%%%%%%%%%%%%%%%%%%%%%%%%%%%%%%%%%
\newpage
\section{Interpolating R\&D expenditures at a higher frequency via Perturbed/Corrupted input}
\label{app:interpolate_appendix_corruptedinput}
\begin{figure}[!htb]
    \centering
    \caption{Cumulative Estimated Monthly R\&D Expenditures over time for the US.}
    \label{fig:us_bem}    
     {\includegraphics[width=0.75\linewidth]{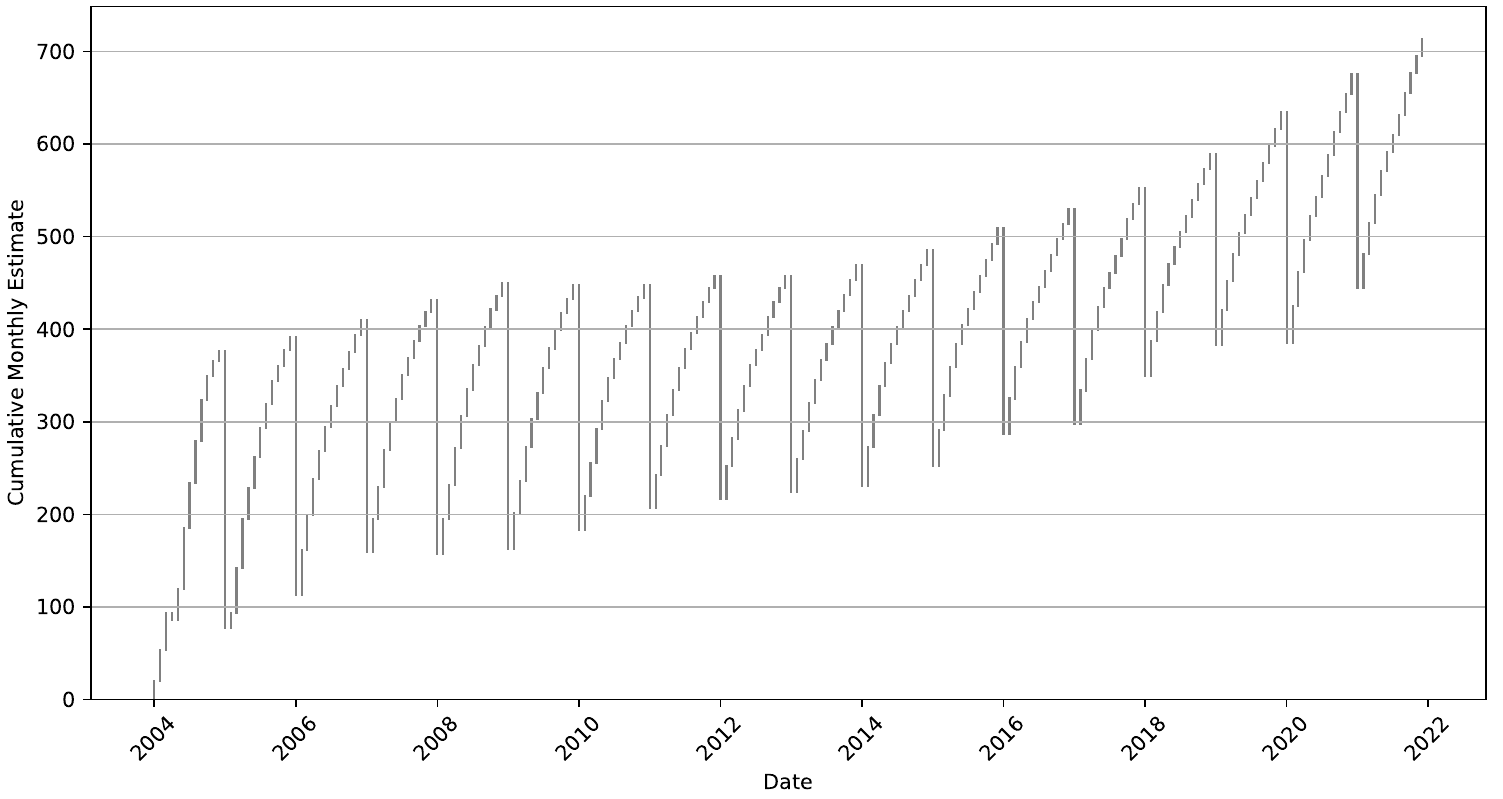}}
\end{figure}

% We introduce a series of approaches for estimation of R\&D expenditure values at higher frequency, i.e. on a monthly-basis, and compare their results to evaluate their performance in different scenarios. It should be noted that for accomplishing the nowcasting at higher frequency, only use of supervised machine learning would not be sufficient since we do not have labeled data anymore – no true values available.
As an alternative approach for interpolating R\&D expenditures, we elaborate on broadening our model by perturbing the input. We leverage on a trained model for yearly prediction, as a forward pass to have distribution of estimated values with respect to various input levels, and then develop a backward pass to breakdown yearly values and assign the resulted values to each month as its estimated contribution to the full-year expenditure. 
The model on the forward pass, is similar to the nowcasting model developed for yearly frequency in Section \ref{sec:model_annual}, in particular the \textit{AGT} configuration: 
\begin{equation*}
y_{t, i} = f^{(m)}\left(\boldsymbol{\mathscr{S}}_{t-\tau, i}^{(m)},\ \mathbf{C}_{j, i}^{(m)} ; \ \boldsymbol{\theta}^{(m)}\right) + \ \varepsilon_{t, j, i}^{(m)}
\end{equation*}
% \begin{equation}
% y_{t, i} = f^{(m)}\left(X_{t, j, i}^{(m)},\ C_{j, i}^{(m)} ; \ \theta\right) + \ \varepsilon_{t, j, i}^{(m)}
% \end{equation}
Since the \textit{AGT} model stands out as the top performer, according to the results in Section \ref{sec:results}, we build up on that with one modification on the aggregation approach for the annual GT values, as follows:  
\begin{equation*}
\boldsymbol{\mathscr{S}}_{t-\tau, i} = \left[\begin{array}{lll}
\mathbf{\Tilde{s}}_{t, i} \ \ldots \ \mathbf{\Tilde{s}}_{t-\tau, i}
\end{array}\right]
\end{equation*}
\begin{equation*}
\mathbf{\Tilde{s}}_{t-\tau, i} = \left[\begin{array}{lll}
\Tilde{s}_{1, t-\tau, i} \ \ldots \ \Tilde{s}_{k_{s}, t-\tau, i}
\end{array}\right]
\end{equation*}
\begin{equation*}
\Tilde{s}_{k_{s}, t-\tau, i} = \sum_{j=0}^{11} {s}_{k_{s}, t-\tau, j, i}^{(m)}
\end{equation*}
After training the model in this setting, we run the model over different input levels, which can be considered as corrupted versions of the original inputs, to get a distribution of outputs  that are diverged from the original predictions, accordingly. In other words, we create different sub-sets for $\mathbf{\Tilde{s}}_{t, i}$ in format of $\mathbf{\Tilde{s}}_{t-j/12, i}^{(m)}$ in which we sum only GT values up to the values of a particular month (the lower bound varies accordingly). $j$ indicates a particular month in which the nowcasting is being done, and it takes values of $j = 0, ..., 11$ accordingly; and (m) superscript reflects the monthly dynamics and sampling frequency, in accordance to the yearly frequency model (Section \ref{sec:model_annual}).
Following the forward pass across all months and for each country, we proceed to the backward pass phase. This phase begins with the last month of the year (December, where $j=0$) and involves comparing its estimated target value with that of the preceding month. This comparison is unique in that the preceding month's input vector contains one less month of data. This backward pass continues sequentially for each month, effectively covering the entire year and also all available years. The difference observed for each month, referred to as the discrepancy, is interpreted as the estimated contribution of that month to the overall model output.
\newline
As a preliminary result to the higher-frequency approach, figure \ref{fig:us_bem} depicts estimated monthly R\&D expenditure over time for the US.

\end{appendix}